\documentclass[10pt,english]{article}
\usepackage[T1]{fontenc}
\usepackage[latin1, utf8]{inputenc}
\usepackage{geometry}
\usepackage{babel}
\usepackage{mathtools}
\usepackage{amsmath}
\usepackage{amssymb}
\usepackage{caption}
\usepackage{subfigure}
\newcommand{\sout}[1]{\unskip}
\usepackage[table]{xcolor}
\usepackage{comment}

\usepackage{listings} 
\usepackage{here}
\usepackage[authoryear]{natbib}
\usepackage[unicode=true,
 bookmarks=false,
 breaklinks=false,pdfborder={0 0 1},backref=section,colorlinks=false]
 {hyperref}
\allowdisplaybreaks
\makeatletter

\usepackage{psfrag}\usepackage{epsf}\usepackage{enumerate}
\usepackage{url}
\usepackage{array}
\usepackage{bm}
\usepackage{multirow}

\usepackage{makecell} 

\usepackage{xr-hyper}

\usepackage{pifont}
\newcommand{\cmark}{\ding{51}}%
\newcommand{\xmark}{\ding{55}}%

\newcommand*{\addFileDependency}[1]{
  \typeout{(#1)}
  \@addtofilelist{#1}
  \IfFileExists{#1}{}{\typeout{No file #1.}}
}

\newcommand*{\myexternaldocument}[1]{%
    \externaldocument{#1}%
    \addFileDependency{#1.tex}%
    \addFileDependency{#1.aux}%
}

\myexternaldocument{supplement}

\def\spacingset#1{\renewcommand{\baselinestretch}%
{#1}\small\normalsize} \spacingset{1}

\usepackage{babel}

\usepackage{tikz}
\usetikzlibrary{fit,positioning}

\usepackage{amsthm, amsmath, amssymb}

\newtheorem{assumption}{Assumption}
\newtheorem{theorem}{Theorem}
\newtheorem*{theorem-non}{Theorem}

\newtheorem{definition}{Definition}

\newcommand\independent{\protect\mathpalette{\protect\independenT}{\perp}}
\def\independenT#1#2{\mathrel{\rlap{$#1#2$}\mkern2mu{#1#2}}}

\newcommand{\pr}{P}

\newcommand{\R}{\mathbb{R}}

\newcommand{\T}{\mathrm{\scriptscriptstyle T}}

\newcommand{\ipw}{\mathrm{IPW}}

\newcommand{\eff}{\mathrm{eff}}

\newcommand{\ipsw}{\mathrm{IPSW}}
\newcommand{\strat}{\mathrm{strat}}

\newcommand{\bg}{\bm{g}}
\newcommand{\bEta}{\bm{\eta}}

\newcommand{\EE}[1][]{\mathbb{E}\left[#1\right]}

\usepackage{colortbl, hhline} 

\usepackage{xcolor}

\newcommand{\modif}[1]{{\color{black} #1}}
\newcommand{\modiff}[1]{{\color{black} #1}}

\usepackage[colorinlistoftodos,textwidth=2.1cm]{todonotes}

\makeatother
\begin{document}
\title{\textbf{Causal inference methods for combining randomized
trials and observational studies: a review}}
\author{B\'{e}n\'{e}dicte Colnet$^{1,2,7 \star}$,
Imke Mayer$^{3,\star}$,
Guanhua Chen$^{4}$, Awa Dieng$^{5}$, Ruohong Li$^{6}$,\\ Ga\"{e}l Varoquaux$^{1}$, Jean-Philippe Vert$^{5}$, Julie Josse$^{7, \diamond}$, Shu Yang$^{8, \diamond}$}

\date{
$^1$ Soda project-team, INRIA Saclay, France.\\ 
$^2$ Centre de Math\'{e}mathiques Appliqu\'{e}es, Institut Polytechnique de Paris, Palaiseau, France.\\
$^3$ Institute of Public Health, Charit\'e -- Universit\"atsmedizin Berlin.\\
$^4$ Department of Biostatistics and Medical Informatics, University of Wisconsin-Madison.\\
$^5$ Owkin, Paris, France.\\
$^6$ Department of Biostatistics, Indiana University School of Medicine and Richard M. Fairbanks School of Public Health.\\
$^7$  Premedical project team, INRIA Sophia-Antipolis, Montpellier, France.
\\
$^8$ Department of Statistics, North Carolina State University.\\
$\star$ are first co-authors and $\diamond$ are corresponding authors
}

\maketitle
\noindent 







With increasing data availability, causal effects can be
evaluated across different data sets, both randomized controlled trials (RCTs) and
observational studies. RCTs isolate the effect of the
treatment from that of unwanted (confounding) co-occurring effects but
they may \modiff{suffer from unrepresentativeness}\sout{struggle with inclusion biases}, 
and thus lack external
validity. On the other hand, large observational samples are often more
representative of the target population but can conflate confounding
effects with the treatment of interest. 
In this paper, we review the growing literature on methods for causal
inference on combined RCTs and observational studies,
striving for the best of both worlds.
We first discuss identification and estimation methods that
improve generalizability of RCTs using the representativeness
of observational data. Classical estimators include weighting, difference between conditional outcome models, and doubly robust estimators.
We then discuss methods that combine RCTs and observational data to \modiff{either ensure uncounfoundedness of the observational analysis or to} improve (conditional) average treatment effect estimation\sout{, handling possible unmeasured confounding in the observational data}. 
We also connect and contrast works developed in both the potential outcomes literature and the structural causal model literature. 
Finally, we compare the main methods using a simulation study and real world data to analyze the effect of tranexamic acid on the mortality rate in major trauma patients.\sout{Code to implement many of the methods is provided.} \modiff{A review of available codes and new implementations is also provided.}
\vspace{12pt} \vfill{}


\section{Introduction\label{sec:Introduction}}

\modif{\textit{Experimental data}, collected through carefully designed
and randomized protocols, are usually considered the gold standard approach
for assessing the causal effect of an intervention or a treatment on an
outcome of interest. In particular, 
the intensive use of randomized controlled trials (RCTs) grounds
the so-called ``evidence-based medicine'', a keystone of modern medicine. 
In an RCT, the treatment allocation is under control, ensuring a
\emph{balanced} distribution of treated and control individuals; as a consequence, simple estimators can be used to
measure the treatment effect, e.g., with the difference in mean effect between the treated and control individuals \citep{imbens2015causal}. Still, RCTs come with practical drawbacks such as cost and time, but also with methodological issues such as restrictive inclusion/exclusion criteria which can lead to a trial sample that differs markedly from the population potentially eligible for the treatment. Therefore, the findings from RCTs can lack generalizability to a target population of interest. This concern is related to the aim of \emph{external validity}, central in medical research \citep{concato2000rctlimits, rothwell2005external, glasgow2006generalization, frieden2017beyondrct} policy research \citep{martel2010theory,deaton2018understanding,deaton2019randomization, Hongseok2022worstcase}, and other fields such as advertising \citep{gordon2019advertising}.}

\modif{In contrast, \textit{observational data} -- collected without
systematically designed interventions, such as disease registries,
cohorts, biobanks, epidemiological studies, or electronic health records
-- are promising as they are readily available, include large and
representative samples, and are less cost-intensive than RCTs. However,
there are often concerns about the quality of these ``big data'', given
that the lack of a controlled experimental intervention opens the door to
\textit{confounding bias}. This concern is referred to as a lack of
\textit{internal validity}. Under assumptions such as unconfoundedness it is possible to estimate a causal treatment effect from observational data. In practice, methods such as matching, inverse propensity weighting (IPW), or augmented IPW (AIPW) are used \citep{imbens2015causal}.
Even when a confounder is unobserved, solutions exist at the price of additional assumptions, for example the \textit{front-door criterion} \citep{pearl1993bayesian}, instrumental variables \citep{angrist_etal_JASA1996, hernan2006instrumental, imbens2014instrumental}, and sensitivity analysis \citep{cornfield1959smoking, RosenbaumRubin1983sensitivity, imbens2003sensitivity}.}

Combining information gathered from experimental and
observational data opens the door to new tools for, 
\modif{
\begin{enumerate}
    \item \modiff{accounting for the lack of representativeness of RCT}, as observational data can constitute an external representative sample of a target population of interest;
    \item \modif{making}\sout{the potential to make} observational evidence more credible using RCT \modiff{to ground observational analysis, such as detecting a} confounding bias; 
    \item improving statistical efficiency, for example to better estimate heterogeneous treatment effects as RCTs are often under-powered in such settings.
\end{enumerate}
}

As of today, there is abundant 
literature about the different ways and purposes of combining both sources of information.  
Terms used to refer to similar problems are \textit{generalizability}
\citep{cole2010generalizing,stuart2011use,hernan2011compound,tipton2013improving,o2014generalizing,stuart2015assessing,keiding2016perils,dahabreh2019extending,dahabreh2018generalizing, buchanan2018generalizing, CinelliGeneralizing2019}, \textit{representativeness} \citep{campbell1957validity},
\textit{external validity} \citep{rothwell2005external, stuart2018generalizability, westreich2018hierarchy}, \textit{transportability}
\citep{pearl2011transportability,rudolph2017robust, westreich2017transportability}, \modiff{\textit{recoverability} \citep{Bareinboim2012Controlling, Bareinboim2014Recovering}} and finally \textit{data fusion} 
\citep{bareinboim2016causal}; \modiff{this review will explain the commonalities or differences between the terminologies}. They have connections to inference from non-probability samples in survey sampling \citep{yang2020doubly,yang2020statistical} and to the covariate shift problem in machine
learning \citep{sugiyama2012machine}. 
 This problem of data integration for causal inference is tackled by two main bodies of
literature, namely the potential outcomes (PO)
framework \citep{splawa1990application,rubin1974estimating}, and the work
on structural causal models (SCM) using directed acyclic graphs (DAGs),
pioneered by \citet{pearl1995causal} and his collaborators.\\
 
 \modif{The present paper reviews this literature on combining experimental and observational data.
Section~\ref{sec:motivatingexample} introduces the notations from the PO literature, as well as the \modiff{common designs}. Section~\ref{sec:noWnoY} details how an observational sample can be used to generalize RCT findings to another population \modiff{point (a)}. We detail the corresponding identifiability assumptions and present the main estimation methods that have been suggested to account for distributional shifts. In this section, only baseline covariates are required in the observational data.
In Section~\ref{sec:specialcase}, we consider the case where observational data also contain treatment and outcome data. This setting in particular provides the opportunity to tackle different scientific questions such as hidden confounding or statistical efficiency (\modiff{points (b) and (c)}).
In Section~\ref{sec:SCM}, we present the SCM
literature, using different notations and ways to formulate assumptions, thus capturing richer and more diverse identifiability scenarios.
In Section~\ref{sec:soft}, we first present existing implementations and software and then we illustrate the properties of the generalization estimators on simulated data with new implementations. 
In  Section~\ref{sec:dataanalysis}, we apply the various methods presented in Section~\ref{sec:noWnoY} on a medical application involving major trauma patients. The aim of this study is to assess the effect of the drug tranexamic acid on mortality in head trauma patients. Both an RCT (the CRASH-3 trial) and an observational database (the Traumabase registry) are available. In this section, we also review methods for addressing data quality issues such as missing values.}

\section{Problem setting} \label{sec:motivatingexample}


\subsection{\label{sec:basic setup} Notations in the PO framework}

Each individual in the RCT or observational population is described by $(X,Y(0),Y(1),A,S)$, a random tuple with distribution $P$, where $X$ is a $p$-dimensional vector of covariates, $A$ the binary treatment assignment (with $A=0$ for the control and $A=1$ for the treated individuals), $Y(a)$ is the binary or continuous outcome had the subject been given treatment $a$ (for $a\in\{0,1\}$), and $S$ a binary variable indicating trial eligibility and willingness to participate\footnote{Note that in the literature, $S$ can have a slightly different meaning, for example other works use two separate indicators, one for participation and one for eligibility \citep{nguyen2018sensitivitybis, dahabreh2018generalizing}.}. 
We model the individuals belonging to an RCT sample of size $n$ and to an
observational data sample of size $m$ by $n+m$ independent random tuples:
$\{X_i,Y_i(0),Y_i(1),A_i,S_i\}_{i=1}^{n+m},$ where the RCT samples
$i=1,\ldots,n$ are identically distributed according to
$P(X,Y(0),Y(1),A,S
\mid S=1)$, and the observational data samples $i=n+1,\ldots,n+m$ are
identically distributed according to $P(X,Y(0),Y(1),A,S)$. The sampling mechanisms of the RCT
and observational samples are assumed to be independent, which corresponds
to a so-called non-nested design as explained in
Section~\ref{subsec:nested-vs-nonnested}. We also denote
$\mathcal{R}=\{1,\ldots,n\}$ the set of indices of units observed in the
RCT study, and $\mathcal{O}=\{n+1,\ldots,n + m\}$ the set of indices of units observed in the observational study. For each RCT sample $i\in \mathcal{R}$, we observe $(X_i,A_i,Y_i,S_i=1)$, while for observational data $i\in \mathcal{O}$, we consider two settings: \textit{(i)} we only observe the covariates $X_i$ (Section~\ref{sec:noWnoY}), \textit{(ii)} we also observe the treatment and outcome $(X_i,A_i,Y_i)$ (Section~\ref{sec:specialcase}).\\

\modiff{In this review we consider the absolute difference, and do not consider other contrast measures\footnote{Considering other measures such as the ratio or odds ratio can have an impact on the assumptions considered, for example in generalization \citep{Huitfeldt2019EffectHeterogeneity}. As the large majority of the literature is focused on the absolute difference, this review reflects the practices, and therefore considers the absolute difference.}.} Doing so, we denote respectively by $\tau(x)$ and  $\tau_1(x)$ the conditional average treatment effect (CATE) in the observational population and the RCT population:
$$
\forall x\in\R^p\,,\quad \tau(x) = \mathbb{E}[Y(1) - Y(0) \mid X=x]\,,\quad  \tau_1(x) = \mathbb{E}[Y(1) - Y(0) \mid X=x, S=1]\,.
$$
We also denote $\tau$ and $\tau_1$ the population average treatment effect (ATE) in the observational population and the RCT \modiff{one}:
$$
\tau = \mathbb{E}[Y(1) - Y(0)] = \mathbb{E}[\tau(X)]\,,\quad \tau_1= \mathbb{E}[Y(1) - Y(0) \mid S=1]\,,
$$
where the population ATE can be different from the RCT ATE, i.e.,  $\tau \neq \tau_1$ \modiff{in general}.
We denote respectively by $e(x)$ and $e_1(x)$ the propensity score in the observational population and in the RCT population:
$$
e(x) = P(A=1 \mid X=x)\,,\quad e_{1}(x) = P(A=1 \mid X=x, S=1)\,,
$$
\modif{where $e_{1}(x)$ is usually known in an RCT.}
We also denote by $\mu_a(x)$ and $\mu_{a,1}(x)$ the conditional mean outcome under treatment $a\in\{0,1\}$ in the observational population and in the RCT population, respectively:
$$
\mu_a(x) = \EE[Y(a) \mid X=x]\,,\quad \mu_{a,1}(x) = \EE[Y(a) \mid X=x, S=1]\,.
$$

Finally, we denote by $\alpha(x)$ the conditional odds that an individual with covariates $x$ is in the RCT or in the observational sample:
$$
\alpha(x) = \frac{\mathbb{P}(i\in\mathcal{R}\mid \exists i\in\mathcal{R}\cup\mathcal{O}, X_i=x)}{\mathbb{P}(i\in\mathcal{O}\mid \exists i\in\mathcal{R}\cup\mathcal{O}, X_i=x)} = \frac{\pi_{\mathcal{R}}(x)}{\pi_{\mathcal{O}}(x)} = \frac{\pi_{\mathcal{R}}(x)}{1 - \pi_{\mathcal{R}}(x)}\,,
$$

where $\pi_{\mathcal{R}}(x)$ (resp. $\pi_{\mathcal{O}}(x) $) is the probability that an individual with covariates $x$ known to be in the concatenated data (RCT sample and observational sample) is in the RCT (resp. in the observational sample).
In the literature another widely used quantity is the selection score -- or sampling propensity score \modiff{(in particular this name was proposed by \cite{tipton2013improving})} -- denoted $\pi_S(x)$ and defined as 
$$
\pi_S(x) = \mathbb{P}(S=1 \mid X=x)\,.
$$
Because $\pi_S(x)$ is the probability of \modiff{being \emph{sampled} in the trial given covariates} values $x$, it is different from
$\pi_{\mathcal{R}}(x)$. $\pi_{\mathcal{S}}(x)$ is often used 
with a nested design (see Section~\ref{subsec:nested-vs-nonnested} for a definition), but is not of interest in our setup \modiff{(non-nested design)} because it cannot be identified. Indeed,
\begin{equation*}
    \pi_S(x) 
    = \mathbb{P}(S=1)\times \frac{\mathbb{P}(X=x \mid S=1)}{\mathbb{P}(X=x)}
    = \mathbb{P}(S=1)\times\frac{\mathbb{P}(X_i = x \mid i\in\mathcal{R})}{\mathbb{P}(X_i = x \mid i \in\mathcal{O})}
    = \underbrace{\mathbb{P}(S=1)}_\textrm{Not known}\,\times \frac{n}{m} \underbrace{\frac{\pi_{\mathcal{R}}(x)}{\pi_{\mathcal{O}}(x)}}_\textrm{$= \alpha(x)$}.
\end{equation*}
\modiff{Detailed derivations can be found in the appendix (see Section~\ref{sec:appendixiden}).}
\sout{In our setup (non-nested design)} \modiff{The quantity $\mathbb{P}(S=1)$ is unknown because}, \sout{we do not know whether} individuals in the target population could have participated in the RCT or not:
$S$ can be equal to $1$ and $0$ in the observational sample but this information is not known. Table~\ref{fig:exdesign} illustrates the considered type of dat\modiff{a}, and Table~\ref{tab:notations} summarizes the notations.

\begin{table}[!h]
\caption{\textbf{Illustration of data structure} of RCT data (Set $\mathcal{R}$) and observational data (Set $\mathcal{O}$) with covariates $X$, trial eligibility $S$, binary treatment $A$ and outcome $Y$. Left: with observed outcomes, Right: with potential outcomes. Note that the $S$ covariate can be either $0$ or $1$ in the observational data set \sout{(while usually unknown from the raw data set in the non-nested design)} \modiff{(it is unknown in the non-nested design, hence the NA for not available)}, and is always equal to $1$ for observations in the RCT. In the nested design (cf. Appendix~\ref{sec:appendixnested}), $S=0$ for all individuals in the observational data set.
\label{fig:exdesign}}
\footnotesize
\setlength{\tabcolsep}{5pt}
\begin{tabular}{ |c|c|c|ccc|c|c| } 
 \hline
 &  & &\multicolumn{3}{c|}{\scriptsize Covariates} & \scriptsize Treatment & \scriptsize Outcome \\
& $S$ & Set &$X_1$ & $X_2$ & $X_3$ & $A$ & $Y$ \\ 
 \hline
1&1 & $\mathcal{R}$ &1.1 & 20 & F & 1 &  1 \\ 
& 1 & $\mathcal{R}$ &-6 & 45 & F & 0  & 1  \\ 
$n$& 1& $\mathcal{R}$ &0 & 15 & M & 1 & 0  \\ 
$n+1$&  \texttt{NA} & $\mathcal{O}$ && $\dots$ & & $\dots$ & $\dots$   \\
& \texttt{NA} & $\mathcal{O}$ &-2 & 52 & M & 0 & 1  \\
& \texttt{NA} & $\mathcal{O}$ &-1 & 35 & M & 1 &    1 \\
$n+m$&  \texttt{NA}  & $\mathcal{O}$ &-2 & 22 & M & 0 & 0    \\
 \hline
\end{tabular}
\begin{tabular}{ |c|c|ccc|c|cc|} 
 \hline
   & &\multicolumn{3}{c|}{\scriptsize Covariates} & \scriptsize Treatment & \multicolumn{2}{c|}{\scriptsize Outcome(s)}  \\
  $S$& Set & $X_1$ & $X_2$ & $X_3$ & $A$ & $Y(0)$ & $Y(1)$ \\ 
 \hline
1 & $\mathcal{R}$ &1.1 & 20 & F & 1 & \texttt{NA} & 1  \\ 
 1 & $\mathcal{R}$ &-6 & 45 & F & 0  & 1 & \texttt{NA}  \\ 
 1 & $\mathcal{R}$ &0 & 15 & M & 1 &  \texttt{NA} & 1 \\ 
 \texttt{NA} & $\mathcal{O}$ && $\dots$ & & $\dots$ & $\dots$  & $\dots$ \\
  \texttt{NA} & $\mathcal{O}$ &-2 & 52 & M & 0 & 1 & \texttt{NA} \\
\texttt{NA} & $\mathcal{O}$ &-1 & 35 & M & 1 &  \texttt{NA} & 1 \\
 \texttt{NA} & $\mathcal{O}$ &-2 & 22 & M & 0 & 0 & \texttt{NA}   \\
 \hline
\end{tabular}
\end{table}

\normalsize

\begin{table}[!h]
\caption{List of notations.}
\label{tab:notations} {\footnotesize{}}%
\begin{tabular}{ll}
\hline 
{\footnotesize{}Symbol } & {\footnotesize{}Description }\tabularnewline
\hline 
{\footnotesize{}$X$ } & {\footnotesize{}Covariates (also known as baseline covariates when
measured at inclusion of the patient) }\tabularnewline
{\footnotesize{}$A$ } & {\footnotesize{}Treatment indicator ($A=1$ for treatment, $A=0$
for control) }\tabularnewline
{\footnotesize{}$Y$ } & {\footnotesize{}Outcome of interest }\tabularnewline
{\footnotesize{}$S$ } & {\footnotesize{}Trial eligibility and willingness to participate if invited to ($S=1$ for eligibility,
$S=0$ for non-eligibility)} \tabularnewline
{\footnotesize{}
$n$ } & {\footnotesize{}Size of the RCT study }\tabularnewline
{\footnotesize{}$m$ } & {\footnotesize{}Size of the observational study }\tabularnewline
{\footnotesize $\mathcal{R}$} & {\footnotesize Index set of units observed
in the RCT study; $\mathcal{R}$=\{1,\ldots,n\} }\tabularnewline
{\footnotesize $\mathcal{O}$} & {\footnotesize Index set of units observed
in the observational study; $\mathcal{O}$=\{n+1,\ldots,n+m\} }\tabularnewline
{\footnotesize $\pi_{\mathcal{R}}(x)$} & {\footnotesize  
Probability that a unit in $\mathcal{R}\cup\mathcal{O}$ with covariate $x$ is in $\mathcal{R}$\modif{, defined as $\pi_{\mathcal{R}}(x) = \mathbb{P}(i\in\mathcal{R} \mid \exists i\in\mathcal{R}\cup\mathcal{O}, X_i=x)$}}\tabularnewline
{\footnotesize $\pi_{\mathcal{O}}(x)$} & {\footnotesize
Probability that a unit in $\mathcal{R}\cup\mathcal{O}$ with covariate $x$ is in $\mathcal{O}$\modif{, defined as $\pi_{\mathcal{O}}(x) = 1-\pi_{\mathcal{R}}(x)$}}\tabularnewline
{\footnotesize $\alpha(x)$} & {\footnotesize Conditional odds $\alpha(x) = \pi_{\mathcal{R}}(x) / \pi_{\mathcal{O}}(x)$}\tabularnewline
{\footnotesize{}$\tau$ } & {\footnotesize{}Population average treatment effect (ATE) defined
as $\tau=\EE[Y(1)-Y(0)]$}\tabularnewline
{\footnotesize{}$\tau_{1}$ } & {\footnotesize{}Trial (or sample) average treatment effect defined as $\tau_{1}=\EE[Y(1)-Y(0)\mid S=1]$}\tabularnewline
{\footnotesize{}$\tau(x)$ } & {\footnotesize{}Conditional average treatment effect (CATE) defined
as $\tau(x)=\EE[Y(1)-Y(0)\mid X=x]$}\tabularnewline
{\footnotesize{}$\tau_{1}(x)$ } & {\footnotesize{}Trial conditional average treatment effect defined
as $\tau_{1}(x)=\EE[Y(1)-Y(0)\mid X=x,S=1]$}\tabularnewline
{\footnotesize{}$e(x)$ } & {\footnotesize{}Propensity score defined as $e(x)=\mathbb{P}(A=1\mid X=x)$}\tabularnewline
{\footnotesize{}$e_{1}(x)$ } & {\footnotesize{}Propensity score in the trial defined as $e_{1}(x)=\pr(A=1\mid X=x,S=1)$, known by design} \tabularnewline
{\footnotesize{}$\mu_{a}(x)$ } & {\footnotesize{}Outcome mean defined as $\mu_{a}(x)=\EE[Y(a)\mid X=x]$
for $a=0,1$}\tabularnewline
{\footnotesize{}$\mu_{a,1}(x)$ } & {\footnotesize{}Outcome mean in the trial defined as $\mu_{a,1}(x)=\EE[Y(a)\mid X=x,S=1]$
for $a=0,1$}\tabularnewline
{\footnotesize{}$\pi_{S}(x)$ } & {\footnotesize{}Selection score defined as $\pi_{S}(x)=\pr(S=1\mid X=x)$}\tabularnewline
{\footnotesize{}$f(X)$ } & {\footnotesize{}Covariate distribution in the target population }\tabularnewline
{\footnotesize{}$f(X|S=1)$ } & {\footnotesize{}Covariate distribution conditional to trial-eligible
individuals ($S=1$) }\tabularnewline
\hline 
\end{tabular}
\end{table}

\subsection{\label{sec:studydesign} Study designs and goals}

\subsubsection{Nested and non-nested study designs}\label{subsec:nested-vs-nonnested}

\begin{figure}[H]
\begin{minipage}{.49\textwidth}
\includegraphics[width=\textwidth]{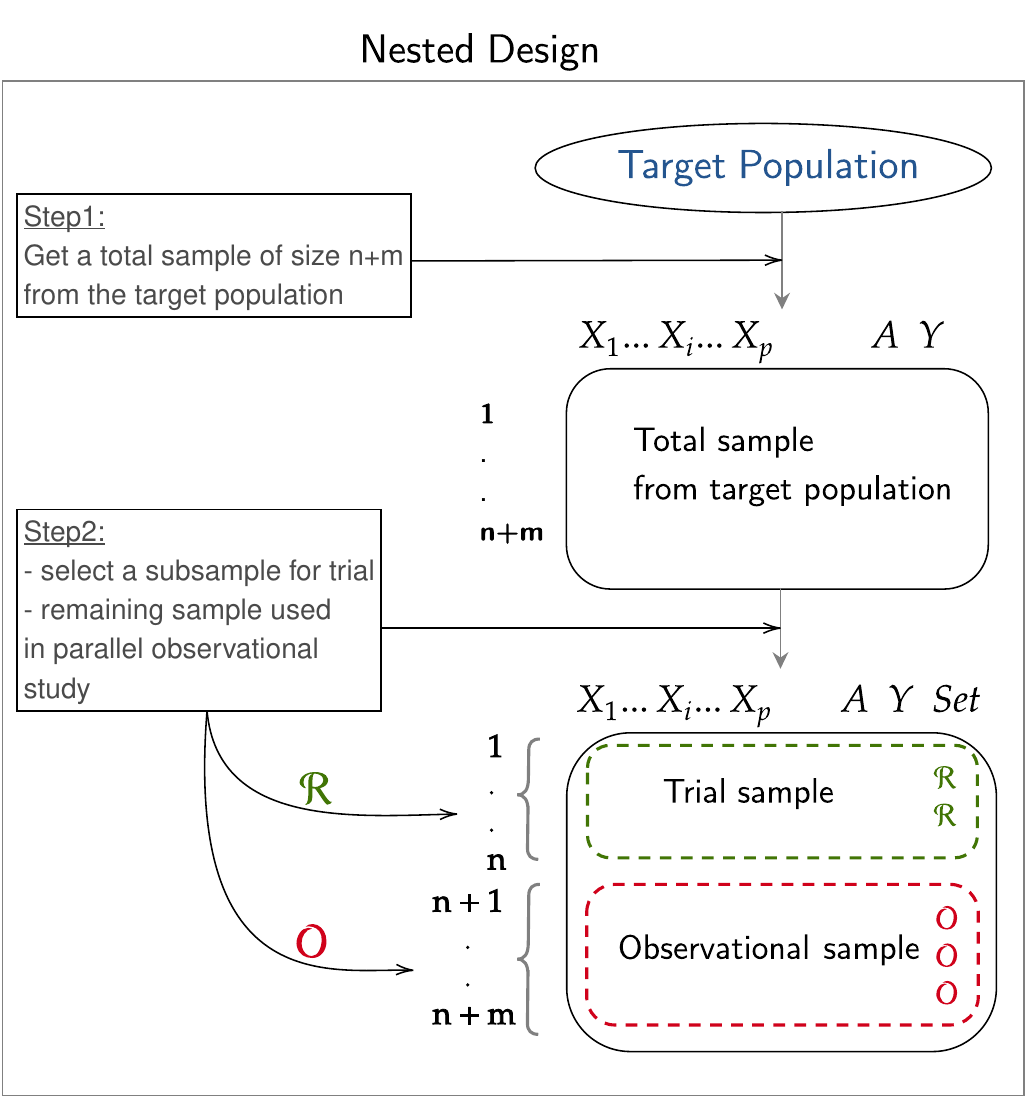}
\end{minipage}%
\hfill%
\begin{minipage}{.49\textwidth}
\includegraphics[width=\textwidth]{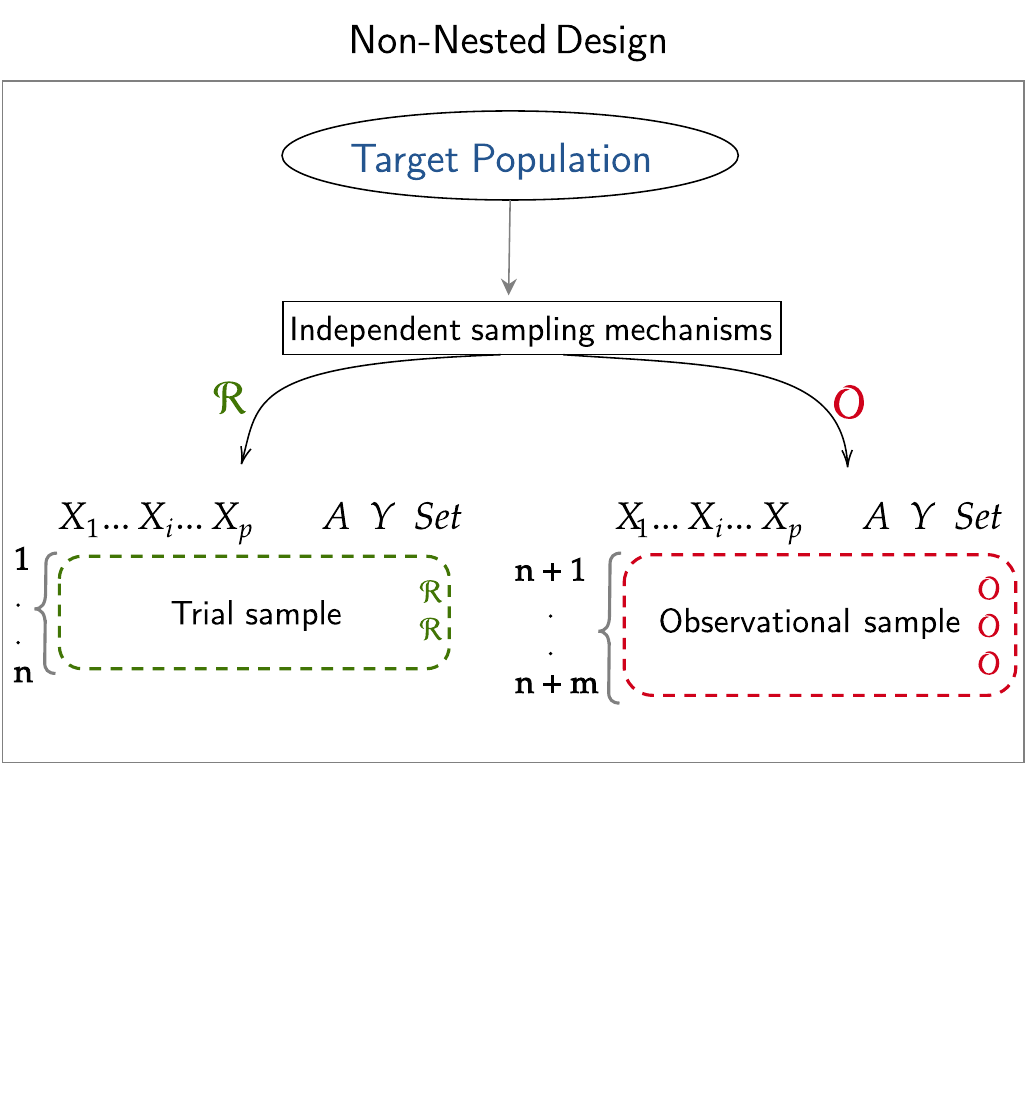}
\end{minipage}%

\caption{Schematics of the nested (left) and non-nested (right) designs, a similar schematic can be found in \cite{josey2021transporting}.} 
\label{fig:designs}
\end{figure}

Following \citet{dahabreh2019study} and \cite{dahabreh2019extending},
the study design to obtain the trial and observational samples can
be categorized into two types: \textit{nested} study designs and \textit{non-nested}
study designs as illustrated on Figure~\ref{fig:designs}. 
Designs imply different identifiability conditions and therefore estimators. \underline{This review focuses}
on what is called the \underline{non-nested design}, as the trial sample and the
observational sample are obtained separately\sout{ from the target
population(s)}. On the contrary the nested design involves a two-stage
nested sampling. For example it can correspond to an embedded trial in a
broader health system. As a concrete example one can mention the Women
Health Initiative, or the recent study on Medicaid where part of the
participants are randomized \citep{degtiar2021conditional}. In this
situation, data are not really combined as the overall data comes from one initial sampling in which two treatment assignment regimes (randomized or not) coexist. 
\modiff{The nested design estimators} are detailed in Appendix~\ref{sec:appendixnested}.

 
\modif{\subsubsection{Transportability, generalizability, and recoverability}}
Several terms are currently present in the literature to describe
the process of predicting the effect of the treatment from an RCT to
another population\modif{: generalization \citep{stuart2011use, buchanan2018generalizing, dahabreh2018generalizing}, transportability \citep{hernan2011compound, bareinboim2016causal, westreich2017transportability},  or recoverability \citep{Bareinboim2014Recovering}}.
\modif{Differences in the definitions can be found in the literature, underlying a specific design such as the existence of a common superpopulation or assumptions such as the support overlap between different populations.}
\modif{For example, \cite{dahabreh2020extending} highlights that several definitions are given,}
\begin{quote}
    \textit{We use the term generalizability when the target population coincides or is a subset of the trial-eligible population
and transportability when the target population includes at least some individuals who are not trial-eligible (and who,
by definition, cannot be trial participants) (others have proposed different definitions). }
\end{quote}

Due to different definitions in the literature, several terms
can be found to describe the same scientific goal. In this review, we
call \emph{generalization} the task that extends the RCT result to \textit{its}
larger population, where it was sampled with a bias (detailed in
Section~\ref{sec:noWnoY}). The SCM literature also uses different
terminologies corresponding to different assumptions --and corresponding
diagrams-- as detailed in Section~\ref{sec:SCM}. For example what is
called transportability refers to two distinct populations, and not
necessarily about different covariate supports as suggested by
\cite{dahabreh2020extending}. In particular, in this literature the task
that we study in Section~\ref{sec:noWnoY} is termed recoverability from a
sampling bias, rather than generalization. \modiff{This terminology has the merit of indicating that generalization can have a much broader coverage, including other types of problems.}
Note that granting
some assumptions about a common support or non-zero probability to be
sampled, then the two problems -- namely recovering from a sampling bias
and transportability -- rely on the same estimators and procedure, as highlighted in Section~\ref{sec:iden} and in \cite{pearl2015findings}.

\section{When observational data have no treatment and outcome information}\label{sec:noWnoY}

We start by considering the case where only the covariates from the observational study are available or used. We consider the observational data as a random sample from the target population. \modiff{Considering this set-up, the question tackled in this section is how to generalize or transport the trial findings toward a target population of interest}.\sout{, as the covariate information from a
larger population-based disease registry (e.g., national cancer database)
can be used to generalize findings of a trial.} Applied examples can be found in
\cite{dong2020integrative, lesko2016generalizingHIV, tipton2016smallsample, li2021generalizing,yang2022rwdintegrative}. In particular \cite{he2020generalizabilitypractice} review current practice, \modiff{revealing that generalization's implementation is still at the stage of prototyping without real usage for clinical and public health decisions yet.}

\subsection{Assumptions needed to identify the ATE on the target population}\label{sec:idenassump}

A fundamental problem in causal inference is that we can observe at most one of the potential outcomes for an individual subject. In order to identify nonetheless
the ATE from RCT and observational covariate data, we require some of the following assumptions.

\subsubsection{Internal validity of the RCT}

\begin{assumption}[Consistency]\label{a:consist} $Y=A\,Y(1)+(1-A)\,Y(0)$.
\end{assumption}
Assumption \ref{a:consist} implies that the observed outcome is the potential outcome under the actual assigned treatment.

\begin{assumption}[Randomization]\label{a:ign}$\{Y(0),Y(1)\}\independent A\mid S=1, X$
\end{assumption}
Assumption \ref{a:ign} corresponds to internal validity. 
It holds by design in a completely randomized experiment, where the treatment is independent of all the potential outcomes and covariates. \sout{Note that a slightly difference assumption such as $\{Y(0),Y(1)\}\independent A\mid S=1, X$ (i.e. stratified randomized trial based on a discrete $X$) would also work.} The more general case of conditional randomization is assumed throughout this review. \\
If Assumptions~\ref{a:consist} and \ref{a:ign} hold, then the RCT is said to be compliant. In addition, in an RCT, it is common that the probability of treatment assignment, 
$e_{1}(x)$, is known. In a complete randomized trial, the propensity score is fixed as a constant, and usually $e_1(x)=0.5$ for all $x$.

\subsubsection{Assumptions ensuring generalizability of the RCT to the target population}

The literature proposes different assumptions to \modiff{generalize trial's findings to a target population}. \sout{We list some assumptions.}\sout{, ranging from more stringent to weaker.}

\begin{assumption}[Ignorability assumption on trial participation]\label{a:trans-2}
 $\{Y(0), Y(1)\}\independent S \mid X$.  \citep{Hotz2005training, stuart2011use,tipton2013improving, hartman2015sate, buchanan2018generalizing, degtiar2021reviewgeneralizability, egami2021covariate}
\end{assumption}
A parallel can be made with the  \textit{strong ignorability condition} in causal inference with observational data (see Assumption \ref{asump:unconfounded} in Appendix), but applied to the sample selection rather than treatment assignment. \modiff{In other words, these assumptions require to control for all covariates being shifted and predictive of $Y$. We call shifted covariates, all the baseline covariates along which the two populations -- trial and target -- do not follow the same distribution.} 
\modiff{A weaker version of Assumption~\ref{a:trans-2}} can be found in \cite{dahabreh2018generalizing, dahabreh2020extending}:
\begin{assumption}[Mean exchangeability]\label{a:trans-3}
\sout{$\mathbb{E}[Y(a)\mid  X= x,S=1,A=a]=\mathbb{E}[Y(a)\mid  X=x,S=1]$ (mean exchangeability over treatment assignment) and }$\mathbb{E}[Y(a)\mid  X=x,S=1]=\mathbb{E}[Y(a)\mid  X=x]$ (mean exchangeability over trial participation), for all $x$ and $a=0,1 $.
\end{assumption}

Another assumption can be found, relying on the transportability of treatment effect rather than the potential outcomes.
\begin{assumption}[Sample ignorability for treatment effects - \cite{kern2016assessing, nguyen2018sensitivitybis}]
\label{a:trans-diff}
$Y(1) - Y(0) \independent S \mid  X$.
\end{assumption}
\modiff{A weaker version} can be found:
\begin{assumption}[Transportability of the CATE]\label{a:trans} $\tau_1(x)=\tau(x)$ for all $x$.
\end{assumption}
\modiff{To meet these last two assumptions}, one requires variables that are both \textit{treatment effects modifiers} and \textit{shifted}. \modiff{Epidemiologists often use the term ``effect modification'' to indicate that the treatment effect varies across strata of baseline covariates, such baseline covariates being treatment effect modifiers.} %
These assumptions are implied by Assumption~\ref{a:trans-2}, but this is not reciprocal \modiff{as all covariates predictive of the outcome are not necessarily treatment effect modifiers}. 
Note that a treatment effect modifier depends on the chosen scale, here we focus on the absolute difference, but if we had considered a risk ratio the variables being treatment effects modifiers would not be the same.
\modiff{Mathematical definitions of a treatment effect modifier are hard to find, but we quote one from \cite{VanderWeele2007FourTypes} for the absolute scale.

\begin{definition}[Treatment effect modifier]\label{def-VanderWeele2007FourTypes}
We say that a variable $X$ is a treatment effect modifier for the causal risk difference of $A$ on $Y$ if $X$ is not affected by $A$ and if there exist two levels of $A$, $a_0$ and $a_1$, such that $\mathbb{E}\left[ Y^{(a_1)}   \mid X = x\right]-\mathbb{E}\left[ Y^{(a_0)}   \mid X = x\right]$ is not constant in $x$.
\end{definition}
}

\sout{Assumption~\ref{a:trans-2} implies Assumption~\ref{a:trans-3} and Assumption~\ref{a:trans}, while Assumption~\ref{a:trans-diff} implies Assumption~\ref{a:trans}.} In this work, we only rely on Assumption~\ref{a:trans-diff} for identification formula. 
Finally a last assumption is needed, the \textit{positivity of trial participation} assumption.
\begin{assumption}[Positivity of trial participation\modif{, also called overlap}]\label{a:pos} \modiff{There exists a constant
$c>0$ such that, almost surely, \mbox{$\mathbb{P}(S=1\mid X)\ge c$.}}
\end{assumption}
Assumption~\ref{a:pos} requires adequate overlap of the covariate distribution
between the trial sample and the target population (in other words, all members of the target population have non-zero probability of being selected into the trial). 
Other formulation of this assumption can be found under the assumption of the target population's support included in the trial sample support \citep{nie2021covariate, Colnet2022Reweighting} 

\subsubsection{\modif{Identifications} formulas}\label{sec:iden}
Under Assumptions \ref{a:consist}, \ref{a:ign}, \ref{a:trans}, and \ref{a:pos} the ATE can be identified based on the following formulas (derivations in Appendix \ref{sec:appendixiden}): 
\begin{enumerate}
\item Reweighting formulation:
\begin{equation}
    \tau = \EE[\frac{n}{m \alpha(X)} \tau_1(X) \mid S=1] = \mathbb{E}\left[\frac{n}{m \alpha(X)} \left( \frac{A}{e_{1}(X)} - \frac{1-A}{1-e_{1}(X)} \right) Y\mid S=1 \right]\,. \label{eq:idweight}
\end{equation}
\sout{In a fully randomized RCT where $e_1(x)=0.5$ for all $x$, this formula further simplifies to}
Note that Equation~\ref{eq:idweight} can be understood as a transportability problem considering two distributions $P_1$ and $P$, and transporting evidence from population $P_1$ to population $P$, 
$$\tau = \mathbb{E}_{P}\left[\tau(X) \right]  = \underbrace{\int_{\mathcal{X}}\tau(x) \,f(x) \,dx}_\textrm{Integral on $P$} = \underbrace{\int_{\mathcal{X}}\tau_1(x) \frac{f(x)}{f_1(x)}\,f_1(x)\,dx}_\textrm{Integral on $P_1$} = \int_{\mathcal{X}}\tau_1(x) \frac{n}{m} \frac{1}{\alpha(x)}\,f_1(x)\,dx, $$
noting that $\alpha(x) =  \frac{\pr(i\in\mathcal{R}\mid \exists i\in\mathcal{R}\cup\mathcal{O}, X_i=x)}{\pr(i\in\mathcal{O}\mid \exists i\in\mathcal{R}\cup\mathcal{O}, X_i=x)} = \frac{\pr(i\in\mathcal{R})}{\pr(i\in\mathcal{O})}\times \frac{\pr(X_i = x \mid i\in\mathcal{R})}{\pr(X_i = x \mid i \in\mathcal{O})} = \frac{n}{m}\times \frac{f_1(x)}{f(x)},$ \modiff{and using the transportability assumption (see Assumption~\ref{a:trans}) stating that $\tau(x) = \tau_1(x)$.}

\item Regression formulation: 
\begin{equation} 
\label{eq:idereg}
\tau=\mathbb{E}\left[\mu_{1,1}\left(X\right)-\mu_{0,1}\left(X\right)\right]  \modiff{= \mathbb{E}\left[\tau_1\left(X\right)\right]  \,.}
\end{equation} 
\end{enumerate}

\begin{figure}[!h]
    \centering
    \includegraphics[width=0.55\textwidth]{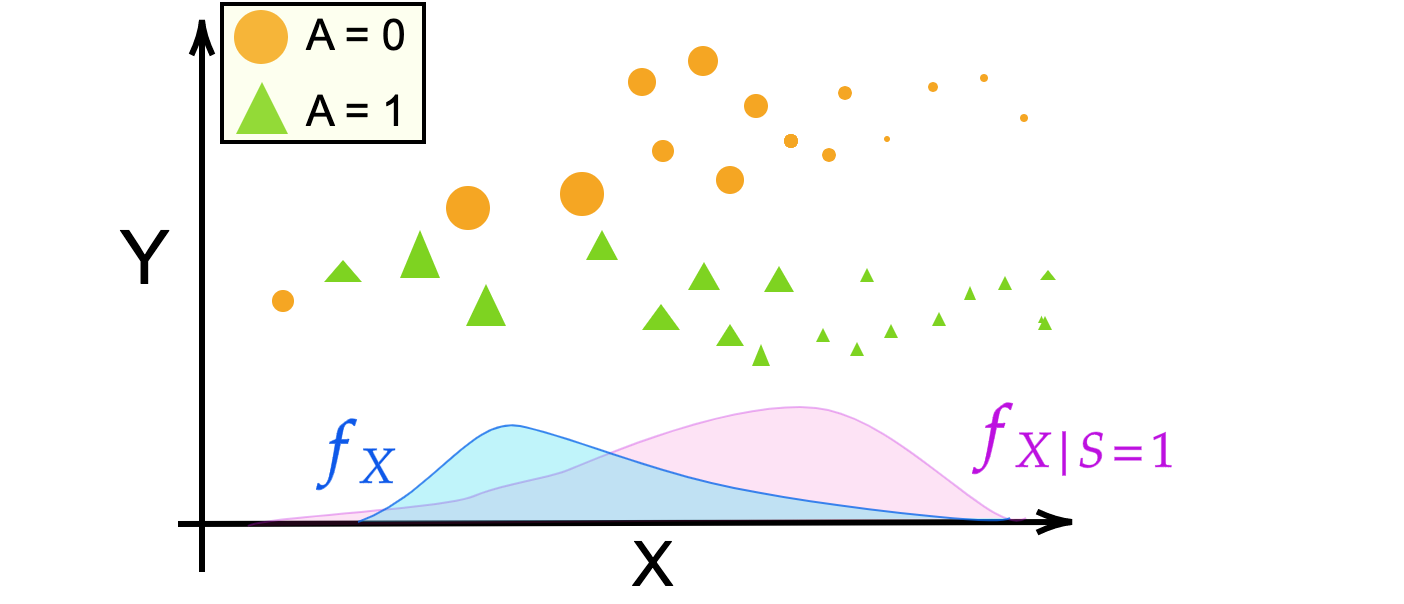} \hspace{-2cm}
    \includegraphics[width=0.55\textwidth]{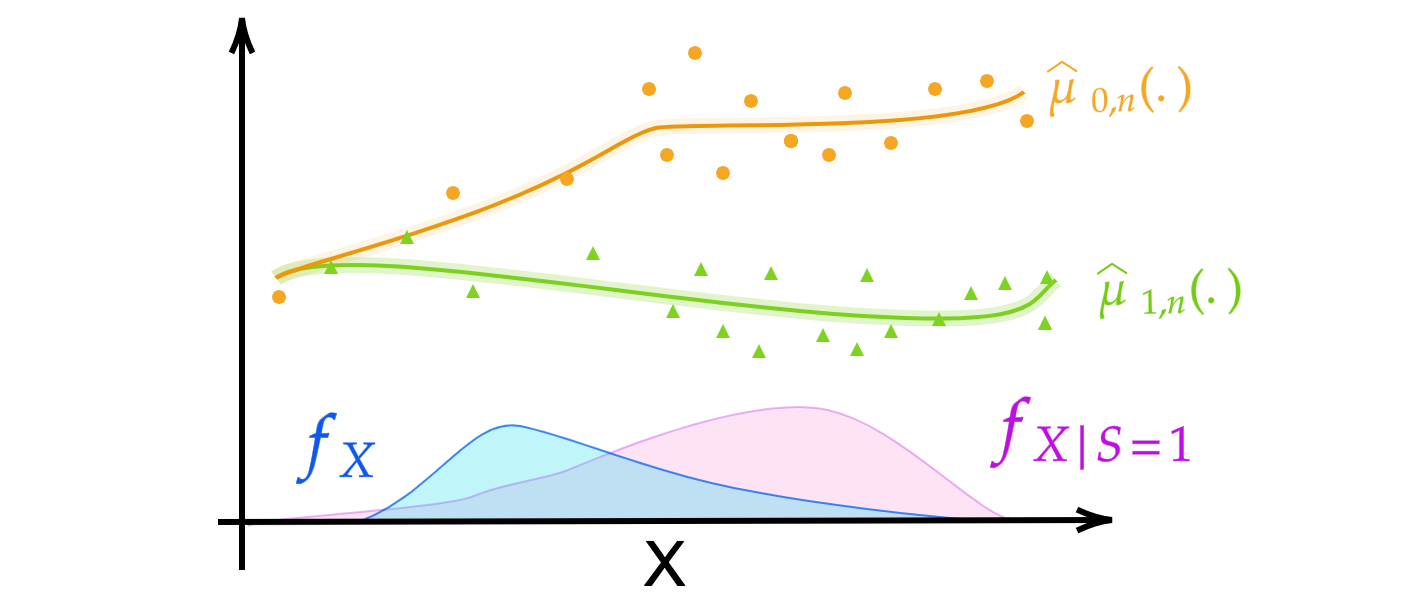} 
    \vspace{0.1cm}
    {\textbf{IPSW} \hspace{5.7cm} \textbf{G-formula}}
    \caption{\modif{\textbf{Illustrative schematics for the estimation strategies}: On this drawing the trial findings $\hat \tau_{1,n}$ would over-estimate the target treatment effect $\tau$ (on an absolute scale). On the left, the IPSW (Definition~\ref{def:ipsw}) strategy, relying on weighting the RCT observations; on the right, the plug-in g-formula (Definition~\ref{def:g-formula}) strategy, relying on modeling the response using the RCT observations. Notations are the same as introduced in Table~\ref{tab:notations}, i.e., $f_X$ ($f_{X\mid S = 1}$) denotes the density of the target (resp. trial) population, and $\hat \mu_{a,n}(\cdot)$ denotes the fitted response surface using the $n$ trial observations.}}
    \label{fig:friendly-drawing}
\end{figure}

Different identification formulas motivate different estimation strategies as discussed next. \modif{These strategies are illustrated in Figure~\ref{fig:friendly-drawing}.}


\subsection{Estimation methods to \modiff{generalize trial findings to a target population of interest}} \label{sec:estimationPO} 

All along this review, estimators are indexed with the number of observations used for estimation. For example, $\hat \tau_n$ indicates that the finite sample estimator only relies on the RCT individuals, or $\hat \tau_{n,m}$ if it depends on both data sets. 

\subsubsection{IPSW and stratification: modeling the probability of trial participation} \label{sec:ipsw}


To overcome the bias due to covariate shift between populations, most existing methods rely on direct modeling of the selection score
previously introduced. The selection score adjustment methods include IPSW \citep{cole2010generalizing, stuart2011use, lesko2017generalizing, buchanan2018generalizing, Colnet2022Reweighting} and
stratification \citep{stuart2011use,tipton2013improving,o2014generalizing}. 

\paragraph{Inverse probability of sampling weighting (IPSW).}

The IPSW approach can be seen as the counterpart of IPW methods for estimating the ATE from observational studies by controlling for confounding (see Appendix~\ref{sec:cate} for details on IPW). 
Based on the identification formula \eqref{eq:idweight}, the IPSW estimator of the ATE is defined as the weighted difference of average outcomes between the treated and control group in the trial. The observations are weighted by the inverse odds $1/\alpha(x) = \pi_{\mathcal{O}}(x)/\pi_{\mathcal{R}}(x)$ to account for the shift of the covariate distribution from
the RCT sample to the target population. The larger $\alpha(X_i)$, the smaller the weight of the observation $i$ \modiff{(as illustrated on Figure~\ref{fig:friendly-drawing})}. The shape of the IPSW estimator is slightly different from the shape of the IPW estimator. In the latter, each observation is weighted by the inverse of the probability to be treated whereas in the former it is weighted by the inverse of the odds  of the probability to be in the trial sample. This is due to the non-nested sampling design (see the IPSW estimator for the nested design \eqref{eq:ipswnested}), as highlighted by \cite{kern2016assessing} and \cite{nguyen2018sensitivitybis}.

\begin{definition}[Inverse probability of sampling weighting - IPSW]\label{def:ipsw}
The IPSW estimator is defined as follows: 
\begin{equation*}
\hat \tau_{\ipsw, n, m}= \frac{1}{n} \sum_{i=1}^{n} \frac{n}{m}\frac{Y_i}{\hat \alpha_{n,m}(X_i)} \left( \frac{A_i}{e_{1}(X_i)} - \frac{1-A_i}{1- e_{1}(X_i)} \right)\,,
\end{equation*}
where $\hat \alpha_{n, m}$ is an estimate of the odds of the indicatrix of being in the RCT.

 
\end{definition}



The IPSW estimator is consistent when the quantity $\alpha$ is consistently estimated by $\hat \alpha_{n,m}$ \citep{buchanan2018generalizing, colnet2021causal}. \modiff{In practice, v}arious methods are used to estimate $\alpha$: for e.g. by logistic regression \sout{fitted to discriminate RCT from observational samples in }\citep{stuart2010matching}, while recent works \modiff{rely on non-parametric} methods such as random forest and Gradient boosting \citep{kern2016assessing}
\modif{or H\'ajek-style estimator to target the density ratio \citep{huang2021leveraging, nie2021covariate}.}
Similar to IPW estimators, IPSW estimators are known to be highly unstable, especially when the \modiff{weights} are extreme. This can occur if the observational study contains units with very small probabilities of being in the trial. Normalized weights can be used to overcome this issue \citep{dahabreh2019extending}. Still, the major challenge remains that IPSW estimators require a correct model specification of the \modiff{weights}.  
Avoiding this problem requires either very strong domain expertise or turning to doubly robust methods (Section~\ref{sec:DR}). \modiff{Current theoretical guarantees and theorems are detailed in Appendix (see Section~\ref{appendix:extension-with-theorems}). For example } \cite{buchanan2018generalizing} propose a derivation of the asymptotic variance under parametric assumptions in the nested case, \modiff{while \cite{Zivich2022Delicatessen} extends this to a non-nested design}. 
\cite{dahabreh2018generalizing} propose the use of sandwich-type
variance estimators (for both nested and non-nested design) or non-parametric bootstrap approaches, and note that the latter may be preferred in practice.
\modiff{\cite{colnet2021causal} has formalized consistency results for any consistent estimator of $\alpha$, including non-parametric estimators.}

\modiff{\begin{assumption}[Consistency assumptions for $\alpha$]\label{a:consistency-alpha}
Denoting by $\frac{n}{m  \hat \alpha_{n,m}(x)}$ the estimated weights on the set $X$, 
the following conditions hold,
\begin{itemize}
    \item $ \sup _{x \in \mathcal{X}}|\frac{n}{m  \hat \alpha_{n,m}(x)} - \frac{f_{X}(x)}{f_{X \mid S = 1}(x)} |=\epsilon_{n,m} \stackrel{a . s .}{\longrightarrow} 0 \text{ , when } n, m \rightarrow \infty$,
    \item  for all $n,m$ large enough $\mathbb{E}[\varepsilon_{n,m}^2]$ exists and $\mathbb{E}[\varepsilon_{n,m}^2] \stackrel{a . s .}{\longrightarrow} 0 \text{ , when } n, m \rightarrow \infty$,
    \item $Y$ is square integrable.
\end{itemize}
\end{assumption}}

\modiff{
\begin{theorem}[IPSW consistency - \cite{colnet2021causal}]
\label{lem:Consistency_IPSW} 
Under causal assumptions (Assumptions~ \ref{a:consist}, ~\ref{a:ign}, \ref{a:trans}, \ref{a:pos}), (identifiability), and Assumption~\ref{a:consistency-alpha} (consistency),
then, $\hat \tau_{\text{\tiny IPSW}, n, m}$ converges toward $\tau$ in $L^1$ norm,
\begin{equation*}
     \hat{\tau}_{\text{{\tiny IPSW}}, n, m} \underset{n,m \to \infty}{\overset{L^1}{\longrightarrow}} \tau.
\end{equation*}
\end{theorem}}

\modiff{More recently \cite{Colnet2022Reweighting} has proposed a finite sample characterization of IPSW when $X$ only contains categorical covariates.}

\paragraph{Stratification.}

The stratification approach -- or subclassification -- is introduced by \citet{cochran1968subclassification} \modiff{for a single observational data set, and has been further extended by \cite{stuart2011use}, \cite{tipton2013improving}, and \cite{o2014generalizing} for the generalization's context}. It is proposed as a solution to mitigate the risks of extreme weights in the IPSW formula.
\modiff{First, one has to estimate the conditional odds $\hat \alpha_{n,m}$ in the same manner as for the IPSW detailed above. Then, based on the values of the conditional odds obtained, $L$ strata are defined (usually $5$ as reported in \citep{o2014generalizing}, following the empirical seminal work of \citep{cochran1968subclassification}). In the trial, for each strata $l$ one has to compute the average effect on this strata defined as $\overline{Y(1)}_l - \overline{Y(0)}_l$, where $\overline{Y(a)}_l$ denotes the average value of the outcome for units with treatment $a$ in stratum $l$ in the RCT. The generalized ATE is defined by the aggregation of the treatment effect estimates on each strata $l$ weighted by the proportion of the strata in the target population $ \frac{m_l}{m} $, where $m_l$ is the number of individuals in strata $l$ in the target sample.}
\sout{The principle is to define $L$ strata based on the covariate values via the selection score, which in practice consists in grouping observations with similar values of $\hat \alpha_{n,m}$. Then, in each stratum $l$, the average treatment effect is estimated as 
$\overline{Y(1)}_l - \overline{Y(0)}_l$, where $\overline{Y(a)}_l$
denotes the average value of the outcome for units with treatment $a$ in stratum $l$ in the RCT.  }
\begin{definition}[Stratification]\label{def:strat}
The stratification estimator denoted $\hat \tau_{\strat,n,m}$ is defined as,
 \begin{equation*}
   \hat \tau_{\strat,n,m} =  \sum_{l=1}^{L} \frac{m_l}{m} \underbrace{\left ( \overline{Y(1)}_l - \overline{Y(0)}_l \right )}_\textrm{from RCT}. 
\end{equation*}
\sout{ where $m_l/m$ is the proportion of population cases in stratum $l$ inside the target sample.}
\end{definition}
\cite{buchanan2018generalizing} has proposed asymptotic normality result for this estimator.
\modiff{Theoretical results for the stratification estimator are detailed in the appendix (Section~\ref{appendix:extension-with-theorems}).}

\subsubsection{\modiff{Plug-in }g-formula estimators: modeling the conditional outcome in the trial} \label{sec:gformula}
Other estimators to generalize RCT findings to a target
population \modif{leverage} the regression formulation \eqref{eq:idereg}, \modiff{in the inspiration of\citep{robins1986new}}.
Known as plug-in $g$-formula estimators,
they fit a model of the conditional outcome mean
among trial participants, rather than modeling the probability of trial participation \modiff{(as illustrated on Figure~\ref{fig:friendly-drawing})}. 
\sout{Applying these models to generalization,} Then a marginalization is done over the empirical covariate distribution of the target population\sout{, gives the corresponding expected outcome}. 
\begin{definition}[Plug-in g-formula]\label{def:g-formula}
The plug-in g-formula (or outcome model-based) estimator is then defined as:
\begin{equation*}
\hat \tau_{\text{\tiny G},n,m} = \frac{1}{m}\sum_{i=n+1}^{n+m}\left(\hat \mu_{1, 1, n}(X_i) - \hat \mu_{0, 1, n}(X_i)\right),
\end{equation*}
where $\hat \mu_{a, 1, n}(X_i)$ is an estimator of $\mu_{a, 1}(X_i)$ fitted using the RCT data.
\end{definition}

\modiff{In practice, any model can be use to fit $\mu_{a, 1}(X_i)$, for e.g. standard ordinary least squares (OLS).} \modiff{\cite{dahabreh2020extending} announce\footnote{see their Appendix, Section A, pages 6-7.} consistency of the plug-in g-formula for parametric estimator of the response model $\mu_a(X)$. Note that derivations are made in the context of a nested design but said to extend to a non-nested design.} \modiff{They also recommend the use of sandwich-type variance for confidence intervals estimation when correctly specified parametric models are used.} \modiff{Machine-learning algorithms such as random forests can also be used to estimate  $\mu_{a, 1}(X_i)$ \citep{kern2016assessing}.
As shown by \cite{colnet2021causal} if the model is correctly specified (see Assumption~\ref{a:consistency-mu} below), the estimator is consistent.}

\modiff{\begin{assumption}[Consistency of surface response estimators]\label{a:consistency-mu}
Denote $\hat \mu_{0,n}$ (respectively $\hat \mu_{1,n}$) an estimator of $\mu_{0}$ (respectively $\mu_{1}$). Let $\mathcal{D}_n$ the RCT sample, so that 
    \item For $a \in \{0,1\}$, $\mathbb{E}\left[ | \hat \mu_{a,n}(X) -  \mu_{a}(X) | \mid \mathcal{D}_n\right] \stackrel{p}{\rightarrow}  0$ when $n \rightarrow \infty$,
    \item  For $a \in \{0,1\}$, there exist $C_1, N_1$ so that for all $ n \geqslant N_{1}$, a.s., $\mathbb{E}[\hat{\mu}_{a,n}^2(X) \mid \mathcal{D}_n] \leqslant C_1$.
\end{assumption}

\begin{theorem}[Consistency of the plug-in g-formula - \cite{colnet2021causal}]
    Under causal assumptions (Assumptions~ \ref{a:consist}, ~\ref{a:ign}, \ref{a:trans}, \ref{a:pos}), and Assumption~\ref{a:consistency-mu} the plug-in g-formula  converges toward $\tau$ in $L^1$ norm,
    \begin{equation*}
    \hat{\tau}_{\text{{\tiny G}}, n, m} \underset{n, m \to \infty}{\overset{L^1}{\longrightarrow}} \tau.
\end{equation*}
    
\end{theorem}}
\subsubsection{Calibration weighting: balancing covariates\label{sec: CWmethod}}

Beyond propensity scores, other schemes use sample reweighting. \citet{dong2020integrative}
propose a calibration weighting approach, similar to the idea
of entropy balancing weights introduced by \citet{hainmueller2012entropy}.
They calibrate subjects in the RCT sample in such a way that after calibration,
the covariate distribution of the RCT sample empirically matches the
target population.

\begin{definition}[Calibration weighting - CW]\label{def:cbss0}
Let $\bg(X)$ be a vector of functions of $X$
to be calibrated, e.g., the moments, interactions, and non-linear
transformations of components of $X$.
Then, assign a weight $\omega_{i}$ to each subject $i$ in the RCT sample by solving
the following optimization problem: 
\begin{align*}
\underset{\omega_1,\ldots,\omega_n}{\text{min}} & \sum_{i=1}^{n}\omega_{i}\log \omega_{i},\\
\text{subject to}\quad & \omega_{i}\ge0,\;\text{for all}\;i,\nonumber \\
\sum_{i=1}^{n}\omega_{i}=1, & \sum_{i=1}^{n} \omega_{i}\,\bg(X_{i})=\widetilde{\bg},\;\text{(the balancing constraint) \nonumber}\label{eq:calibration constraints}
\end{align*}
where $\widetilde{\bg}=m^{-1}\sum_{i=n+1}^{m+n}\bg(X_{i})$
is a consistent estimator of $\mathbb{E}[\bg(X)]$ from the
observational sample. 
Based on the calibration weights, the CW estimator is then 
\begin{equation*} 
\hat \tau_{\text{\tiny CW},n,m}=\sum_{i=1}^{n}\hat \omega_{n,m}(X_i) Y_i \left ( \frac{A_{i}}{e_{1}(X_i)}-\frac{1-A_{i}}{1-e_{1}(X_i)}\right ) ,
\end{equation*}
\modif{where $\hat \omega_{n,m}(.)$ is the estimated $\omega(.)$ using the RCT and observational data.}
\end{definition}
The optimization problem in Definition~\ref{def:cbss0} corresponds to the negative entropy of the calibration
weights; thus, minimizing this criterion ensures that the empirical
distribution of calibration weights is not too far away from the
uniform distribution. This aims at minimizing the variability due to heterogeneous
weights. This optimization problem can be solved using convex optimization
with Lagrange multipliers. For an intuitive understanding of the calibration weighting
framework, consider $\bg(X) = X$. In such a setting, the balancing constraint is forcing the
means of the observational data and of the RCT to be equal after reweighting. More complex
constraints can enforce balance on higher-order moments.
The calibration algorithm is inherently imposing a log-linear model on the sampling propensity score and solving the corresponding parameters by a set of estimating equations induced by covariate balance. 
\modiff{Other objective functions of the weights correspond to different models for the sampling propensity score \citep{chu2022targeted}.
\cite{wu2021transfer} propose a cross-validation procedure to select the calibration weights that target at the smallest mean squared error of the resulting estimator.} 
The CW estimator $\hat \tau_{\text{\tiny CW},n,m}$ is doubly robust in that
it is a consistent estimator for $\tau$ if the selection score
of RCT participation follows a log-linear model, i.e., $\pi_{ S}(X)=\exp\{\bEta_{0}^{\top}\bg(X)\}$
for some $\bEta_{0}$, or if the CATE is linear in $\bg(X)$, i.e.,
$\tau(X)=\boldsymbol{\gamma}_{0}^{\top}\bg(X)$, though not necessarily
both. The authors suggest a bootstrap approach to estimate its variance.

\subsubsection{Doubly-robust estimators} \label{sec:DR}

The model for the expectation of the outcomes among randomized individuals
(used for the plug-in $g$-formula estimator in Definition~\ref{def:g-formula}) and the model for the probability of trial
participation (used in the IPSW estimator in Definition~\ref{def:ipsw}) can be combined to form
an Augmented IPSW estimator (AIPSW).

\begin{definition}[Augmented IPSW -AIPSW]\label{def:aipsw}
 The augmented IPSW estimator, denoted $\hat \tau_{\text{\tiny AIPSW},n,m}$, is defined as
 \begin{align*}
\begin{split}
\hat \tau_{\text{\tiny AIPSW},n,m}= \frac{1}{n} \sum_{i=1}^{n}  \frac{n}{m \, \hat \alpha_{n,m}(X_i)}\left(\frac{A_{i}\left (Y_{i}-\hat\mu_{1, 1, n}(X_{i})\right ) }{e_{1}(X_i)}-\frac{(1-A_{i})\left ( Y_{i}-\hat\mu_{0,1,n}(X_{i})\right ) }{1-e_{1}(X_i)}\right ) \\
+\frac{1}{m}\sum_{i=n+1}^{m+n}\left( \hat\mu_{1, 1, n}(X_{i})-\hat\mu_{0, 1, n}(X_{i})\right ),
\end{split}
\end{align*}
where $\hat \mu_{a,1,}$ are estimated on the RCT sample (see Definition~\ref{def:g-formula}), and $\hat \alpha_{n,m}$ (see Definition~\ref{def:ipsw}) on the concatenated RCT and observational samples.
\end{definition}

It can be shown
that this estimator is doubly robust, i.e., consistent when either one of the two models for $\hat \alpha_{n,m}(\cdot)$ and $\widehat\mu_{a,1}(\cdot)$ $(a=0,1)$ is correctly
specified.
\modiff{\cite{dahabreh2020extending} has proposed a proof in the nested-case (see their appendix, Section A) said to follow the same principle in the non-nested design (Section B page 25).}
\modiff{In the plain text we recall the results from \cite{colnet2021causal}.}
\modiff{
\begin{assumption}[Consistency assumptions - AIPSW]\label{a:consistency-aipsw}
The nuisance parameters are bounded, and more particularly

\begin{itemize}
      \item There exists a function $\alpha_0$ bounded from
above and below (from zero), satisfying $$ \lim\limits_{m,n \to \infty}
\sup_{x \in \mathcal{X}}\biggr| \frac{n}{m\hat \alpha_{n,m}(x)} -
\frac{1}{\alpha_0(x)} \biggl| = 0,$$

    \item 
    There exist two bounded functions $\xi_1, \xi_0: \mathcal{X} \to \R$, such that $\forall a \in \{0, 1\},$ 
    $$\lim\limits_{n \rightarrow +\infty} \sup_{x \in \mathcal{X}}|\xi_{a,1}(x) - \hat \mu_{a,1,n}(x)| = 0.$$

\end{itemize}

\end{assumption}

\begin{theorem}[AIPSW consistency - \cite{colnet2021causal}]
\label{lem:Consistency_AIPSW}
Assuming causal assumptions (Assumptions~ \ref{a:consist}, ~\ref{a:ign}, \ref{a:trans}, \ref{a:pos}), and Assumption~\ref{a:consistency-aipsw} (consistency), and considering that estimated surface responses $\hat \mu_{a,1,n}(.)$ where $a \in \{0, 1\}$ are obtained following a cross-fitting estimation, then if Assumption~\ref{a:consistency-mu} \textbf{or} Assumption~\ref{a:consistency-alpha} also holds then, $\hat \tau_{\text{\tiny AIPSW}, n, m}$ converges toward $\tau$ in $L^1$ norm,
\begin{equation*}
     \hat{\tau}_{\text{{\tiny AIPSW}}, n, m} \underset{n,m \to \infty}{\overset{L^1}{\longrightarrow}} \tau.
\end{equation*}
\end{theorem}

}
\modiff{This estimator is also shown to be asymptotically normal when both the outcome mean and
conditional odds model are consistently estimated at least at rate $n^{1/4}$ in
\cite{dahabreh2019extending} and \cite{Li2021SemiparametricEfficientGen}.} 
 \modiff{Note that machine-learning tools are tempting to avoid model mis-specification
when estimating nuisance parameters. Still, this practice requires
specific caution, such as using cross-fitting, due to overfitting and regularization. These issues are well described in the situation of a single observational data set. We refer to \cite{chernozhukov2018double} for a detailed explanation, and to \cite{Kennedy2021Package,DoubleML2021R,DoubleML2022Python} for implementations.}\\
 
More recently, \citet{dong2020integrative} 
propose an augmented calibration weighting (ACW) estimator.

\begin{definition}[Augmented CW - ACW]\label{def:taudc-2}
The ACW estimator, denoted $\hat \tau_{\text{\tiny ACW}, n, m}$, is defined as
 \begin{align*}
\begin{split}
\hat \tau_{\text{\tiny ACW}, n, m}=\sum_{i=1}^{n} \hat \omega_{n,m}(X_i)\left(\frac{A_{i}\left
(Y_{i}-\hat\mu_{1, 1, n}(X_{i})\right) }{e_{1}(X_i)}-\frac{(1-A_{i})\left( Y_{i}-\hat\mu_{0, 1, n}(X_{i})\right) }{1-e_{1}(X_i)}\right)\\
+\frac{1}{m}\sum_{i=n+1}^{m+n}( \hat\mu_{1, 1,n}(X_{i})-\hat\mu_{0, 1, n}(X_{i})),
\end{split}
\end{align*}
where the estimation of $\hat \omega_{n,m}(.)$ is detailed in Definition~\ref{def:cbss0}, and where $\hat \mu_{a,1,n}$ are estimated on the RCT sample (see Definition~\ref{def:g-formula}).
\end{definition}

They show that $\hat \tau_{\text{\tiny ACW},n,m}$ achieves double robustness
and local efficiency, i.e., its asymptotic variance achieves the semiparametric
efficiency bound \modif{when both the calibration weights and the outcome mean model are correctly specified}. 
Moreover, 
the convergence rate of the ACW estimator corresponds to the product of the convergence rates of the nuisance estimators, enabling
the use of \sout{double }machine-learning estimation
of nuisance functions while preserving the $\sqrt{n}$-consistency of the
ACW estimator\modif{, when both the outcome mean and calibration weights model are consistently estimated at rate $n^{1/4}$  \citep{dong2020integrative}.} 
Furthermore, \cite{lee2022generalizable} and \cite{lee2022transporting} extend the framework for handling survival outcomes. 

\sout{Other doubly robust estimators include targeted maximum likelihood estimators
(TMLE) \citep{rudolph2017robust}. They provide a semiparametric
efficiency score for transporting the ATE from one study site to another,
where one site is regarded as a population. } 

\subsubsection{\modiff{Practical issues: non-parametric estimation, overlap and unobserved covariates}} 
\sout{\paragraph{Relying on machine learning for nuisance parameters estimation.}}
\sout{As mentioned before, flexible regression tools coming from the
machine learning literature are tempting to avoid model mis-specification
when estimating nuisance parameters. Still, this practice requires
specific caution, such as using cross-fitting, due to overfitting and regularization. These issues are well described in the situation of a single observational data set. We refer to \cite{chernozhukov2018double} for a detailed explanation, and to \cite{Kennedy2021Package,DoubleML2021R,DoubleML2022Python} for implementations.}
\sout{\citet{chernozhukov2018double} propose double/debiased machine learning methods to consistently estimate the ATE by using flexible machine learning methods for the nuisance parameters estimation to avoid model mis-specification. Their approach uses Neyman-orthogonal scores to reduce sensitivity of ATE estimation with respect to approximation errors of nuisance parameters. In addition, they use cross-fitting \citep{zheng2011cross, chernozhukov2018double} to provide an efficient and unbiased form of data-splitting. } 
\paragraph{\sout{Consequences of a l}Lack of overlap.}
\modiff{The overlap assumption (see Assumption~\ref{a:pos}) is restrictive because RCT inclusion and exclusion criteria can be strict as the goal of RCTs (at least in early stages) is to show a clear effect even on a restricted population. Whenever Assumption~\ref{a:pos} does not hold, it is still possible to generalize on a different target population, such as the subset of the target population for which eligibility criteria of the trial are ensured. 
This has also been suggested before, for e.g. by \cite{tipton2013improving} (p.245).
The question asked would rather be ``What would have been the estimated treatment effect in a situation where the trial has sampled individuals from the target population who meet the trial eligibility criteria?"}. Another approach has been proposed by \citet{chen2021generalizability}. Similarly to the idea of trimming propensity scores for dealing with limited overlap between treated and control groups, they propose a generalizability score: a function of participation probability and propensity score, to select subpopulations from the observational data for causal generalization when the overlap is limited.
\sout{Note that when the overlap assumption between RCT and observational distributions does not hold, we may have to select a subpopulation from the observational data for generalization. In addition, and similarly to the idea of trimming propensity scores for dealing with limited overlap between treated and control groups \citep{crump2009dealing, yang2018asymptotic}, \citet{chen2021generalizability} propose a generalizability score, a function of participation probability and propensity score, to select subpopulations from the observational data for causal generalization when the overlap is limited.}

\paragraph{Unobserved treatment effect modifiers.}
Finally, we point out the important caveat that all methods assume the ignorability conditions \modif{(see Assumptions~\ref{a:trans-2}, \ref{a:trans-3}, \ref{a:trans-diff}, or \ref{a:trans})}: given the covariates $X$, the conditional treatment effect must be the same in the observational data and the RCT.
In particular, this assumption could be violated if some shifted treatment effect
modifiers are not captured in the concatenated data, which is a plausible scenario given that data are seldom collected jointly and thus typically
measure different covariates.

In case of a richer set of covariates in the RCT than in the
observational study \modiff{(which doesn't necessarily mean that a sufficient set of pre-treatment covariates can be chosen, see for e.g. M-bias, see \cite{Pearl2000Book}, page 186)}, \citet{egami2021covariate} propose a method to
select a sufficient set of covariates. But in the case of a
low number of common covariates, standard practice is to
consider the subset of covariates present in both data sets, \modiff{but this violates the identifiability condition}. 
Recently, sensitivity analyses have been proposed to mitigate the
consequences of missing covariates \modiff{in the RCT, or in the observational sample or even in both data sets} \citep{nguyen2017sensitivity, Andrews2018ValidityBias, nguyen2018sensitivitybis, dahabreh2019sensitivity, colnet2021causal, nie2021covariate, huang2022sensitivity}.


\section{When observational data contain treatment and outcome information}
\label{sec:specialcase}

\modiff{Section~\ref{sec:noWnoY} studied how to 
correct RCT selection bias (with respect to the target population) while leveraging covariate distribution of an
observational sample. When the observational sample
also contains treatment and outcome information $(Y,A)$,  efficiency improvements can be obtained \citep{huang2021leveraging}. But beyond the generalization question, such additional covariates enable different questions of interest. These questions are the purpose of Section~\ref{sec:specialcase}.} Indeed, RCTs can make causal conclusions from the observational sample
more trustworthy, either by removing confounding bias (detailed in Section~\ref{sec:unmeasured confoundersATE}) or via more
efficient estimation (detailed in Section~\ref{sec:toward-more-efficient}). 
For completeness, we recall in Appendix~\ref{sec:cate} how to perform causal inference from purely observational data.


\subsection{Dealing with unmeasured confounders in observational data} \label{sec:unmeasured confoundersATE}

\paragraph{Motivation.}

Unmeasured confounding implies that $\left\{ Y(1), Y(0)
\right\} \not \independent A \mid X$, where $X$ are the observed
covariates. In such situations, standard causal inference estimators 
$\hat \tau_m^\mathcal{O}(x)$ \modiff{(resp. $\hat \tau_m^\mathcal{O}$)} of the CATE $\tau(X)$ \modiff{(resp. ATE $\tau$)}, that are designed for purely
observational data of size $m$, face a so-called hidden confounding bias for these quantities, i.e.,
$$\lim\limits_{m \rightarrow +\infty}  \hat \tau_m^\mathcal{O}(x) \neq
\tau(x), \quad\text{ and }\quad \lim\limits_{m \rightarrow +\infty}  \hat \tau_m^\mathcal{O} \neq \tau.$$

In practice, former RCTs can be used as \emph{negative controls}\footnote{\modiff{The term negative controls comes from usual routine precaution in biological laboratory experiments, where such controls are used to -- at least partially -- check that the experiment is not undermined. For example it can test the absence of reagents or components that are necessary for a detection of something particular. For example one of the two bars of the covid antigenic test is one of these controls. The analogy of this principle in causal inference is detailed in \citep{Lipsitch2010NegativeCA}.}}, to ensure the observational study does not suffer from confounding. 
For example, in a recent observational study on 
a COVID-19 vaccine, \citet{dagan2021covidvaccine} use such approach \sout{checking that their adjustment set ensure that vaccinated and non-vaccinated individuals did not differ in terms of outcomes seven days after injection between the former trial and the real-world analysis}\modiff{to ensure that previous trial results conclusion could be retrieved}. 
When confounding remains, solutions such as sensitivity analysis have been developed to handle such
situations \citep{rosenbaum2002sensitivity, imbens2003sensitivity}, but they
typically rely on sensitivity parameters which are
difficult to set. Including additional experimental data brings
interesting promises to handle such identification bias. Recent works \modiff{described below}
propose to use an RCT to \textit{ground} the observational analysis and debias the estimator that
would be obtained on purely confounded observational data. 

\paragraph{Using an assumption on secondary outcomes or surrogates.} 
The use of surrogate outcomes arises in different contexts, for example in clinical studies \citep{prentice1989surrogate, begg2000surrogate}, where it may be difficult to observe long-term outcomes, e.g., the effect of early childhood medical or economic interventions.  
\cite{athey2020combining, athey2020surrogate} observe that the effect of class size reduction leads to a decrease in children $3^{rd}$ grades in the observational data, while a famous RCT, the Tennessee Student/Teacher Achievement Ratio (STAR) study \citep{krueger1999experimental}, concludes on a positive effect. This difference could come from the fact that the two populations are different, but they assume the apparent difference can be entirely explained by confounding\footnote{Assuming the bias comes from an unobserved confounder and not from inherent differences between populations can be stated as, $S \independent  \left\{ Y(1), Y(0) \right\},$ which means that the two samples come from comparable populations (see Section~\ref{sec:noWnoY}).}.
\modif{In their set-up, they consider two outcomes, a primary long-term outcome $Y^{1^{st}}$ ($8^{th}$ grades)
and a secondary short-term outcome $Y^{2^{nd}}$ ($3^{rd}$ grades). 
The RCT contains information on the surrogate but not the long-term outcome while this is the opposite for the observational sample. 
Their central assumption to recover identifiability is called \textit{latent unconfoundedness}, i.e.,
\begin{equation*}
    A \independent  Y^{1^{st}}(a) \mid  Y^{2^{nd}}(a),   i \in \mathcal{R}, \text{ , for } a=0,1,
\end{equation*}
which corresponds to the assumption that hidden confounders violating identification of the effect on $Y^{1^{st}}$ are the same than for $Y^{2^{nd}}$.} In other words, their method consists in adjusting the estimates of the treatment effects on the primary outcome using the differences observed on the secondary outcome. 
Their assumptions can be understood as a missing data problem, i.e., the missing data in the \sout{potential}\modiff{primary} outcomes are missing at random in the concatenated data \citep{rubin1976inference}. 
For estimation, they suggest three methods, namely, \emph{i)} imputing the missing primary outcome in the RCT, \emph{ii)} weighting the units in the observational sample, and \emph{iii)} using control function methods. 

\paragraph{Deconfound using the bias/confounding function.}
\cite{kallus2018removing} propose to use an RCT sample to deconfound the CATE estimated on a single observational data set, denoted $\hat \tau_m^\mathcal{O}(x)$. 
Due to possible unmeasured confounding, $\hat \tau_m^\mathcal{O}(x)$ may be biased for $\tau(x)$, that is $\eta(x) \neq 0$ where $\eta(x):= \tau(x) - \hat{\tau}_m^\mathcal{O}(x)$ is the bias function.
To correct for this bias, they assume they have at hand a narrow RCT \modiff{(as it is usually the case with strict eligibility criteria in trial)} with high internal validity, and with covariate support included in the observational sample support. 
Given that $\hat \tau_m^\mathcal{O}(x)$ is obtained from the observational data,
one can estimate $\eta(\cdot)$ on the common support between the RCT and the observational data using the (unconfounded) RCT data. Another assumption is required, being that the bias can be well approximated by a function with low complexity, e.g., a linear function of the covariates $x$: $\eta(x) = \theta^T x$. 
\cite{kallus2018removing} then propose to estimate the bias as $\hat{\eta}_{m,n}(x) = \hat \theta_{m,n}^T x$  by solving the following minimization:
$$\hat{\theta}_{m,n}=\rm{argmin}_{\eta} \sum_{i=1}^n \left(Y_i^{*}-\hat \tau_m^{\mathcal{O}}(X_i)-\eta(X_i)\right)^2 = \rm{argmin}_{\theta} \sum_{i=1}^n \left(Y_i^{*}-\hat \tau_m^{\mathcal{O}}(X_i)-\theta^T X_i\right)^2\,,$$

where $Y_i^{*}=\left({e(X_i)}^{-1}A_i-\{1-e(X_i)\}^{-1}(1-A_i)\right)Y_i$, which satisfies $\mathbb{E}[Y_i^{*}\mid X_i]= \tau(X_i)$.

Note that the linear assumption guarantees the validity of the framework even if the observational data does not fully overlap with the experimental data as the bias, \modiff{i.e, the confounding error is assumed to be extrapolable.}
\modif{
\sout{In other words, the confounding error is assumed to be extrapolatable.}
Finally, $\hat{\tau}_{m,n}(x)=\hat \tau_m^\mathcal{O}(x)+\hat{\eta}_{m,n}(x)$ is the estimated conditional average treatment effect. They prove that under conditions of parametric identification of $\eta$, $\hat{\tau}_{m,n}(x)$ is a consistent
estimate of $\tau(x)$ which converges at a rate governed by the rate of estimating $\mathbb{E}[\hat \tau_m^\mathcal{O}(x)]$ by $\hat \tau_m^\mathcal{O}(x)$. }

\modif{More recently, \citet{yang2020improved} proposed another approach. Rather than $\eta(x)$, they consider what they call the \textit{confounding function} $\lambda(x)$, 
\[
\lambda(x)=\mathbb{E}[Y(0)\mid A=1,X=x]-\EE[Y(0)\mid A=0,X=x],
\]
summarizing the impact of unmeasured confounders on the potential outcome distribution between the treated and untreated patients. In the absence of unmeasured confounding, $\lambda(x)$ is zero for any $x\in\mathcal{X}$, while if there is unmeasured confounding, $\lambda(x)\neq0$ for some $x$.
Assuming a parametric model assumption for the CATE $\tau(x):=\tau_{\varphi_{0}}(x)$  with $\varphi_0\in\R^{p_1}$, and for $\lambda(x):=\lambda_{\phi_{0}}(x)$ with $\phi_0\in\R^{p_2}$, the coupling of RCT and observational data allows identifiability of $\tau(x)$ and $\lambda(x)$.
The key insight is to introduce the following random variable
\begin{equation*}
H_{\psi_{0}}=Y-\tau_{\varphi_{0}}(X)A-(1-S)\lambda_{\phi_{0}}(X)\{A-e(X)\}\,,\label{eq:def of H(k)-1}
\end{equation*}
where $\psi_0=(\varphi_0^{\T},\phi_0^{\T})^{\T}$
is the full vector of model parameters in the CATE and confounding
function\modiff{, and where here $S = 1$ (resp. $S=0$) denotes trial participation (resp. observational study participation)}. By separating the treatment effect $\tau_{\varphi_{0}}(X)A$
and $(1-S)\lambda_{\phi_{0}}(X)\{A-e(X)\}$ from the observed $Y$,
$H_{\psi_{0}}$ mimics the potential outcome $Y(0)$. They then derive
the semiparametric efficient score of $\psi_{0}$:
\begin{equation}
S_{\psi_{0}}(V)=\left(\begin{array}{c}
\frac{\partial\tau_{\varphi_{0}}(X)}{\partial\varphi_0}\\
\frac{\partial\lambda_{\phi_{0}}(X)}{\partial\phi_0}(1-S)
\end{array}\right)\left(\sigma_S^{2}(X)\right)^{-1}(H_{\psi_{0}}-\mathbb{E}\left[H_{\psi_{0}}\mid X,S \right])\left(A-e(X)\right)\,,\label{eq:simpleEE}
\end{equation}
where $\sigma_{S}^{2}(X)=\mathbb{V}[Y(0)\mid X,S]$. A semiparametric
efficient estimator of $\psi_{0}$ can be obtained by solving the
estimating equation based on (\ref{eq:simpleEE}). If the predictors
in $\tau_{\varphi_0}(X)$ and $\lambda_{\phi_0}(X)$ are not linearly
dependent, they show that the integrative estimator of the CATE
is strictly more efficient than the RCT estimator. As a by-product,
this framework can be used to generalize the ATEs from the RCT to
a target population without requiring an overlap covariate distribution
assumption between the RCT and observational data. \citet{Wu2022Integrative} propose an integrative R-learner that extends the framework of \citet{yang2020improved} to allow flexible machine learning methods for approximating CATE, confounding function, and nuisance functions.
}

\subsection{Toward more efficient estimation}\label{sec:toward-more-efficient}

Under Assumptions \ref{a:consist}, \ref{a:ign}, and \ref{a:trans}, 
\modif{the CATE can be estimated based on the RCT}, \modif{while under the classical unconfoundedness assumption} (see Appendix \ref{asump:unconfounded}), \modif{the CATE can be estimated using the observational sample.} Therefore when both \modiff{sets of} assumptions are met, the two data sources can be pooled to improve estimation efficiency. 
\modiff{Toward this end, \citet{yang2020elastic} use the semiparametric efficiency theory to derive the semiparametrically
efficient integrative estimator of $\varphi_{0}$ for the CATE $\tau_{\varphi_0}(X)$.
{However, if the unconfoundedness assumption is violated, integrating the observational
sample would bias the CATE estimation. }Leveraging the design advantage
of RCTs, \citet{yang2020elastic} derive a preliminary test statistic for the comparability
and reliability assessment of the observational data and decide whether
to use it in an integrative analysis. Denote the efficient score based
solely on the RCT and observational data as $S_{{\rm rct,\varphi_{0}}}(V)$
and $S_{{\rm os,\varphi_{0}}}(V)$, respectively, where $V$ is a
full vector of variables. Their basic idea is to derive an RCT estimator
$\widehat{\varphi}_{{\rm rct}}$ for $\varphi_{0}$ and construct
the preliminary test statistics based on $S_{{\rm os},\widehat{\varphi}_{{\rm rct}}}(V)$.
The rationale is that if the observational sample is comparable to
the RCT sample for estimating $\varphi_{0}$, $S_{{\rm os},\widehat{\varphi}_{{\rm rct}}}(V)$
is expected to be close to zero; otherwise, $S_{{\rm os},\widehat{\varphi}_{{\rm rct}}}(V)$
is expected to deviate from zero. This thought process leads to the
test statistics 
\begin{equation}
T=\left\{ n^{-1/2}\sum_{i=n+1}^{n+m}S_{{\rm os},\widehat{\varphi}_{{\rm rct}}}(V_{i})\right\} ^{\mathtt{T}}\widehat{\Sigma}_{SS}^{-1}\left\{ n^{-1/2}\sum_{i=n+1}^{n+m}S_{{\rm os},\widehat{\varphi}_{{\rm rct}}}(V_{i})\right\} ,\label{eq:T}
\end{equation}
where $\widehat{\Sigma}_{SS}$ is a consistent estimator for the asymptotic
variance of $n^{-1/2}\sum_{i=n+1}^{n+m}S_{{\rm os},\widehat{\varphi}_{{\rm rct}}}(V_{i})$.
Under ${\rm H}_{0}$ that the observational sample is comparable to
the RCT sample, $T\rightarrow\chi_{p}^{2},$ a Chi-square
distribution with degrees of freedom ${\rm dim}(\varphi_{0})$, as
$n\rightarrow\infty$. This result serves to detect the violation
of the assumption required for the observational data.

\citet{yang2020elastic} propose the elastic integrative estimator by solving
\begin{equation}
\sum_{i=1}^{n}\widehat{S}_{{\rm rct},\varphi}(V_{i})+\mathbb{I}(T<c_{\gamma})\sum_{i=n+1}^{n+m}\widehat{S}_{{\rm os},\varphi}(V_{i})=0,\label{eq:EE}
\end{equation}
where $c_{\gamma}$ is the $100(1-\gamma)$th percentile of $\chi_{p}^{2}$,
serving as a switch to decide combining or not. The methodological
contribution of \citet{yang2020elastic} is to derive a data-adaptive selection
of $c_{\gamma}$ such that the resulting estimator has the smallest
mean squared error and thus performs at least similar to the RCT-only
estimator, if not better. Moreover, the elastic integrative estimator
is non-regular and belongs to pre-test estimation by construction.
The theoretical contributions of \citet{yang2020elastic} include characterizing
the distribution of the elastic integrative estimator under local
alternatives, which better approximates the finite-sample behaviors,
and providing data-adaptive confidence intervals that are uniformly
valid. } 
\sout{In this case, the covariate shift problem for the RCT does not bias the estimator of the CATE; furthermore, we do not require the covariate distribution to overlap between the RCT and observational data.  }
\sout{Methods mentioned above are applicable for general cases of integrative
analysis, other approaches have been tailored for special applications.
For example, \cite{peysakhovich2016combining} consider time series as
observational data, used to estimate a (confounded) 
mapping from observed treatments to unit-level effects. 
Assuming the confounding preserves unit-level relative rank ordering, the 
experimental data is then used to identify a monotonic transformation 
from biased estimates to actual treatment effects.}

\subsection{Other use cases} 
 Beyond generalizability or overcoming confounding, there are other
purposes \sout{for combining}\modiff{motivating the combination of} experimental and observational data. 
\modif{We provide a brief list of these purposes and methodologies. A
detailed or exhaustive survey is beyond the scope of this review.}
 
\paragraph{Using hybrid controls.}

A hybrid control arm is a control arm constructed from a combination of randomized patients and patients receiving usual care in standard clinical practice, as introduced by \cite{pocock1976hybrid} and pursued by \cite{carlin2012hybrid, schmidli2014hybrid}. Recently the FDA has detailed their usage in the regulatory purposes \citep{fda2018framework}.
Using hybrid controls has the potential to decrease the cost of randomized trials, and to reduce ethic constraints on control groups.


\paragraph{Case-control studies.}
In certain applications, e.g., in epidemiology, the observational data at
hand comes from a case-control study where the selection of observations
is driven by the outcome of interest $Y$. Thus, the RCT and observational
data differ in terms of the outcome distribution, typically a
preferential selection on the outcome for the observational data set. Several solutions have been proposed to handle this
type of selection bias. \citet{robins2000marginal} and
\citet{hernan2005structural} propose marginal structural model approaches
to eliminate this bias \modif{given} sufficient knowledge of the selection model given treatment. \citet{guo2021multi} propose a control variates
technique \citep{tan2006distributional,yang2018combining} identifying and estimating an estimand that is sufficiently correlated with the target estimand of interest for the observational cohort.

\modiff{
\paragraph{Encouragement design intervention}
An encouragement design intervention is a design in which some individuals or groups are randomly assigned to receive encouragement to take up the program.
\citep{rudolph2017robust} provide a semiparametric efficiency score for transporting the ATE from one study following an encouragement design, to another population. Due to the design, their set-up is a variant of the generalization work from Section~\ref{sec:noWnoY}, but with treatment allocation information in the target population.
}
\section{Structural causal models (SCM) and transportability}\label{sec:SCM}

\modif{Within the SCM framework \citep{pearl1995causal,pearl2009causality}, 
\citet{bareinboim2016causal} have proposed
answers for transportability and combination of different data-sources --
also called \textit{data fusion}. This section is split off from the
previous section as it builds on additional concepts.}

Let us first briefly introduce the SCM framework, using as much as possible the notations of Section~\ref{sec:basic setup} that we introduced for the PO framework
(Appendix~\ref{sec:appendixscm} gives a more general primer on the SCM
framework\modif{, and in particular the \textit{do}-operator}). The
covariates $X$, treatment $A$, and response $Y$ are modeled in the SCM
framework as random variables with joint distribution $P(X,A,Y)$. Each
intervention, such as setting $A$ to $a=0$ or $a=1$, defines an
alternative distribution over $(X,A,Y)$ that can be systematically
deduced from the no-intervention (or observational) distribution $P$
using the SCM model, and which is written $P(X,A,Y\,|\,do(A=a))$. In this
framework, the CATE is written:
\begin{equation*}\label{eq:cate-scm}
\tau(x) = \EE[Y\,|\,do(A=1), X=x] - \EE[Y\,|\,do(A=0), X=x];
\end{equation*}
and the ATE: 
\begin{equation*}\label{eq:ate-scm}\tau = \EE[Y\,|\,do(A=1)] - \EE[Y\,|\,do(A=0)].
\end{equation*}
These expressions \modif{mirror} the corresponding expressions in the PO framework (Table~\ref{tab:notations}) when one identifies the variable $Y(a)$ in the PO framework to the variable $Y$ under the intervention $do(A=a)$ in the SCM framework, namely when we set $P(Y(a),X) = P(Y,X\,|\,do(A=a))$. In fact this analogy is valid in the sense that any theorem that holds for SCM counterfactuals holds in the PO framework, and vice-versa (\citealp{pearl2009causality}, Chapter 7; \citealp{Pearl2009Causal}, Chapter 4). In spite of this formal equivalence, the two frameworks differ in how they allow practitioners to express causal assumptions, and to derive corresponding estimands of causal effects. The SCM framework provides a convenient graphical representation known as causal diagram to encode potentially complex causal assumptions between variables, and provides a complete language known as $do$-calculus to express causal effects (i.e., some expectation under the $do(A=a)$ probability) as a function of observational data (i.e., some expectation under the no-intervention distribution)  \citep{pearl1995causal,pearl2009causality}. When this reduction is possible, the causal effect is called \textit{identifiable}. In addition, the $do$-calculus is complete in the sense that a causal effect is identifiable if and only if it can be reduced to a function of observational data using $do$-calculus \citep{Huang2006Pearl,Shpitser2006Identification}. Interestingly, this provides a variety of formulas to correctly infer causal effects even in the presence of unmeasured confounders, which cannot be handled by the PO framework (without additional structural and modeling assumptions), such as the front-door adjustment formula \citep{pearl1995causal}. 

\subsection{Formulating transportability in the SCM framework}

The SCM \modif{literature} and $do$-calculus naturally cover the problem of generalizing an RCT experiment to a different target population. Following our notations  in the PO setting (Section~\ref{sec:basic setup}), we again denote by $S$ a binary random variable that indicates which individuals can be in the RCT. The RCT population then follows the distribution $P(X,Y,A\,|\,S=1)$, and by design the RCT allows estimating the conditional distributions $P(Y\,|\,do(A=a), X, S=1)$ for $a=0,1$. The problem of generalization to the target population in this setting is then to deduce the distributions of $P(Y\,|\,do(A=a), X)$ for $a=0,1$ from these two distributions and the observed distribution of the covariates $P(X)$ in the target distribution (as in Section~\ref{sec:noWnoY}), or of the covariates, treatments and responses $P(X,A,Y)$ in the target population (as in Section~\ref{sec:specialcase}). If this deduction (using $do$-calculus) is possible, then the causal effect on the target population is identifiable, and the deduction provides a formula for the causal effect that can then be estimated from a finite population using some consistent estimator.

Interestingly, this formalism covers two important situations: (i) the \emph{sample selection bias} problem, when the RCT population is a subset of the target population that fulfills some eligibility criterion\footnote{This setting has been termed as \textit{generalizability} in the introduction of the different study designs in Section~\ref{sec:studydesign}.}, and (ii) the \emph{transportability} problem, where the RCT population differs more drastically from the target, e.g., when one wants to generalize an RCT conducted in one country to a population in another country \citep{pearl2015findings}. To model sample selection bias, on the one hand, one typically adds a node $S$ with incoming edges to a causal graph in order to capture the eligibility conditions that may depend on pre- or post-treatment variables. It is then possible to derive conditions under which one can recover from selection bias when the probability of selection is available \citep{cooper1995causal,Lauritzen2008Discussion,Geneletti2008Adjusting} or when no quantitative knowledge is available about probability of selection \citep{Didelez2010Graphical,Bareinboim2012Controlling}. We provide examples of such conditions in Appendix~\ref{eq:SCMselectionbias}. To model transportability to a different population, on the other hand, the node $S$ has typically no incoming edge, and instead points to variables that differ between the RCT and the target population, either in their functional dependency to their parents in the causal graph, or in the distribution of their exogenous variables. The resulting graph is called a \emph{selection diagram} and allows to encode graphically detailed assumptions about the differences between populations \citep{pearl2011transportability,bareinboim2012completeness,Pearl2014External,Bareinboim2013General}. 
Note that even if the two situations imply different causal diagrams, the problem of selection bias ``\textit{has some unique features, but can also be viewed as a nuance of the transportability problem, thus inheriting all the theoretical results of transportability}'' \citep{pearl2015findings}; this remark is connected to the discussion from Section~\ref{sec:studydesign}.

\begin{figure}[H]
\centering
(a)
\includegraphics[width= 0.25\textwidth]{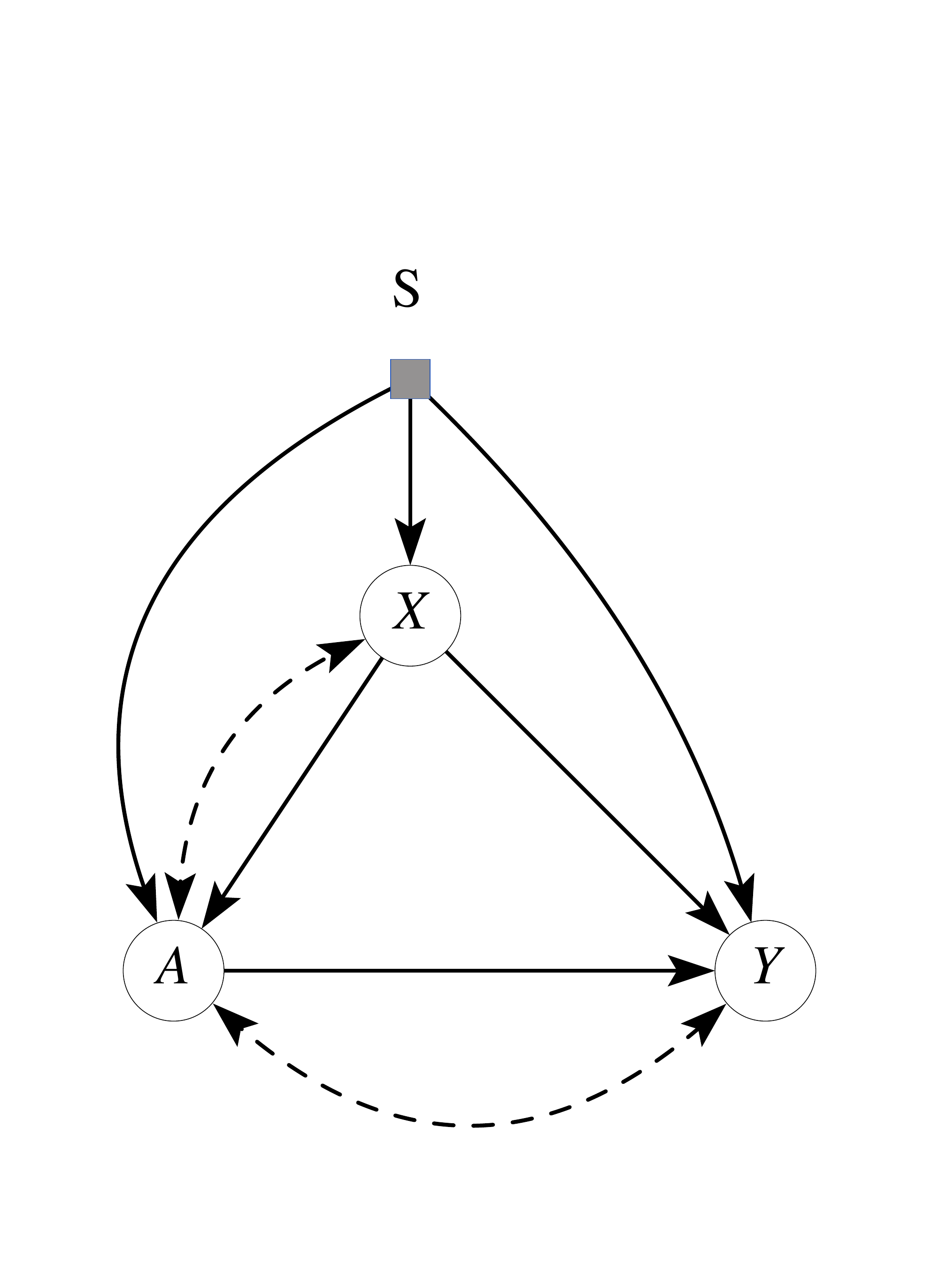} 
\hspace{1cm}
(b)
\includegraphics[width= 0.25\textwidth]{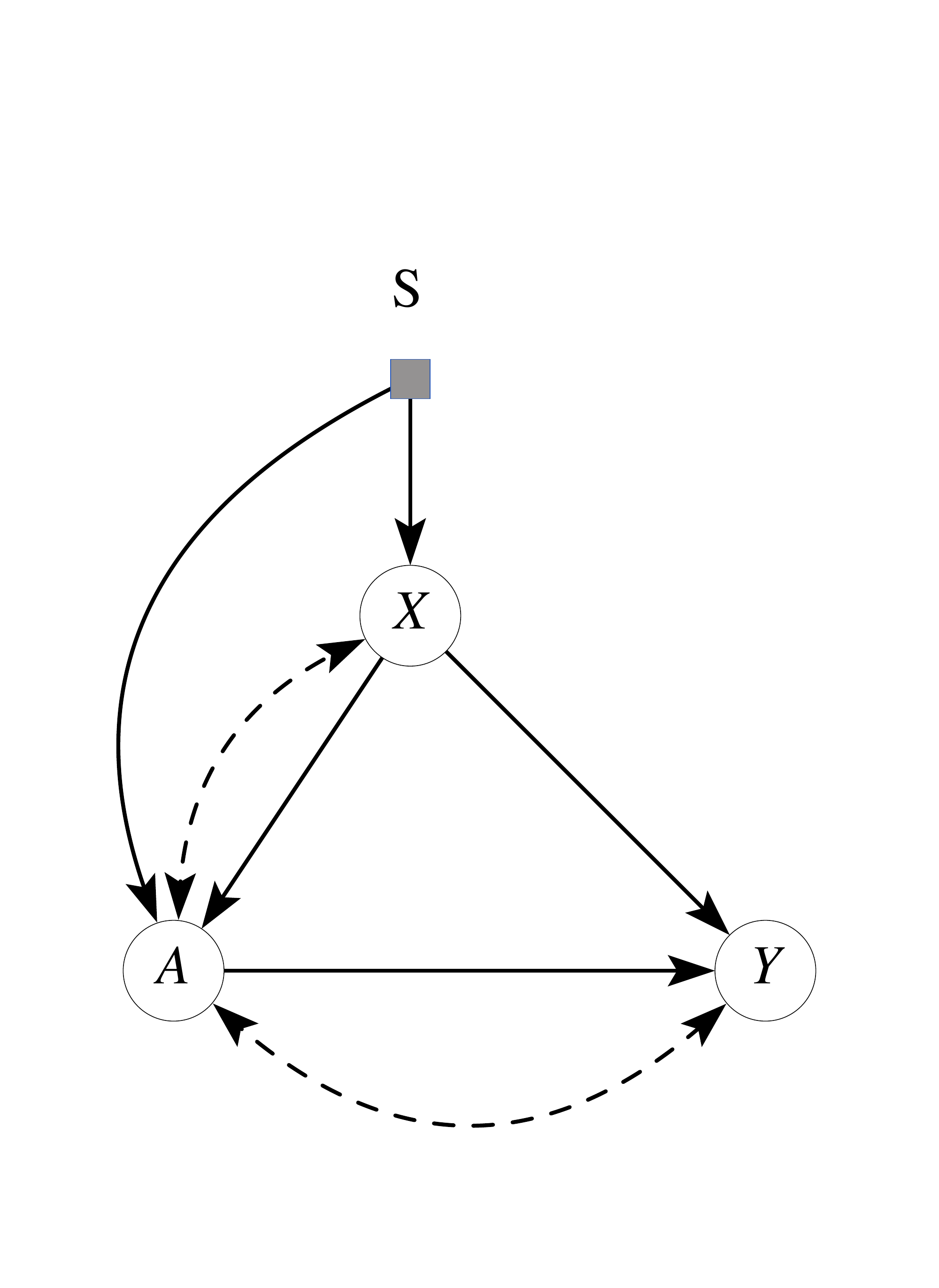} 
\caption{\textbf{Illustration of selection diagrams depicting differences between source and target populations}: In (a) and (b), the two populations differ by covariate distributions (indicated by $S$ pointing to $X$) and the two populations differ in their interventional nature ($S$ pointing to $A$). Assumption~\ref{a:trans} (transportability assumption) is assumed on (b), but not on (a) (since $S$ points to $Y$ in (a)). These two examples are inspired by \citet{pearl2011transportability}.}
\label{fig:selectiondiagram}
\end{figure}

\begin{figure}[!h]
	\begin{minipage}{.3\linewidth}
        \includegraphics[width=0.95\textwidth]{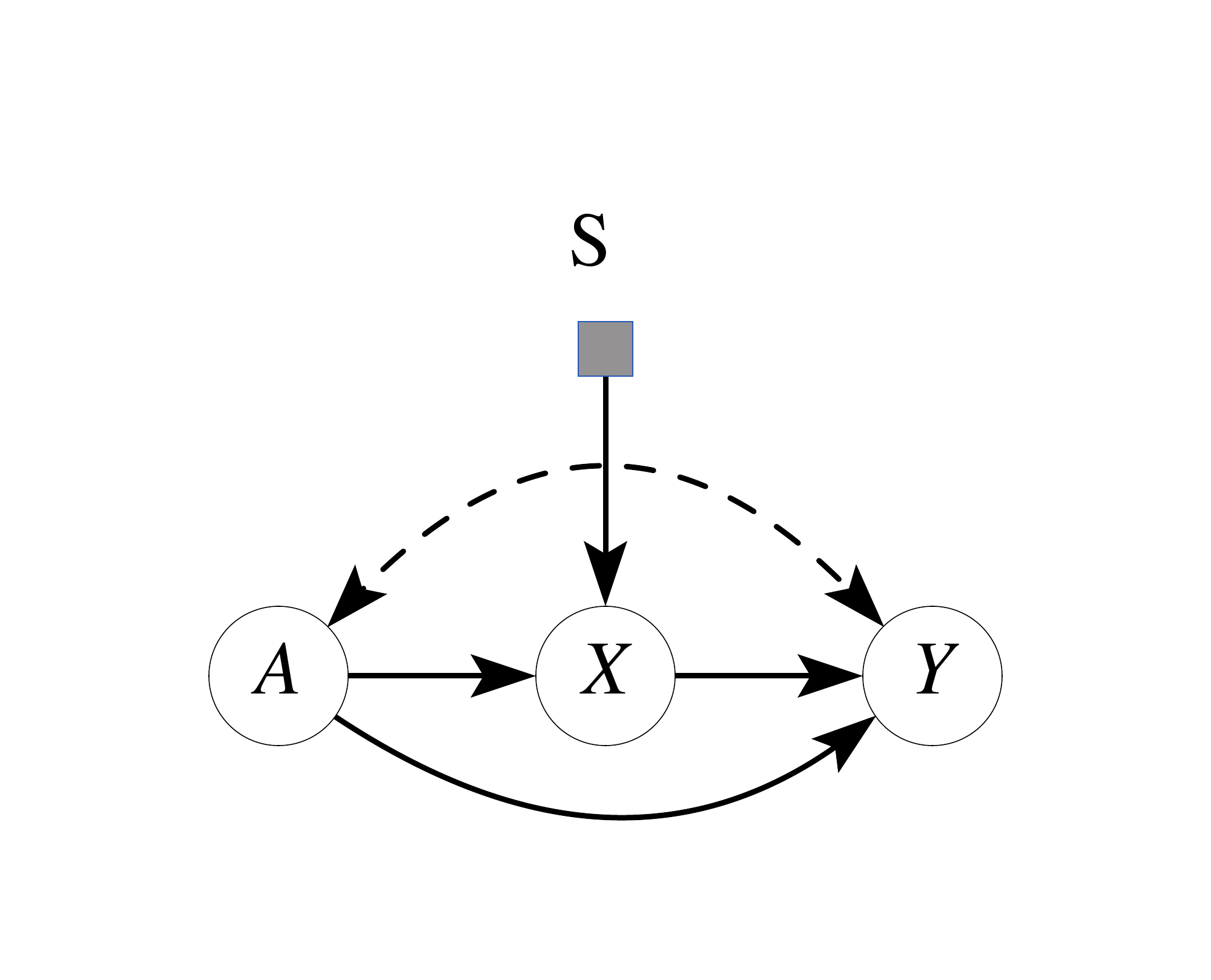}
    \end{minipage}
    \begin{minipage}{.69\linewidth}
	\caption{\textbf{Post-treatment covariate adjustment}: On this selection diagram the arrow from $S$ to $X$ indicates the assumption of different effect of $A$ on $X$ in the two populations. Here, $X$ is S-admissible but not S-ignorable, and the corresponding transport formula is $P(Y \mid do(A=a)) = \sum_{x} P(Y \mid do(A=a), X=x, S = 1)P(X=x \mid A = a),$
	where it invokes an unconventional average of the CATE weighted by a conditional probability in the target population. This example is taken from \citet{pearl2015findings}.}
    \label{fig:dag_posttreatment}
	\end{minipage}
\end{figure}

The SCM approach thus provides powerful machinery to generalize causal effect across populations, and entails a detailed description of the causal assumptions between variables in the selection diagram, including the selection variable $S$. The two selection diagrams of Figure~\ref{fig:selectiondiagram} represent for example transportability problems with a distributional change of covariates $X$ between the RCT and target populations (with an arrow from $S$ to $X$), and where the interventional nature of the RCT versus the target population is also represented with an arrow from $S$ to $A$. 
In addition, in Figure~\ref{fig:selectiondiagram}(a) the arrow from $S$ to $Y$ indicates that the conditional distribution of $Y$ given $X$ and $A$ differs between the two populations, which in general prevents any transportability of causal effect, while the lack of arrow from $S$ to $A$ in Figure~\ref{fig:selectiondiagram}(b) encodes the independence assumption $\mathbb{P}(Y\,|\,X,A) = \mathbb{P}(Y\,|\,X,A,S=1)$, which implies the transportability assumption $\mathbb{P}(Y\,|\,do(A=a),X,S=1) = \mathbb{P}(Y\,|\,do(A=a), X)$ (which itself implies Assumption~\ref{a:trans} in the PO framework).
In that case, one easily deduces by simple conditioning on $X$ that the distribution of $Y$ under intervention on the whole population is given by
\begin{equation}\label{eq:transport}
   \mathbb{P}(Y \mid do(A=a)) = \sum_{x} \underbrace{\mathbb{P}(Y \mid do(A=a), X=x, S=1)}_{RCT}\underbrace{\mathbb{P}(X=x)}_{Obs.}\,.
\end{equation}
This transport formula, also known as \emph{re-calibration}, \emph{re-weighting} or \emph{post-stratification} formula \citep{pearl2015findings}, thus combines experimental results obtained in the RCT population and the observational description of the target population to estimate the causal effect in the target population. In particular, we easily deduce the ATE on the target population by integrating~(\ref{eq:transport}) in $Y$ to get
\begin{equation}\label{eq:recalibration}
\tau = \sum_{x} \underbrace{\tau_1(x)}_{RCT}\underbrace{\mathbb{P}(X=x)}_{Obs.}\,,
\end{equation}
where 
$\tau_1(x)$ is by design identifiable by conditioning on treatment in the RCT population. This formula (\ref{eq:recalibration}) is equivalent to the regression formula (\ref{eq:idereg}) in the PO framework, which is valid under Assumption~\ref{a:trans}.
Interestingly, \citet{pearl2011transportability} show that the transport formula (\ref{eq:transport}) holds more generally as soon as $X$ is a set of pre-treatment variables which is \emph{$S$-admissible}, i.e., if $S \independent Y\mid X, do(A=a)$ for $a=0,1$. Graphically, $S$-admissibility holds whenever $X$ blocks all paths from $S$ to $Y$ after deleting from the graph all incoming arrows into $A$. We note that S-admissibility implies the mean exchangeability assumption (Assumption \ref{a:trans-3}) and is equivalent to the $S$-ignorability assumption $S \independent Y(a) \mid X$ (Assumption \ref{a:trans-2}) used in the PO literature  when $X$ and $S$ are pre-treatment variables, and entails similar transport formula in that situation.
However, the two notions differ for treatment-dependent selection and covariates, as discussed by~\citet{pearl2015findings}, where several examples illustrate how the $S$-admissibility assumption can lead to different transport formulas when both pre- and post-treatment variables are leveraged. Such an example is presented on Figure~\ref{fig:dag_posttreatment}, where the covariate $X$ is a post-treatment variable, for example a biomarker, believed to mediate between treatment and outcome.

\modif{Here, we presented how Assumptions~\ref{a:ign}, \ref{a:trans-2} and \ref{a:trans-3} are translated in the SCM literature and how another scenario with post-treatment covariates can be identified. More identifiability scenarios have been discussed in the SCM literature \citep{huang2012pearls, Bareinboim2013limitedExperiments, pearl2015findings, Lee2020arbitrarySurrogate}, and to our knowledge we have found no similar identifiability scenario in the PO literature.} It is worth mentioning that the transportation problem discussed so far, to export a causal effect estimated in an RCT to a general population is only one specific instance of the more general problem of \emph{data fusion} \citep{pearl2011transportability,bareinboim2012completeness, bareinboim2016causal, hunermund2019causal, Lee2020synthesizing}, which simultaneously accounts for confounding issues of observational data, sample selection issues, as well as extrapolation of causal claims across heterogeneous environments. The SCM framework, with its elegant way of formalizing the problem, helps practitioners  formulate and discuss causal assumptions across variables and environments. In particular, subject to a good knowledge of the graph, it helps selecting sets of variables that are sufficient to establish identifiability and exclude variables that would bias the analysis. As we will see in Section~\ref{sec:dataanalysis}, already in the early phase of a study, the causal and selection diagrams offer a very convenient tool to discuss with clinicians and explicitly lay out conditional independence assumptions. Once a diagram encodes assumptions about a system, algorithmic solutions implementing the $do$-calculus are available to determine whether non-parametric identifiability holds, and to provide correct formula if it holds \citep{correa2018generalized, tikka2019causal}.

While the SCM literature provides powerful and versatile sets of concepts and tools to \emph{identify} causal effects, practical estimators with publicly available implementations and detailed consistency, convergence rates or robustness results are still scarce. Some recent work has proposed solutions for this estimation task in the context of either experimental or observational data by extending weighting-based methods developed for the back-door case to more general settings \citep{jung2020estimating,Jung2020Learning}, or extending the double/debiased machine learning (DML) approach proposed by \citet{chernozhukov2018double} under ignorability assumption to any identifiable causal effect \citep{Jung2021Estimating}. 
In the same spirit, \cite{karvanen2020dosearch} propose combination of data from a survey and a meta-analysis of 34 trials, where identifiability and transport formula are the output of the algorithm \texttt{do-search} (see Section~\ref{sec:soft}), and estimation is performed with the real data at hand.
Additionally, even if a causal effect is not identifiable,  partial-identifiability techniques have been proposed for
deriving bounds for the causal effect \citep{Tian2000Probabilities,dawid2019bounding}.
\cite{CinelliGeneralizing2019} give an example illustrating partial identifiability on real data, with experiments assessing the effect of the Vitamin A supplementation. In this setting the existence of experimental data from one
source population leads to identify bounds on the transported causal effect, but the availability of two trials instead of one leads to a point estimate. 
Finally, \citet{dahabreh2018generalizing, dahabreh2020benchmarking} propose an alternative approach for generalizability and integrative analyses of trials and observational studies using structural equation models under weaker error assumptions and represented using single world intervention graphs \citep{richardson2013single}.

\section{Software for combining RCT and observational data}
\label{sec:soft}

\subsection{Review of available implementations}

An \modif{important} point to bridge the gap between theory and practice is the availability of software. In recent years, there have been more and more solutions for users interested in causal inference and causation, see \cite{tikka2017identifying, guo2018survey, yao2020survey} for surveys \modiff{and \cite{mayer2022cran} for a task view of \texttt{R} implementations}. 
Regarding the specific subject of this article, we present in Table~\ref{tab:software} the implementations available about both identifiability and estimators.
The available implementations are often dedicated to specific sampling designs and, as mentioned, estimators are different from nested and non-nested framework.
As a consequence, a new user has to pay attention to all of these practical -- but fundamental -- details.
\begin{table}[!h]
\caption{Inventory of publicly available code for generalization (top: software for identification; bottom: software for estimation). }
\label{tab:software}
\footnotesize
\begin{tabular}{lll}
\hline
 & &  \\[0cm]
Name &  Method - Setting & Source \& Reference \\ [0.5cm] \hline \hline
\rowcolor[HTML]{EFEFEF}  & & \\
\rowcolor[HTML]{EFEFEF} \multirow{-2}{*}{\textit{Identification}}& & \\\hline
causaleffect &  \makecell[l]{Identification and transportation of\\causal effects, e.g., conditional \\causal effect identification algorithm} & \makecell[l]{R package on CRAN,\\\cite{tikka2017identifying}} \\ [0.5cm]
dosearch &  \makecell[l]{Identification of causal effects \\from arbitrary observational and\\ experimental probability distributions \\via do-calculus} & \makecell[l]{R package on CRAN,\\\cite{tikka2019causal}} \\[0.5cm] 
Causal Fusion &  \makecell[l]{Identifiability in data fusion \\framework,   (Section \ref{sec:SCM})} & \makecell[l]{Browser beta version upon request \\\cite{bareinboim2016causal}}   \\[0.5cm]
 & & \\[0.1cm]

\hline \hline

\rowcolor[HTML]{EFEFEF}  & & \\
\rowcolor[HTML]{EFEFEF} \multirow{-2}{*}{\textit{Estimation}}& & \\\hline
ExtendingInferences & \makecell[l]{IPSW (Definition~\ref{def:ipsw}), \\  plug-in g-formula equation \eqref{eq:gformulanested} - Nested \\AIPSW \eqref{eq:aipsw-nested}  - Nested \\ Continuous outcome} & \makecell[l]{R code on \href{https://github.com/serobertson/ExtendingInferences}{GitHub},\\ \cite{dahabreh2020extending}} \\[1cm]
generalize &  \makecell[l]{IPSW (Definition~\ref{def:ipsw}),\\ TMLE (Section \ref{sec:DR})} & \makecell[l]{R package on \href{https://github.com/benjamin-ackerman/generalize}{GitHub} \\ \cite{ackerman2020generalization}} \\[0.5cm]
genRCT & \makecell[l]{IPSW (Definition~\ref{def:ipsw}),\\ calibration weighting (Section \ref{sec:DR}) \\ Continuous and binary outcome} & \makecell[l]{R package \\  \cite{dong2020integrative}} \\[0.5cm]

IntegrativeHTE  & Integrative HTE (Section \ref{sec:unmeasured confoundersATE}) & \makecell[l]{R package on \href{https://github.com/shuyang1987/IntegrativeHTE}{GitHub}, \\ \cite{yang2020elastic}}\\[0.5cm]
IntegrativeHTEcf  &   \makecell[l]{Includes confounding functions \\(Section \ref{sec:unmeasured confoundersATE})} & \makecell[l]{R package on \href{https://github.com/shuyang1987/IntegrativeHTEcf}{GitHub}, \\ \cite{yang2020elastic}} \\[0.5cm]
generalizing  & \makecell[l]{SCM with probabilistic graphical \\model for Bayesian inference \\ Binary outcome} & \makecell[l]{R package on \href{https://github.com/carloscinelli/generalizing}{GitHub}, \\ \cite{CinelliGeneralizing2019}} \\[1cm]
RemovingHiddenConfounding  &  \makecell[l]{Unmeasured confounder \\(Section \ref{sec:unmeasured confoundersATE})} & \makecell[l]{R package on \href{https://github.com/carloscinelli/generalizing}{GitHub}, \\ \cite{kallus2018removing}}
\\[1cm]
\modif{senseweight}  &  \makecell[l]{Sensitivity analysis \\(IPSW Definition~\ref{def:ipsw})} & \makecell[l]{R package on \href{https://github.com/melodyyhuang/senseweight}{Github} \\ \cite{huang2022sensitivity}}
\\[1cm]
transport  & \makecell[l]{Targeted maximum likelihood estimators (TMLEs)  \\ Transport} & \makecell[l]{R package on \href{https://github.com/kararudolph/transport}{GitHub}, \\ \cite{rudolph2018transport}} 
\\[1cm]
combine-rct-rwd-review  & \makecell[l]{Generalization estimators of Section~\ref{sec:noWnoY}  \\ } & \makecell[l]{R code on \href{https://github.com/BenedicteColnet/combine-rct-rwd-review}{GitHub}} 
\\[1cm]

\\[0.5cm]

\hline \hline
\end{tabular}
\end{table}

\subsection{Simulation study of \sout{the main approaches}\modiff{generalization estimators} }
\label{sec:simulations}

This part presents simulations results to illustrate the different
estimators introduced in Section~\ref{sec:noWnoY} and their behavior
under several mis-specifications patterns. The code to reproduce the
results is available on
\href{https://github.com/BenedicteColnet/combine-rct-rwd-review}{Github}\footnote{\texttt{https://github.com/BenedicteColnet/combine-rct-rwd-review}}.
\modif{We implement in R \citep{CRAN} our own version of the estimators to match}  exactly the formulas introduced in the review (\texttt{IPSW} and \texttt{IPSW.normd} see
Definition~\ref{def:ipsw}, \texttt{stratification}; Definition~\ref{def:strat},
\texttt{plug-in g-formula}; Definition~\ref{def:g-formula}, and \texttt{AIPSW};
Definition~\ref{def:aipsw}), \modif{except for the CW and ACW estimators
(Definitions ~\ref{def:cbss0}) and~\ref{def:taudc-2})
for which we use the \texttt{genRCT} package}.

\paragraph{Scenario 1: well-specified models.}
\modif{Similarly to \cite{dong2020integrative}, 
We generate non-nested trial settings as follow.
First, we draw a sample of size $50,000$ from a covariate distribution
with four covariates are generated independently as with $X_{j} \sim
\mathcal{N}(1,1)$ for each $j=1, \ldots, 4$. From this sample, we then
select an RCT sample of size $n \sim 1000$ with 
trial selection scores defined using a logistic regression model:} 
\begin{equation}
\label{eq:rctmodel}
    \operatorname{logit}\left\{\pi_S(X)\right\}=-2.5-0.5\,X_{1}-0.3\, X_{2}-0.5\, X_{3}-0.4\, X_{4}.
\end{equation}
\modif{Then, we generate the treatment according to a Bernoulli distribution
with probability equals to 0.5, $e_1(x) = e_1 = 0.5$ and the 
outcome according to a linear model:}
\begin{equation}
\label{eq:Ymodel}
    Y(a) =-100+27.4\, a\, X_{1}+13.7\, X_{2}+13.7\, X_{3}+13.7\, X_{4}+\epsilon \mbox{ with } \epsilon \sim \mathcal{N}(0,1).
\end{equation}
This outcome model implies a target population ATE of $\tau=27.4$, and $\mathbb{E}\left[X_1\right] = 27.4$.
\modif{Finally, we generate an observational sample by drawing a new
sample of size $m=10,000$ from the distribution of the covariates.}

\begin{center}
\begin{figure}[tb!]
    \begin{minipage}{.3\linewidth}
    \caption{\textbf{Well-specified model} Estimated ATE with the inverse propensity of sampling weighting with and without weights normalization (IPSW and IPSW.norm; Definition~\ref{def:ipsw}), stratification (with 10 strata; Definition~\ref{def:strat}), plug-in g-formula (Definition~\ref{def:g-formula}), calibration weighting (CW; Definition~\ref{def:cbss0}), augmented IPSW (AIPSW; Definition~\ref{def:aipsw}) \modif{and ACW (Definition~\ref{def:taudc-2}))} over 100 simulations.}
    \label{fig:sim-simple-100}
    \end{minipage}%
    \hfill%
    \begin{minipage}{.8\linewidth}
    \begin{center}
       \includegraphics[width= 0.85\textwidth]{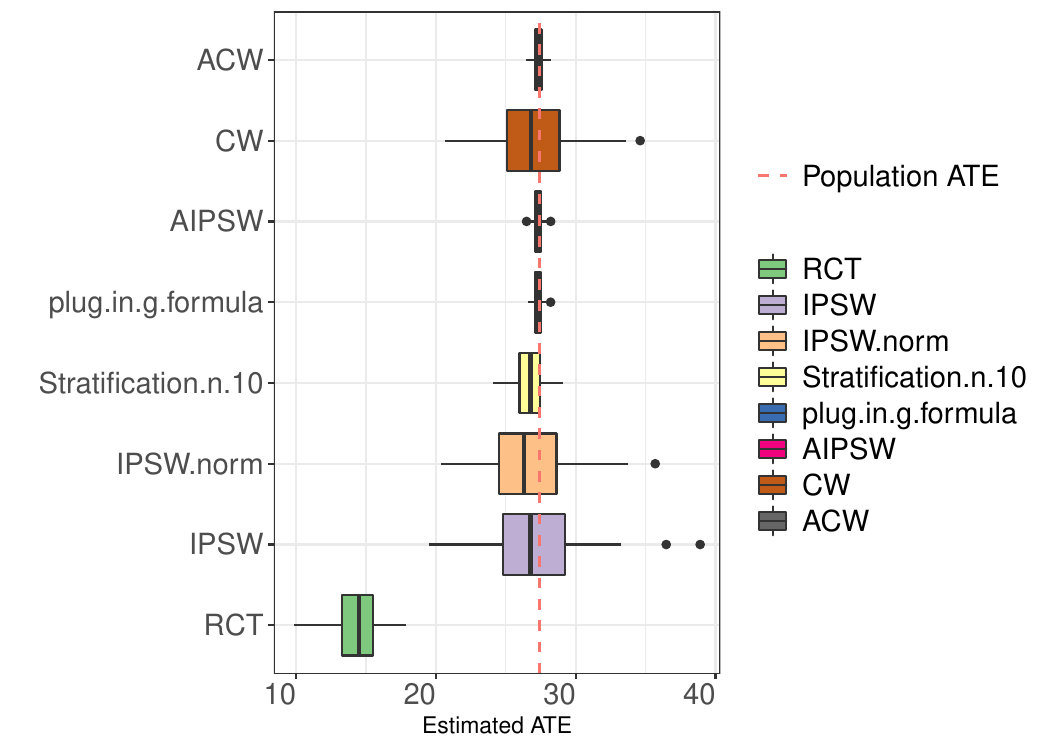}
    \end{center}
    \end{minipage}
\end{figure}
\end{center}

Figure \ref{fig:sim-simple-100} presents estimated ATE  
over 100 simulations.
The true ATE is represented with a dash line. The ATE estimated only with the RCT sample is also displayed as a baseline. As expected it is biased downward (its mean is equal to 14.24) as the distribution of the covariates and in particular the treatment effect modifiers such as $X_1$ is not the same in the trial sample and in the population (as illustrated in Table~\ref{fig:simbaseline} in Appendix~\ref{sec:appendixsimulations}). \modif{Note that in this simulation all the estimators are unbiased.} 
The variability of the two IPSW estimators are larger than the others. 
The number of strata in the stratification estimator plays an important role. 
As shown in Figure \ref{fig:simstrata} in Appendix ~\ref{sec:appendixsimulations}, the results are biased when the number of strata is smaller than $10$. 

\begin{figure}[h]
    \centering
    \includegraphics[width= 0.8\textwidth]{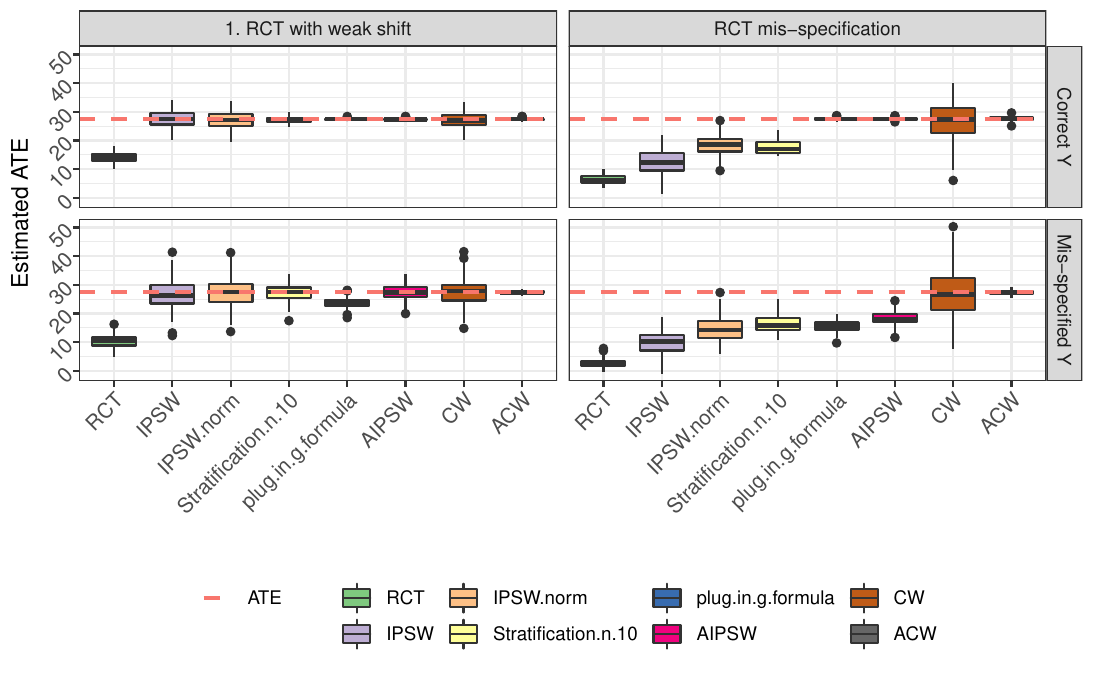}
    \caption{\textbf{Mis-specified models} Estimated ATE when selection in RCT and/or outcome models are mis-specified. Estimators used being IPSW (IPSW and IPSW.norm; Def.~\ref{def:ipsw}), stratification (with 10 strata; Def.~\ref{def:strat}), plug-in g-formula (Def.~\ref{def:g-formula}), calibration weighting (CW; Def.~\ref{def:cbss0}), augmented IPSW (AIPSW; Def.~\ref{def:aipsw}), \modif{and ACW (Def.~\ref{def:taudc-2}}) over 100 simulations.}
    \label{fig:sim-RCT-outcome-mis}
\end{figure}
\paragraph{Scenario 2: mis-specification of the sampling propensity score or outcome model.}

To study the impact of mis-specification of the sampling propensity score model, we generate the RCT selection according to the model
$$ \operatorname{logit}\left\{\pi_S(X)\right\}=-2.5-0.5\, e^{X_{1}}-0.3\, e^{X_{2}}-0.5\, e^{X_{3}}-0.4\, e^{X_{4}} + 3, $$
and  outcome according to the model
$$ Y(a) = -100 + 27.4\, a\, X_{1} X_{2}+13.7\, X_{2}+13.7\, X_{3}+13.7\, X_{4} + \epsilon. $$
The analysis is then performed using classical logistic and linear estimators on the four covariates. 
As shown in Figure \ref{fig:sim-RCT-outcome-mis}, when the sampling propensity score model is mis-specified, the IPSW estimators are biased; whereas when the outcome model is mis-specified, the \modiff{plug-in }g-estimator is biased. In both settings, the double robust estimator (AIPSW) is unbiased and robust to mis-specification. In the case where both models are mis-specified, all estimators are biased except the CW and ACW estimators. This demonstrates some robust properties of calibration against slight model mis-specification.

Appendix~\ref{sec:appendixsimulations} investigates the effect of a missing covariate, homogeneous treatment effect, and the impact of a stronger covariate shift, i.e., poorly satisfied Assumption~\ref{a:pos}.
\sout{A practical summary of these estimators and their properties can be found in Appendix~\ref{appendix:table-estimators}.
}

\section{Application: Effect of Tranexamic Acid\label{sec:dataanalysis}}



To illustrate the methodological question of combining experimental and observational data and demonstrate some of the previously discussed methods, we consider an open medical question about major trauma patients. 
We focus on trauma patients suffering from a traumatic brain injury (TBI)\modif{: brain damage caused by a blow or jolt to the head.}
Tranexamic acid (TXA) is an antifibrinolytic agent that limits excessive bleeding, commonly given to surgical patients.
Previous clinical trials showed that TXA decreases mortality in patients with traumatic \emph{extracranial} bleeding \citep{crash2}.
Such prior result raises the possibility that it might also be effective in TBI, because \emph{intracranial} hemorrhage is common in TBI patients\modif{, with risks of raised intracranial pressure, brain herniation, and death.} 
Therefore the aim here is to assess the potential decrease of mortality in patients with intracranial bleeding when using TXA. 
To answer \modif{this question}, we have at our disposal both an RCT, \textit{CRASH-3}, and an observational study, the \textit{Traumabase}. Both data have previously been analyzed separately in \cite{crash32019,crash3} (for the RCT) and in \cite{mayer2019doubly} (for the observational study) and the medical teams of both studies want to share their respective data to answer both medical and methodological questions. 
Such initiatives allow to: \modif{\modif{(1) collate the results from the observational study with the RCT findings; (2) assess the generalizability methods, considering the Traumabase as the target population, and \modif{assess} the estimators presented in this review \modif{in a real application.}}}
We first present the two data sources, treatment effect analyses and findings from these, before turning to the combined analysis in Section \ref{sec:combine-crash3-traumabase}. The code to reproduce all \modif{these} analyses is available on  \href{https://github.com/BenedicteColnet/combine-rct-rwd-review}{Github}\footnote{\texttt{https://github.com/BenedicteColnet/combine-rct-rwd-review}}, however the medical data cannot be publicly shared for privacy \modif{concerns}.



\subsection{The observational data: Traumabase}
\label{sec:traumabase_study}

\subsubsection{Context}

\modif{The \href{http://www.traumabase.eu/fr_FR}{Traumabase} regroups 23 French Trauma centers that collect detailed clinical data from major trauma patients from the scene of the accident to hospital discharge in form of a registry.} The data\modif{, currently counting over 30,000 patient records,} are of unique granularity and size in Europe. However, they are highly heterogeneous, with both categorical -- sex, type of illness, ...-- and quantitative -- blood pressure, hemoglobin level, ...-- features, multiple sources, and many missing data (98\% of the records are incomplete). 
\modif{Here, we use
8,270 patients suffering from TBI extracted from the Traumabase}. 
\modif{\citet{mayer2019doubly} performed a first, purely observational, study to assess the effect of TXA on mortality for traumatic brain injury patients from this data: 
the treatment variable is the administration of TXA during pre-hospital care or on admission to a Trauma Center\footnote{More precisely, to the resuscitation room of a hospital equipped to treat major trauma patients.} within three hours of the initial trauma.}
\modif{The Traumabase analysis contains many missing values (see Appendix~\ref{appendix:sup-TB}), which implies additional assumptions to perform causal inference}.

\subsubsection{Purely-observational results from two different estimation strategies} \label{sec:restrauma}
\modif{The direct causal effect of TXA on 28-day intra-hospital TBI-related mortality and on all cause intra-hospital mortality among traumatic brain injury patients is estimated by} \modif{a}djusting for confounding using 17 confounding variables. In addition, 21 variables predictive of the outcome but not related to the treatment are included  (see \citet{mayer2019doubly} for the detailed \sout{list and DAG}\modiff{adjustment set}). 
We recall the results from this study \modiff{which put a focus on how to estimate treatment effects in the presence of incomplete data.} \sout{the treatment effect estimation based on} \modiff{The presented methods rely} either \modiff{on} logistic regressions or generalized random forests \citep{athey_etal_AS2019} for the nuisance components, denoted respectively by \textit{GLM} and \textit{GRF} in Table~\ref{tab:tbi_txa_results}. 
The doubly robust results (AIPW) in Table~\ref{tab:tbi_txa_results} show that from this study there is no evidence for an effect of TXA on mortality of TBI patients. 
These findings ---obtained prior to the publication of CRASH-3---are consistent with the main conclusion of the CRASH-3 study.
However, the results from IPW conclude on a possible deleterious effect. In such a situation, the possibility to generalize the treatment effect from the RCT is also a step to comfort the results.
 \modif{I}n Appendix~\ref{app:trauma_strata}\modif{, we additionally recall results on sub-groups obtained by stratifying along trauma severity}.

\begin{table}[H] 
\caption{\textbf{ATE estimations from the Traumabase} for TBI-related 28-day mortality. Red cells conclude on deteriorating effect, white cells can not reject the null hypothesis of no effect. GLM stands for Generalized Linear Models and GRF for Generalized Random Forests to estimate nuisance components. Two estimators of the treatment effect are considered: IPW and AIPW, as well as two methods to deal with missing values: multiple imputation or missing incorporated in attribute (MIA) in GRF.}
\label{tab:tbi_txa_results}
\centering
\footnotesize
\begin{tabular}{|l|cccc|cc||c|}
\cline{2-8}
\multicolumn{1}{l|}{\multirow{3}{*}{}}     & \multicolumn{4}{c|}{\small Multiple imputation (MICE)} & \multicolumn{2}{c||}{\small GRF-MIA} & \multirow{3}{*}{\makecell{ Unad-\\justed\\ATE\\ $\times 10^{2}$}} \\ \cline{2-7}
\multicolumn{1}{l|}{}                      & \multicolumn{2}{c|}{\makecell{\small IPW\\(95\% CI)\\ $\times10^{2}$}} & \multicolumn{2}{c|}{\makecell{\small AIPW\\(95\% CI)\\ $\times10^{2}$}} & \multicolumn{1}{c|}{\makecell{\small IPW\\(95\% CI)\\$\times10^{2}$}} & \makecell{\small AIPW\\(95\% CI)\\$\times10^{2}$}                    &                   \\ \cline{2-5}
\multicolumn{1}{l|}{}                      & \multicolumn{1}{c|}{GLM} & \multicolumn{1}{c|}{GRF} & \multicolumn{1}{c|}{GLM} & \multicolumn{1}{c|}{GRF} &         \multicolumn{1}{c|}{}        &                &   \\ \hline
\makecell[l]{Total\\ ($n=8248$)} & \multicolumn{1}{l|}{\cellcolor{red!05}\makecell{\small15\\(6.8, 23)}} & \multicolumn{1}{l|}{\cellcolor{red!05}\makecell{\small11\\(6.0, 16)}}  & \multicolumn{1}{l|}{\makecell{\small3.4\\(-9.0, 16)}} & \makecell{\small-0.1\\(-4.7, 4.4)} & \multicolumn{1}{l|}{\cellcolor{red!05}\makecell{\small9.3\\(4.0, 15)}} & \makecell{\small-0.4\\(-5.2, 4.4)}                      &  \cellcolor{red!05}  16               \\ \hline
\end{tabular}
\end{table}



\subsection{The RCT: CRASH-3}

\subsubsection{Context}

CRASH-3 is a multi-centric randomized and placebo-controlled trial launched over 175 hospitals in 29 different countries \citep{crash3protocol}. This trial recruited 9,202 adults --unusually large for a medical RCT--, all suffering from TBI with only intracranial bleeding, i.e., without major extracranial bleeding. All participants were randomly administrated TXA \citep{crash32019,crash3}. 
The primary outcome studied is head-injury-related death in hospital within 28 days of injury in patients included and randomized within 3 hours of injury. 
The study concludes that the risk of head-injury-related death is 18.5\% in the TXA group versus 19.8\% in the placebo group. The causal effect, measured as a Risk Ratio (RR) was not significant (RR = 0.94 [95\% CI 0.86 - 1.02])).  \modif{Note that CRASH-3 revealed a positive effect of TXA only when considering mild and moderate cases. In the Appendix~\ref{app:trauma_strata}, we provide a complementary analysis to study this sub-group.} 


\subsubsection{RCT selection}
\begin{figure}[h!]
\centering
\includegraphics[width=.6\textwidth]{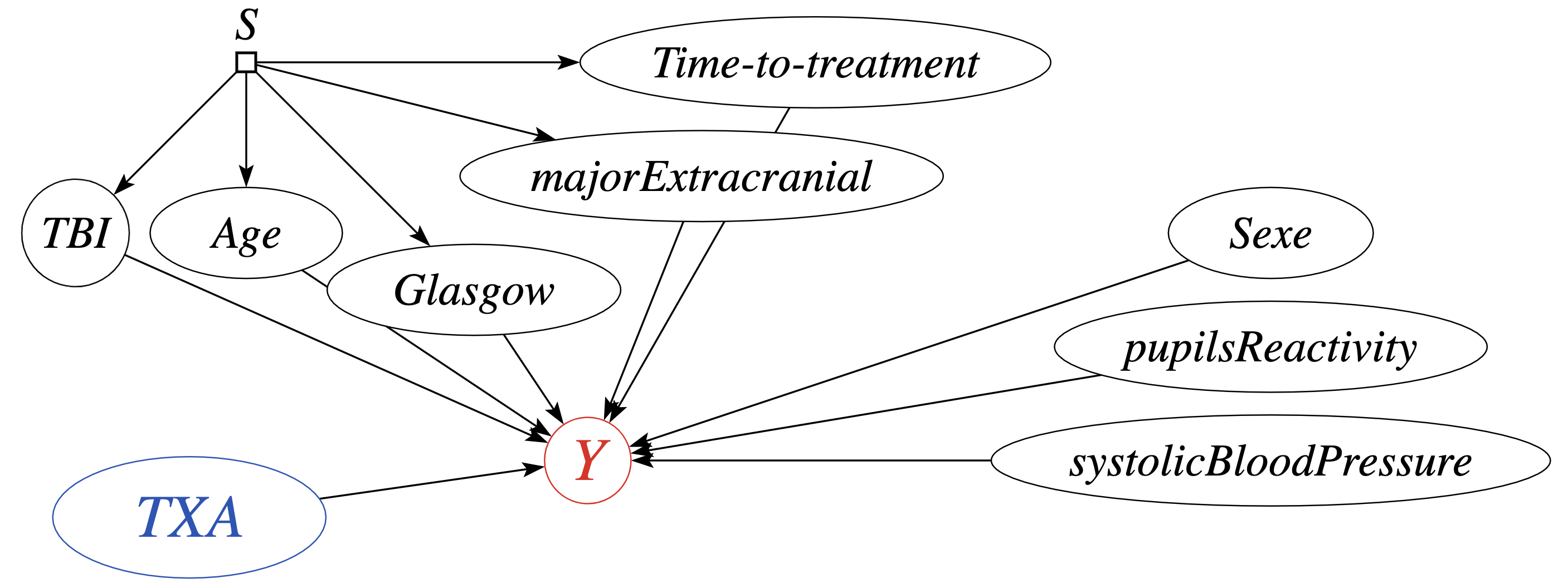}
\caption{\textbf{Structural causal diagram} representing treatment, outcome, inclusion criteria with $S$ and other predictors of outcome (Figure generated using the Causal Fusion software presented in Section~\ref{sec:soft} from \cite{bareinboim2016causal}).}
\label{fig:crash3-tbi-txa-dag}
\end{figure}

Six covariates are present at baseline, being age, sex, time since injury, systolic blood pressure, Glasgow Coma Scale score (GCS)\footnote{The Glasgow Coma Scale (GCS) is a neurological scale which aims to assess a person's consciousness. The lower the score, the higher the severity of the trauma.}, and pupil reaction. 
The inclusion criteria of the trial are patients with a GCS score of 12 or lower or any intracranial bleeding on CT scan (computed tomography), and no major extracranial bleeding. 
We provide a DAG summarizing the trial selection and predictors of the outcome present in CRASH-3 in Figure~\ref{fig:crash3-tbi-txa-dag}. 

\subsection{Transporting the ATE on the observational data \label{sec:combine-crash3-traumabase}}

With the two separate analyses in mind, we \modif{can now turn to} the combined analysis, more specifically, the generalization from the RCT results to the target population defined by the observational Traumabase registry. Before any analysis aiming to compare and combine two data sets an important step is to assess that baseline covariates, treatment, and outcome are the same \modif{(for details,} see Appendix~\ref{sec:combine-crash3-traumabase-covariates}). 



\subsubsection{Descriptive analyses}

\paragraph{Missing values.}
\modif{T}he RCT contains almost no missing values, whereas the variables for determining eligibility in the observational data contain important fractions of missing values, ranging from 0.27 to 29 \%. 
\modif{Thus the methods discussed in this review must be adapted to account for missing values\footnote{If we assumed the missing values being missing completely at random (MCAR), we could ``throw away'' the incomplete observations and perform the analyses on the complete observations, but this would reduce the total sample size to 917 observations. And as explained in Section \ref{sec:traumabase_study}, the MCAR assumption is not plausible for the present observational data, thus such a \textit{complete case analysis} would be biased.}\sout{estimation of trial include and outcome models (Section \ref{sec:estimationPO}) on covariates with missing values, especially on the observational data} 
\sout{For this}\modiff{In order to estimate the nuisance components, i.e., the conditional odds and the outcome model(s)}, despite the missing data, we explore two alternative strategies}:
\modif{(1)} logistic regression with incomplete covariates using an expectation maximization algorithm \citep{dempster1977maximum}, a computationally efficient variant of this method using stochastic approximation is implemented in the R package \texttt{misaem} \citep{jiang2020misaem}; \modif{(2)} generalized regression forest with missing incorporated in attributes \citep{twala_etal_2008, josse2019consistency}, this method is implemented in the R package \texttt{grf} \citep{grf}.



\begin{table}[h!]
\begin{center}
\caption{Sample sizes for both studies.}
\footnotesize
\label{tab:sample_sizes}
\begin{tabular}{|c|c|c|c|c|c|}
\hline
 \multicolumn{3}{|c|}{Traumabase} & \multicolumn{3}{c|}{CRASH-3} \\
       m    &   \#treated    &  \#death   & n & \#treated &   \#death  \\ \hline
    
  8248  &   683    &   1411  & 9168 & 4632 & 1745  \\ \hline
\end{tabular}
\end{center}
\end{table}

\begin{figure}
    \centering
    \includegraphics[width=0.8\textwidth]{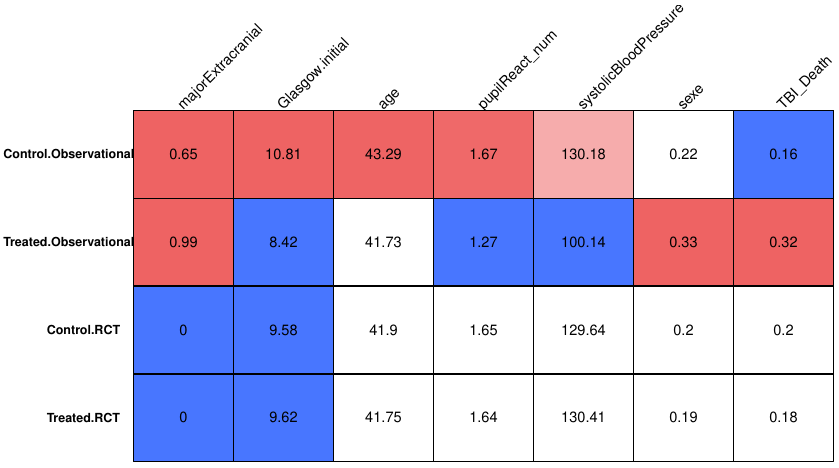}
    \caption{\textbf{Distributional shift} and difference in terms of univariate means of the trial inclusion criteria (red: group mean greater than overall mean, blue: group mean less than overall mean, white: no significant difference with overall mean, numeric values: group mean (resp. proportion for binary variables). Graph obtained with the \texttt{catdes} function of the \texttt{FactoMineR} package \citep{FactoMineR}.} 
    \label{fig:catdes_eligibility}
\end{figure}

\paragraph{Distribution shift.}
\modif{Simple comparisons of the means of the covariates between the treatment groups of the two studies --Figure \ref{fig:catdes_eligibility}-- reveal} the fundamental difference between the two studies, namely the treatment \modif{assignment} bias in the observational study and the balanced treatment groups in the RCT. 
\modif{In Appendix \ref{sec:shift-crash-tb} we further explore the distribution shift with univariate} histograms (Figures \ref{fig:histogramsshiftage}
--\ref{fig:histogramsshiftpupil}).

\subsubsection{Analyses}

\paragraph{Notations and estimator details.}
We use two consistent ATE estimators from the CRASH-3 data, namely the difference in mean estimator (\texttt{Difference in means}; Section \ref{sec:rctonly}) and the difference in conditional mean relying on OLS (\texttt{Difference in conditional means}).
\modif{We also present the results from the purely observational study outlined earlier: AIPW coupled with multiple imputation (\texttt{MI AIPW}) and AIPW based on nuisance parameters estimated via generalized random forest (\texttt{GRF AIPW}) that can directly handle missing values when needed with missing incorporated in attribute strategy.}

To generalize the ATE to the target population, we apply the estimators discussed in this review \modif{while implementing strategies to handle the missing values. The resulting estimators are presented in Table~\ref{tab:estimators-with-missings}.}

\begin{table}[H]
\caption{\modif{Overview of generalization estimators based on different missing values handling strategies used in the data analysis.}}
\label{tab:estimators-with-missings}
\begin{tabular}{cc|cc|}
\cline{3-4}
\multicolumn{2}{l|}{\multirow{2}{*}{}}                               & \multicolumn{2}{c|}{Missing values strategy}                                                     \\ \cline{3-4} 
\multicolumn{2}{l|}{}                                                & \multicolumn{1}{c|}{Logistic regression  with missing values} & Generalized random forests (grf) - MIA\\ \hline
\multicolumn{1}{|c|}{\multirow{3}{*}{$\hat \tau_{n,m}$}} & IPSW      & \multicolumn{1}{c|}{\texttt{EM IPSW}}                                  & \texttt{GRF IPSW}                         \\ \cline{2-4} 
\multicolumn{1}{|c|}{}                                   & Plug-in g-formula & \multicolumn{1}{c|}{\texttt{EM Plug-in g-formula}}                             & \texttt{GRF Plug-in g-formula}                    \\ \cline{2-4} 
\multicolumn{1}{|c|}{}                                   & AIPSW     & \multicolumn{1}{c|}{\texttt{EM AIPSW}}                                 & \texttt{GRF AIPSW}                        \\ \hline
\end{tabular}
\end{table}

The confidence intervals of these estimators are computed with a stratified bootstrap in the RCT and the observational data set in order to maintain the ratio of relative size of the two studies (with 100 bootstrap samples). Note that the Calibration Weighting  estimators (CW and ACW) are not used in this analysis as they would require a specific adaptation to the case of the missing values. 

\begin{figure}[h!]
    \centering
    \includegraphics[width=0.85\textwidth]{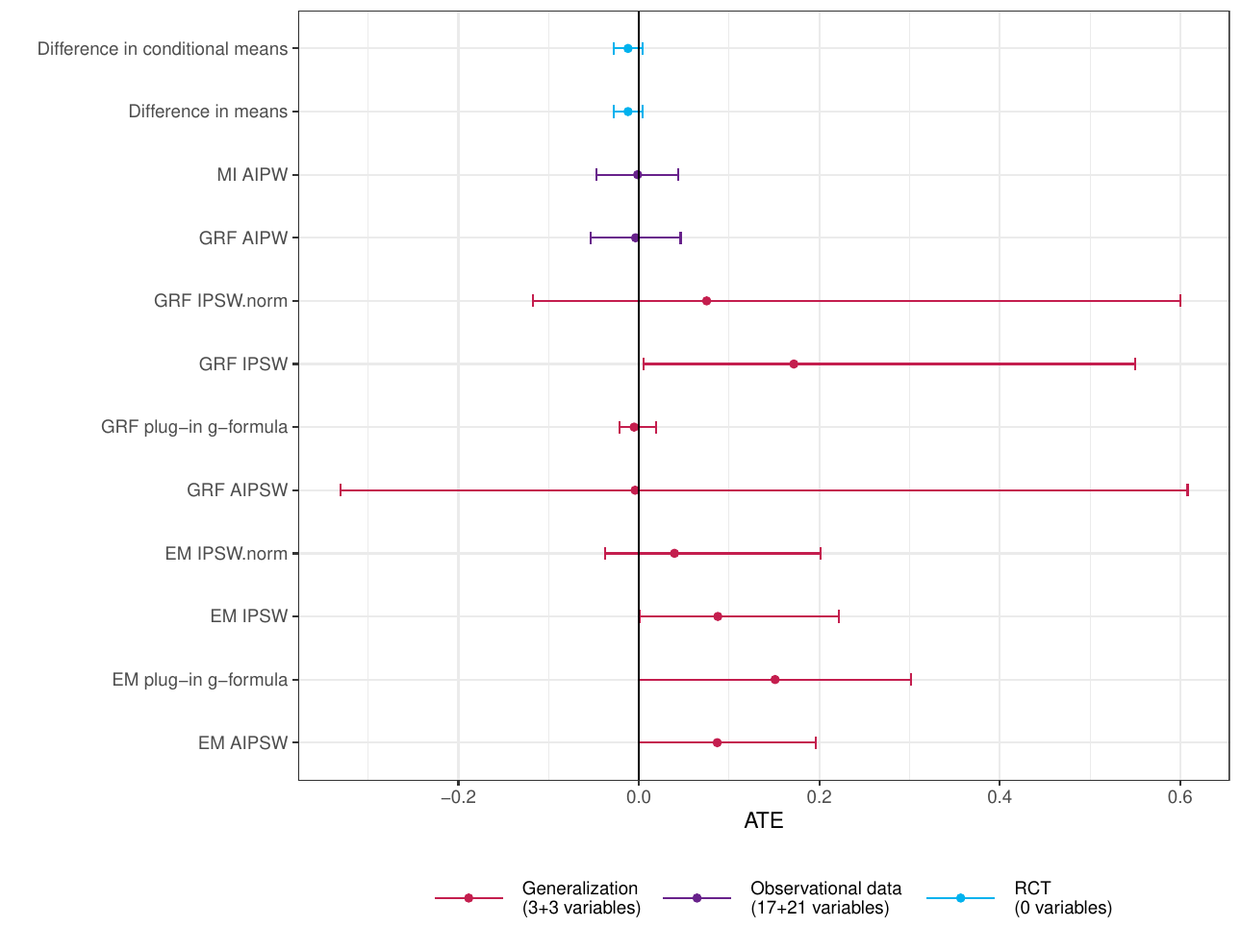}
    \caption{\textbf{Juxtaposition of different estimation results} with ATE estimators computed on the Traumabase (observational data set), on the CRASH-3 trial (RCT), and transported from CRASH-3 to the Traumabase target population. All the observations are used. Number of variables used in each context is given in the legend.}
    \label{fig:results_crash3_traumabase}
\end{figure}

\paragraph{Results of the combined analysis.}
\modif{Figure~\ref{fig:results_crash3_traumabase} gives the generalization from the RCT to the target population using all the observations from both data sets, showing certain discrepancies with respect to the separate analysis results.} On the one hand, one half of the generalization estimators support the CRASH-3 conclusion about the treatment effect: no significant effect. On the other hand, some estimators point towards a deleterious treatment effect. 
Recall that the AIPW ATE estimations from the \modif{purely observational data study} do not reject the null hypothesis of no treatment effect. 
Note that these results are to be interpreted carefully due to the potential impact of missing values on the performance of the chosen estimators. For example, the large confidence intervals for the \texttt{GRF} estimators when used to estimate weights are likely to be due to the imbalanced proportions of missing values in the RCT and the observational data. Indeed, the variance is much smaller using the plug-in g-formula with GRF. Dealing with missing values when generalizing a treatment effect \modif{remains an open research question}.

Here we present the results transported onto the total TBI Traumabase population, but the CRASH-3 study highlights a specific subgroup of patients (mild and moderate
patients) for which a positive effect of the tranexamic acid is measured. The generalization of the CRASH-3 findings onto this subgroup in the Traumabase raises multiple methodological issues that still need to be addressed in future works (detailed in Appendix \ref{sec:limitations-crash-tb}).


Overall this data analysis highlights \modif{the interest of combining two different data sets, but also some challenges}: the need for a good understanding of the common covariates, exposure, and outcome of interest before combining the data sets, different missing data patterns, and poor overlap when considering specific target (sub-)populations.

\section{\sout{Summary and shortcomings} \modiff{Conclusion}\label{sec:Concluding-remarks}}

\modif{Combining observational data and RCTs can improve many aspects of causal inference, from increased statistical power to better external validity. A large part of this review is dedicated to generalizability and transportability of RCT from one population to another. The corresponding rich and prolific literature answers a real practical concern: external validity. Indeed, questions about external validity arise as soon as there are treatment effect heterogeneities in the populations under study. \sout{And yet, most effects of interest present some form of heterogeneity: mortality rates increase with age, different populations are exposed to different risks, have different comorbidities, etc. More recently, the literature has increasingly tackled other goals around data combination, for example in order to improve statistical efficiency of estimation or to handle an unobserved confounder in an observational study. } We find that, as any growing scientific field, the ideas are in flux:  notations differ, implementations are scattered, and the proposed methods proposed still lack real-world benchmarks, generated hand in hand with practitioners. \sout{To conclude, we would like to highlight some open questions that became apparent in this review.} \modiff{In addition, many open questions still remain as detailled below.}}

\paragraph{Discrepancies between RCTs and observational data.} The application on tranexamic acid effect hinted to moderate external validity of the RCT as the generalized ATE is  concordant with the findings from the RCT, at least for half of the estimators. Additionally, the purely observational data study also supports the results from the RCT. 
Determining which analysis to trust depends on the assumptions we are willing to make -- either related to transportability or unconfoundedness -- as well as the suitability of the selected variables.
Beyond these assumptions, caution is needed when interpreting the results, as observing the methods in action reveals threats to validity. 
\sout{In particular the impact of the missing values when combing RCT and observational data requires further research (detailed below).}
The target population of interest and overlap also raise concerns. Considering certain strata  revealed violated positivity, which leads to a non-transportable treatment effect on the strata of interest: mild and moderate patients. Therefore, further discussions and analyses with the medical expert committee are necessary to re-define a target population of interest on which generalization is possible and medically relevant. As it is generally the case, beyond methodological and theoretical guarantees, a major step to be taken before applying a set of methods is to clearly state the causal question and estimand(s) and the associated identifiability requirements. This task is even more complex when combining data sets. A primary and fundamental concern is whether outcome, treatment, and covariates are comparable in the two studies \citep{hernan2019comparingrctandobs}. \sout{Indeed, a plausible explanation of the difference is that the treatment effect depends on the timing, and that the time to treatment differs slightly across both data sets. } 

\modif{\paragraph{Right choice of covariates to answer the question.} Domain expertise can be used to postulate a causal graph: a directed acyclic graph representing the mechanisms (as Figure~\ref{fig:crash3-tbi-txa-dag}). The SCM framework is then convenient to assess whether the
question of interest can be formulated in an identifiable way. This approach offers a principled way of
selecting variables needed for identification of the causal effect and to avoid biased causal effect estimates. Without such an approach, identifiability claims are limited. A common practical recommendation is to include as many variables as possible to avoid violation of any assumption as proposed for e.g. by \cite{stuart2017CaseStudyDifficulties, Ling2022CriticalReview} and \cite{dahabreh2019extending}: ``it is probably best to include as many outcome predictors as possible in regression models for the expectation of the outcome or the probability of trial participation''. \modiff{On the contrary, a recent work alerts about the bad consequences of adding covariates that are shifted between the two populations while not being treatment effect modifiers, resulting in variance inflation \citep{Colnet2022Reweighting}.} In its current state, the field probably lacks work on covariate selection and its impact on bias and variance. Some recent works propose the use
of causal graphs to select optimal adjustment sets that allow the reduction of the variance of the final estimation \citep{rotnitzky2019efficient, witte2020efficient, guo2020efficient}\modiff{, but such methods have not yet been developed for generalization or data fusion}. \sout{However these works are not made in the context of combining experimental and observational data.}}

\modif{\paragraph{Challenges in handling missing values.}
In our data analysis, we have seen the need to account for missing values, and in particular different missing value patterns between data sources. Missing values typically occur more often in observational data since in RCTs, investigators typically deploy significant efforts to avoid them.  RCTs may however suffer from participants missing scheduled visits or completely dropping out from the study. The literature for RCT mainly focuses on missing outcome data and calls for sensitivity analysis given
that available strategies to handle such missing data (weighting, multiple imputation) rely on
untestable assumptions about the missing values mechanism
\citep{carpenter2007missing, little2012prevention, ken2013handling,
o2014clinical, li2019best, cro2020four}. 
Missing values may lead to subtle biases in the inferences, as they are seldom 
uniformly distributed across both data sets -- occurring more in one than in the
other. While a recent research work proposes an assessment of the effect of different missing data patterns \citep{mayer2021transporting}, further research is needed to clarify identifiability conditions and estimators in this setting in order to better understand the scope of each method.} 

\section*{Acknowledgment} 
This work is initiated by a SAMSI working group jointly led by JJ and SY in the 2020 causal inference program.
We would like to acknowledge the helpful discussions during the SAMSI working group meetings.
We also would like to acknowledge the discussions and insights from the Traumabase group and physicians, in particular Drs Fran\c{c}ois-Xavier \textsc{Ageron}, Tobias \textsc{Gauss}, and Jean-Denis \textsc{Moyer}. In addition, none of the data analysis part could have been done without the help of Dr. Ian \textsc{Roberts} and the CRASH-3 group, who shared with us the clinical trial data.
Part of this work was performed while JJ was a visiting researcher at Google Brain Paris. Finally, we would like to warmly thank Issa \textsc{Dahabreh} for his comments, suggestions of additional references, and insightful discussions.

\bibliographystyle{apalike}
\bibliography{ci,ci2,ci3,ci4}

\pagebreak{}

\global\long\def\theequation{S\arabic{equation}}%
\setcounter{equation}{0}

\global\long\def\theexample{S\arabic{example}}%
\setcounter{equation}{0}

\global\long\def\thesection{S\arabic{section}}%
\setcounter{section}{0}

\global\long\def\thetheorem{S\arabic{theorem}}%
\setcounter{equation}{0}

\global\long\def\thecondition{S\arabic{condition}}%
\setcounter{equation}{0}

\global\long\def\theremark{S\arabic{remark}}%
\setcounter{equation}{0}

\global\long\def\thestep{S\arabic{step}}%
\setcounter{equation}{0}

\global\long\def\theproof{S\arabic{proof}}%
\setcounter{equation}{0}

\global\long\def\theproposition{S\arabic{proposition}}%
\setcounter{equation}{0}

\global\long\def\theassumption{S\arabic{assumption}}%
\setcounter{assumption}{0}

\appendix

\section{Randomized controlled trial}
\label{sec:rctonly}
This section recalls assumptions and estimators for average treatment estimation in the case of a single RCT. The assumptions for average treatment effect identifiability in RCTs are the SUTVA assumption and assumptions \ref{a:consist} (consistency) and \ref{a:ign} (random treatment assignment within the RCT). These assumptions allow the average treatment effect to be identifiable. The most intuitive estimators coming from these assumptions is the difference-in-means estimators:

\begin{equation}
\label{eq:DMinRCT}
    \hat \tau_{\text{\tiny DM}, n}=\frac{1}{n_{1}} \sum_{A_i=1} Y_{i}-\frac{1}{n_{0}} \sum_{A_i=0} Y_{i}
\end{equation}

With $n_1$ being the number of individuals in the trial that have been treated and $n_0$ the number of individuals in the trial who have not been treated ($n_0 + n_1 = n$). This estimator is unbiased and $\sqrt n$-consistent if the trial is a random sample of the target population. If not, it is a biased estimation of the population average treatment effect. 



\section{Estimation of ATE in observational data}
\label{sec:cate}

Under classical identifiability assumptions, it is possible to estimate
the ATE and CATE based only on the observational data. In what follows, we \modif{briefly} recall the usual
assumptions\modif{, which can be seen as an introduction to Section~\ref{sec:specialcase}}. 

\begin{assumption}[Unconfoundedness]\label{asump:unconfounded}$Y(a)\independent A\mid X$
for $a=0,1$.
\end{assumption}
Assumption \ref{asump:unconfounded} (also called \textit{ignorability} assumption) states that treatment assignment is as good as random conditionally on the attributes $X$. In other words, all confounding factors are
measured. Unlike the RCT, in observational studies, its plausibility relies on whether or not the observed covariates $X$ include all the confounders that affect the treatment as well as the outcome.

\begin{assumption}[Overlap]\label{assump:overlap}
There exists a constant $\eta>0$ such that for almost all $x$, $\eta\,<e(x)\,<\,1-\eta$.
\end{assumption}
Assumption \ref{assump:overlap} (also called \textit{positivity} assumption) states that the propensity score $e(\cdot)$ is  bounded away from $0$ and $1$ almost surely.\\

Under Assumptions \ref{asump:unconfounded} and \ref{assump:overlap},
the ATE can be identified based on the following formulas from the
observational data:
\begin{enumerate}
\item Reweighting formulation:
\begin{equation}
\tau= \EE[\frac{AY}{e(X)}-\frac{(1-A)Y}{1-e(X)}];\label{eq:idipwobs}
\end{equation}
\item Regression formulation:
\begin{equation}
\tau=\EE[\tau(X)]=\EE[\mu_{1}(X)-\mu_{0}(X)].\label{eq:idregobs}
\end{equation}
\end{enumerate}
For example the identification formulas\modif{, and more particularly the reweighting formulation,} motivates the Inverse Propensity Weighting (IPW) estimator \citep{hirano2003efficient},

\begin{equation}
   \hat \tau_{\ipw, m} = \frac{1}{m} \sum_{i=1}^{m} \left \{ \frac{A_{i} Y_{i}}{e(X_i)} - \frac{(1-A_{i}) Y_{i}}{1 - e(X_i)}\right \}, 
   \label{eq:classicalipw}
\end{equation}
where $e(x) = P(A=1 \mid X = x)$ is the propensity score, i.e., the probability to be treated given the covariates. The rationale of IPW is to upweight treated observations with a small propensity score (and the other way around) to balance the two groups, treated and non treated, with respect to their covariates.
These identification formula motivate also the regression estimators or doubly robust estimators based solely on the observational data. \modif{Efficient estimation of the ATE with one single observational data set and non-parametric models is detailed in \cite{vanderlaan2011tmlebook, kennedy2016semiparametric, chernozhukov2018double}}
There are \modif{also many available methods} to estimate the CATE, based on the observational data such as causal forests \citep{causalforest}, causal BART \citep{hill2011bayesian,hahn2020bayesian}, causal boosting \citep{powers2018some}, or causal multivariate adaptive regression splines (MARS) \citep{powers2018some}. There are also meta-learners such as the S-Learner
\citep{kunzel2018causaltoolbox}, T-learner \citep{kunzel2018causaltoolbox},
X-Learner \citep{kunzel2019metalearners}, MO-Learner \citep{rubin2007doubly,kunzel2018causaltoolbox}, modified covariate method (MCM) \citep{tian2014simple,chen2017general}, modified covariate method with efficiency augmentation (MCM-EA)  \citep{tian2014simple,chen2017general}, 
and R-learner \citep{nie2017quasi}, which build upon any base learners for regression or supervised classification.
\citet{knaus2021machine} and \citet{powers2018some} conduct comprehensive simulation studies to compare these methods.

\section{Identification formula}\label{sec:appendixiden}

This part focuses on the non-nested design only, as it corresponds to the central design of this review.

\textbf{Identification by the g-formula or regression formula in the target population}

\begin{proof}
\label{proof:gformulatarget}
  \begin{align*}
    \mathbb{E}[Y(a)] &=  \mathbb{E}\left[ \mathbb{E}[Y(a) \mid X] \right] &&\text{Law of total expectation}\\
    &=\mathbb{E}\left[ \mathbb{E}[Y(a) \mid X,  S = 1] \right] &&\text{Assump. \ref{a:trans-3}} \\
    &= \mathbb{E}\left[ \mathbb{E}[Y(a) \mid X,  S = 1, A = a] \right] &&\text{Assump. \ref{a:ign}}\\
    &= \mathbb{E}\left[ \mathbb{E}[Y \mid X,  S = 1, A = a] \right] &&\text{Assump. \ref{a:consist}} \qedhere
  \end{align*}
\end{proof}

This last quantity can be expressed as a function of the distribution of $X$ in the target population:
$$
\mathbb{E}\left[ Y(a)\right] = \int \mathrm{E}[Y \mid X=x, S=1, A=a] d f(x) \,,
$$
where $f(X)$ denotes the distribution of $X$ in the target population.\\

\noindent \textbf{Identification by weighting}


\begin{proof}
\label{proof:weightingtarget} 
    \begin{align*}
        \tau &= \EE[\tau(X)] &&\text{Law of total expectation}\\
        &= \EE[\tau_1(X)] &&\text{Assump. \ref{a:trans}}\\
        &= \EE[\frac{f(X)}{f(X\mid S=1)} \tau_1(X) \mid S=1] &&\text{Assump. \ref{a:pos}.}\\
    \end{align*}
\end{proof}
Using Bayes' rule, we note that
$$
\frac{f(x)}{f(x\mid S=1)} = \frac{\pr(S=1)}{\pr(S=1 \mid X=x)} = \frac{\pr(S=1)}{\pi_S(x)}\,.
$$
In this expression, however, it is important to notice that neither $\pi_{S}(x)$ nor $\pr(S=1)$ can be estimated from the data, because we do not observe the $S$ indicator in the observational study (Figure \ref{fig:designs}). On the other hand, the conditional odds $\alpha(x)$ can be estimated by fitting a logistic regression model that discriminates RCT versus observational samples, and Bayes' rule gives:
\begin{equation*}
    \begin{split}
\alpha(x) 
&= \frac{\mathbb{P}(i\in\mathcal{R}\mid \exists i\in\mathcal{R}\cup\mathcal{O}, X_i=x)}{\mathbb{P}(i\in\mathcal{O}\mid \exists i\in\mathcal{R}\cup\mathcal{O}, X_i=x)} \\
&= \frac{\mathbb{P}(i\in\mathcal{R})}{\mathbb{P}(i\in\mathcal{O})}\times \frac{\mathbb{P}(X_i = x \mid i\in\mathcal{R})}{\mathbb{P}(X_i = x \mid i \in\mathcal{O})} \\
&= \frac{n}{m}\times \frac{f(x \mid S=1)}{f(x)}\,,
\end{split}
\end{equation*}
and therefore
$$
\tau = \EE[\frac{n}{m \alpha(X)} \tau_1(X) \mid S=1]\,.
$$

This quantity can be further developed, underlying $\tau_1(X)$ identification as presented in the following proof \ref{proof:tau1identif}.

\begin{proof}
\label{proof:tau1identif}
    \begin{align*}
    \tau_1(x) &= \EE[Y(1) -  Y(0) \mid X=x, S = 1] \\
    &= \EE[Y(1) \mid X=x, S = 1] - \EE[ Y(0) \mid X=x, S = 1]\\
    &= \frac{\EE[A \mid X=x, S = 1] \EE[Y(1) \mid X=x, S = 1]}{e_1(x)} \\
    & \qquad -\frac{\EE[1-A \mid X=x, S = 1] \EE[Y(0) \mid X=x, S = 1]}{1-e_1(x)} \\
    &= \frac{\EE[A Y(1) \mid X=x, S = 1]}{e_1(x)} - \frac{ \EE[(1-A) Y(0) \mid X=x, S = 1]}{1-e_1(x)} &&\text{Assump. \ref{a:ign}}\\
    &= \frac{\EE[AY \mid X=x, S=1]}{e_1(x)} - \frac{E[(1-A)Y \mid X=x, S=1]}{1-e_1(x)} &&\text{Assump. \ref{a:consist}} \\
    &= \EE[ \frac{A}{e_1(x)}Y - \frac{1-A}{1-e_1(x)}Y \mid X=x, S=1]\,. \qedhere 
    \end{align*}
\end{proof}

\section{Sources of formal statements of estimators described in Section~\ref{sec:estimationPO}}\label{appendix:extension-with-theorems} 

\modiff{
This section proposes formal statements on the statistical properties of the exposed estimators in the form of theorems. As part of a review work, this section only reports results that are stated along a Theorem environment and with explicit proof in the original papers.

\subsection{Inverse Propensity of Sampling Weighting}

Beyond the result from \cite{colnet2021causal} recalled in plain document, other theoretical results on the IPSW can be found in:

\begin{itemize}
\item \cite{Egami2021CovariateSelection}, which provides finite sample unbiasedness, consistency and asymptotic normality of an oracle version of the IPSW, that is an estimator where the true $\alpha$ is known (see their appendix, Section SM-2).
\item \cite{buchanan2018generalizing}, which provides consistency and asymptotic normality assuming that the \sout{sampling score}\modiff{conditional odds} are well approached by a parametric model (for e.g. a logistic regression). Results are detailed both in the main paper (p.7) and in appendix for detailed derivations. Note that they also obtain asymptotic normality and consistency for an oracle version of the IPSW. Their proof rely on M-estimation methods \citep{Stefanski2002Mestimation, lunceford2004stratification}, writing the estimation problem as a stacked equation, with the specificity that the observations are not necessarily identically distributed. The authors retrieve a well-known result in causal inference: estimating the weights leads to a gain in variance. Note that the proof is done in the context of a nested design, which is not exactly the purpose of the review. Without stating theoretical results, \cite{Zivich2022Delicatessen} extends this work to non-nested design showing how to compute the sandwich type confidence intervals. \cite{buchanan2018generalizing} also propose sandwich-type estimation of variance, while noting that estimation of the variance of the oracle version of IPSW would provide conservative but valid confidence intervals. 
\item \cite{dahabreh2020extending}, which announces consistency of the IPSW for parametric estimator of the RCT selection model $\alpha(X)$, and sketches the proof in Appendix for both a normalized and non-normalized version of the IPSW (see Section A). Note that derivations are made in the context of a nested design but said to extend to a non-nested design.
\item \cite{colnet2021causal}, which provides consistency (i.e. asymptotically unbiased) for any consistent parametric or non-parametric method to estimate $\alpha$.
\item \cite{Colnet2022Reweighting}, which provides finite and large sample bias and variance when the adjustment set is constituted of categorical covariates. The consistency is a by-product of their results. To our knowledge, their results is the only one characterizing different variance regimes depending on the size of the two data sample (RCT and observational).
They also recommend to estimate the probability to be treated in the trial $e_1(X)$ to decrease the asymptotic variance.
\end{itemize}

\subsection{Stratification}

\begin{itemize}
\item \cite{o2014generalizing} provide a formula of the variance under the situation where the strata estimates are assumed independent and the estimation of the strata proportion $m_l/m$ is without error (i.e. infinite target sample).
\item \cite{buchanan2018generalizing} provide asymptotic normality for the stratification estimator, assuming that the estimator is the average of $L$ independent, within-stratum, treatment effect estimators \citep{lunceford2004stratification, tipton2013improving}. They propose a formula for the asymptotic variance.
\end{itemize}

\subsection{Calibration Weighting}
\cite{dong2020integrative} provide regularity conditions and theoretical properties  of the CW and ACW estimators in terms of consistency, asymptotic normality, and inference procedures.  The proof can be found in the supplementary material of \cite{dong2020integrative}. 
}
\section{Nested study design}
\label{sec:appendixnested}

The nested trial design has different impacts on the estimators expressions previously introduced, and even on the causal quantity of interest. In a nested trial design the randomized trial is embedded in a cohort (e.g. a large cohort - considered as a sample from the target population - in which eligible people are proposed to participate in the trial, but if they refuse they are still included in the cohort study). \textbf{As a consequence, $S$ is the binary indicator for trial participation, with $S=1$ for participants and $S=0$ for non-participants.} Therefore the sampling probability of non-randomized individuals is known in nested trial designs \citep{lesko2017generalizing, buchanan2018generalizing, dahabreh2019study}. Mathematically it means that the quantity $\pr(S=1)$ is identifiable. In addition, two causal quantities can be identified: $\EE [Y(1) - Y(0)]$ and $\EE [Y(1) - Y(0) \mid S=0]$. It is important to note that the second quantity can have a scientific interest in order to better understand heterogeneities within the cohort, and variables that influence the sampling selection and/or the treatment effect on the outcome.

\subsection{When observational data have no outcome and treatment information}

Main estimators, such as IPSW, g-formula, and doubly-robust estimators are presented for the specific case of nested trial design.

\subsubsection{IPSW}
 
In this design the weights in the IPSW estimators are different, because the quantity $\pi_S$ can be estimated directly from the observed data as the indicator $S$ is observed. This allows the IPSW formula to be closer to the classic IPW expression without the need to use the odds to weight data. The IPSW expression is the following:

\begin{equation}
\label{eq:ipswnested}
\hat \tau_{\ipsw \text{-nested},n, m}=\frac{1}{n}\sum_{i=1}^{n}\frac{n}{n+m}\frac{A_{i}Y_{i}}{\hat \pi_{ S,n,m}(X_i)e_1(X_i)}-\frac{1}{n}\sum_{i=1}^{n} \frac{n}{n+m}\frac{(1-A_{i})Y_{i}}{\hat \pi_{ S,n,m}(X_i)(1-e_1(X_i))}.
\end{equation}
The normalized version is the following one:

\begin{align}
\label{eq:ipswnestednormalized}
\begin{split}
\hat \tau_{\ipsw \text{-nested norm.},n,m}=&\frac{\sum_{i=1}^{n} (\hat \pi_{ S,n,m}(X_i) e_1(X_i))^{-1}A_{i}Y_{i}}{\sum_{i=1}^{n} (\hat \pi_{ S,n,m}(X_i) e_1(X_i))^{-1}A_{i}}\\& \,-\frac{\sum_{i=1}^{n}(\hat \pi_{ S,n,m}(X_i) (1-e_1(X_i)))^{-1}(1-A_{i})Y_{i}}{\sum_{i=1}^{n}(\hat \pi_{ S,n,m}(X_i)(1-e_1(X_i)))^{-1}(1-A_{i})}.
\end{split}
\end{align}

{\footnotesize{\begin{proof}
\label{proof:ipswnested}
  \begin{align*}
        \tau &= \EE[\tau(X)] &&\text{Law of total expectation}\\
        &= \EE[\tau_1(X)] &&\text{Assump. \ref{a:trans}}\\
        &= \EE[\frac{f(X)}{f(X\mid S=1)} \tau_1(X) \mid S=1] &&\text{Assump. \ref{a:pos}}\\
        &= \EE[\frac{P(S = 1)}{\pi_S(X)} \tau_1(X) \mid S=1] &&\text{Bayes law}\\
        &= \EE[\frac{n}{n+m} \pi_{ S}(X_i)^{-1}\tau_1(X) \mid S=1] &&\text{$P(S = 1) = \frac{n}{n+m}$ in the nested design}
  \end{align*}
\end{proof}}
Where $\pi_{ S}$ can be estimated directly using the randomized and the non randomized data. $\tau_1$ is further derived as presented in proof \ref{proof:tau1identif}.

}

\subsubsection{G-formula}

The g-formula formulation in the case of nested trial design depends on the causal quantity of interest. When the target population is the causal quantity of interest, then the identification expression is the same as in the non-nested design. But, because $f \ne f_{. \mid S =0}$, the estimator's expression is slightly different:
\begin{equation}
\label{eq:gformulanested}
\hat \tau_{g-nested,n,m} = \frac{1}{n+m}\sum_{i=1}^{n+m}\left(\widehat \mu_{1, 1,n}(X_i) - \widehat \mu_{0,1,n}(X_i)\right),
\end{equation}

In the case where the population of interest is the non-randomized one, the identification of the causal quantity of interest is the following:

\begin{equation}
 \mathrm{E}\left[Y^{a} \mid S=0\right] =\mathrm{E}[\mathrm{E}[Y \mid X, S=1, A=a] \mid S=0] = \EE[\mu_{1,1}(X)-\mu_{0,1}(X) \mid S = 0] 
\end{equation}
The Proof \ref{proof:gformulanested} details the calculus. And the estimator is the same as given in Definition~\ref{def:g-formula} as the integration is done on the law $f_{. \mid S =0}$.

{\footnotesize{\begin{proof}
\label{proof:gformulanested}
  \begin{align*}
    \mathbb{E}[Y(a) | S = 0] &= \mathbb{E}[ \mathbb{E}[Y(a) \mid X]  | S = 0] &&\text{Law of total expectation}\\
    &= \mathbb{E}[ \mathbb{E}[Y(a) \mid X,  S = 1]  | S = 0 ] &&\text{Assump. \ref{a:trans-3}} \\
    &= \mathbb{E}[ \mathbb{E}[Y(a) \mid X,  S = 1, A = a]  | S = 0 ] &&\text{Assump. \ref{a:trans-3}}\\
    &= \mathbb{E}[ \mathbb{E}[Y \mid X,  S = 1, A = a] | S = 0 ] &&\text{Assump. \ref{a:consist}} \qedhere
  \end{align*}
\end{proof}}

This last quantity can be expressed as a function of the distribution of $X$ in the non-randomized population:
$$
\mathbb{E}[Y(a)] = \int \mathrm{E}[Y \mid X=x, S=1, A=a] f(x | S = 0) dx
$$

where $f(X | S = 0)$ denotes the density function of $X$ in the non-randomized population. 

}

\subsubsection{Doubly-robust estimator}

Similarly to the doubly-robust estimation in the non-nested case (Section \ref{sec:DR}), the g-formula and the IPSW methods can be leveraged into a doubly-robust estimator. The AIPSW expression for the nested case is the following:

\begin{align}
\begin{split}
\hat \tau_{\text{AIPSW-nested},n,m}= &  \frac{1}{n+m}  \sum_{i=1}^{n+m} \frac{S_i A_{i}}{\hat \pi_{S,n,m}(X_i) e_1(X_i)}
(Y_{i}-\widehat\mu_{1,1,n}(X_i)) \\ &- \frac{1}{n+m} \sum_{i=1}^{n+m} \frac{S_i(1-A_{i})}{\widehat\pi_{S,n,m}(X_i)(1 - e_1(X_i) )}(Y_{i}-\hat \mu_{0,1,n}(X_i)) \\
&+\frac{1}{m+m}\sum_{i=1}^{m+n} \left\{ \hat \mu_{1, 1,n}(X_{i})-\hat \mu_{0, 1,n}(X_{i})\right\} .\label{eq:aipsw-nested}
\end{split}
\end{align}

\subsection{Combining treatment-effect estimates from both sources of data}

Under Assumptions \ref{a:consist}, \ref{a:ign} and \ref{a:trans-2} for the RCT and
Assumptions \ref{asump:unconfounded} and \ref{assump:overlap} for
the observational data, separate estimators of the ATEs from the two
data sources can be constructed.  \citet{lu2019causal} considered the ATEs for the comprehensive cohort studies (CCS) which include participants who would like to be randomized, constituting the RCT, and participants who would like to choose the treatment by their preference, constituting the observational sample. In particular, they considered the ATE over the CCS study population $\tau_2$ and the ATE over the trial population $\tau_1$. Note that $\tau_2$ is different from $\tau$ in our setting because $\tau_2$ is defined with respect to the combined RCT and observational sample; while $\tau$ is defined with respect to the observational sample only.
In order to construct improved estimators
by combining study-specific estimators, they derived the optimal influence functions for 
$\tau_1$ and $\tau_2$, 
which suggest that the
efficient estimators of $\tau_1$ and $\tau_2$ can be obtained by 
\begin{multline*}
\widehat{\tau}_{1,\eff}=\frac{1}{n}\sum_{i=1}^{n+m}\left[\frac{\widehat{\pi}_{S}(X_{i})A_{i}Y_{i}}{\widehat{e}(X_{i})}+\left\{ S_{i}-\frac{A_{i}\widehat{\pi}_{S}(X_{i})}{\widehat{e}(X_{i})}\right\} \widehat{\mu}_{1}(X_{i})\right.\\
-\left.\frac{\widehat{\pi}_{S}(X_{i})(1-A_{i})Y_{i}}{1-\widehat{e}(X_{i})}-\left\{ S_{i}-\frac{(1-A_{i})\widehat{\pi}_{S}(X_{i})}{1-\widehat{e}(X_{i})}\right\} \widehat{\mu}_{0}(X_{i})\right],
\end{multline*}
\[
\widehat{\tau}_{2,\eff}
=\frac{1}{n+m}\sum_{i=n}^{n+m}\frac{A_{i}\{Y_{i}-\widehat{\mu}_{1}(X_{i})\}}{\widehat{e}(X_{i})}-\frac{(1-A_{i})\{Y_{i}-\widehat{\mu}_{0}(X_{i})\}}{1-\widehat{e}(X_{i})}+\{\widehat{\mu}_{1}(X_{i})-\widehat{\mu}_{0}(X_{i})\},
\]
where $\widehat{e}_{1}(X_{i})$, $\widehat{\mu}_{0,1}(X_{i})$, and
$\widehat{\mu}_{1,1}(X_{i})$ for units in the RCT are simplified as
$\widehat{e}(X_{i})$, $\widehat{\mu}_{0}(X_{i})$, and $\widehat{\mu}_{1}(X_{i})$.


\subsection{Softwares: Examples of implementations}

This part completes Section \ref{sec:soft} and proposes specific examples of implementations, such as identifiability questions with the package \texttt{causaleffect}, the beta version of \texttt{causalfusion}, and implementation examples for the nested case.

\subsubsection{R package \texttt{causaleffect}}

The  R packages \texttt{causaleffect}  \citep{tikka2017identifying} and \texttt{dosearch} \citep{tikka2019causal} can be used for causal effect identification,  with the later handling transportability, selection bias and missing values (bivariates) issues simultaneously. 
  In this package, the \texttt{dosearch} function takes the observable  distributions, a query, and a semi-Markovian causal graph as the input and outputs a formula for the query over the input distributions, or decides that it is not identifiable. It is based on a search
algorithm that directly applies the rules of do-calculus. Their general identification procedure is not necessary complete given an arbitrary query and an arbitrary set of input distributions
In order to retrieve the backdoor criterion in theorem \ref{theo:back}, one can write:

\begin{lstlisting}
data <- "P(Y, X,Z)"
query <- "P(Y|do(X))"
graph <- "X -> Y 
           Z -> X 
           Z -> Y"
dosearch(data, query, graph)
\end{lstlisting}

\begin{lstlisting}
$identifiable
[1] TRUE
$formula
[1] "[sum_{Z} [p(Z)*p(Y|X,Z)]]"

\end{lstlisting}

\subsubsection{Beta version of \texttt{causalfusion}}

The beta version of causal fusion \citep{bareinboim2016causal} can be used, with a user-friendly interface requiring no coding skills. For example, if uploading the selection diagrams from Figure~\ref{fig:selectiondiagram} onto this interface, it will state that diagram (a) is not transportable, while (b) is transportable along with the correct transport formula. The authors also propose to load their diagrams from previous publications and research works, some of which have been discussed in this review.

\subsubsection{IPSW for the nested case} 
\label{subsec:ipswnestedsoft}

Te IPSW estimator can be implemented using the available code from \cite{dahabreh2018generalizing}. It requires as input a \texttt{data.frame} (here called \texttt{study}) which columns represent treatment, denoted by $A$ (binary), the RCT indicator, denoted as $S$ (binary), the outcome as $Y$ (continuous), and the quantitative covariates. The current available code for 3 quantitative covariates denoted $X_1$, $X_2$, $X_3$ is presented below. A first function \texttt{generate\_weights()} estimates the sampling propensity score and the propensity score as logistic regressions, and compute the according weights to each data point. The variance is estimated with the \texttt{geex} library \citep{geex2020package} through the \texttt{m\_estimate} function which computes the empirical sandwich variance estimator.

\begin{lstlisting}
# Compute selection score model and propensity score in the trial (logit)
weights <- generate_weights(Smod = S~X1+X2+X3, Amod = A~X1+X2+X3, study)

# Use these scores to compute IPSW 
IOW1 <- IOW1_est(data = weights$dat)

# Compute the empirical sandwich variance
param_start_IOW1 <- c(coef(weights$Smod) , coef(weights$Amod),
                     m1 = IOW1$IOW1_1, m0 = IOW1$IOW1_0, ate = IOW1$IOW1) 
IOW1_mest <- m_estimate( estFUN = IOW1_EE, data  = study,
root_control = setup_root_control(start = param_start_IOW1)) 
  
# Format the output
IOW1_ate <- extractEST(geex_output = IOW1_mest, 
est_name ="ate",
param_start = param_start_IOW1)
\end{lstlisting}

The output is:
\begin{lstlisting}
print(IOW1_ate)
>    ate        SE 
>  -0.16961  0.02751 
\end{lstlisting}

\subsubsection{G-formula for the nested case}
\label{subsec:gformulanestedsoft}

The G-formula can also be implemented in the nested design using the available code from \cite{dahabreh2018generalizing}. It takes a similar entry as the IPSW previously presented. The variance is estimated with the \texttt{geex} library \citep{geex2020package} through the \texttt{m\_estimate} function which computes the empirical sandwich variance estimator.

\begin{lstlisting}
# Linear regression cond. outcome mean as a function of covariates on the RCT
# Compute ATE on the observational data 
OM <- OM_est(data = study)

# Compute the empirical sandwich variance
param_start_OM <- c(coef(OM$OM1mod), coef(OM$OM0mod), 
                   m1=OM$OM_1, m0=OM$OM_0, ate=OM$OM)
OM_mest <- m_estimate( estFUN = OM_EE, data  = study,
  root_control = setup_root_control(start = param_start_OM))

# Format the output
OM_ate <- extractEST(geex_output = OM_mest, est_name = "ate", 
param_start = param_start_OM)

\end{lstlisting}

The output is:
\begin{lstlisting}
>   ate        SE 
> -0.1934  0.0300
\end{lstlisting}

\section{Additional information on the SCM framework\label{sec:appendixscm}}

\subsection{Notations and Assumptions \label{sec:appendixscm-intro}}

This supplementary introduction aims to provide an introduction to the whole SCM framework, and introduce the graphical representation, along with the do-calculus concepts and notations.

\begin{figure}
\centering
     \subfigure{%
         \includegraphics[width=.48\linewidth]{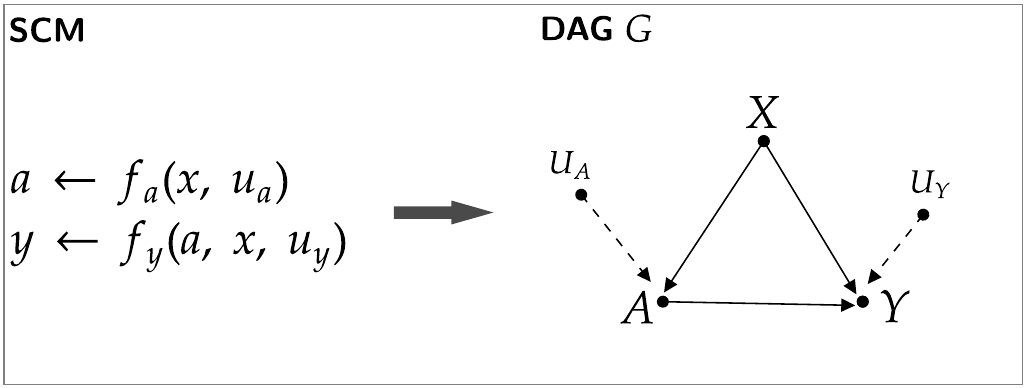}%
         \label{fig:scm}
         
     }\hfill%
     \subfigure{
         \includegraphics[width=.5\linewidth]{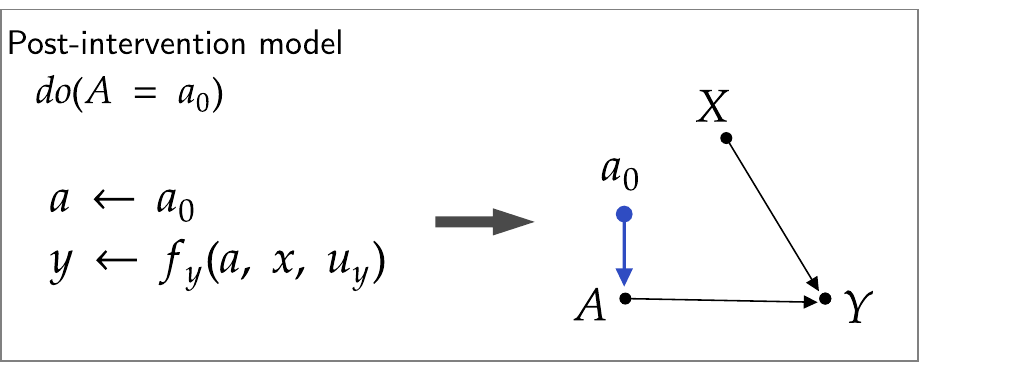}%
         \label{fig:post_int}
     }
    \caption{Left: (a) example of an SCM $M$ and corresponding DAG; right: (b) Post-intervention graph of $M$ for $do(A = a_0)$. }
\end{figure}

\paragraph{Structural Causal Models.} 
Formally \citep[p.203]{pearl2009causality}, an SCM is a 4-tuple $M = (U, V, F, P)$ where:
\begin{enumerate}
    \item $U$ is a set of \emph{background} or \emph{exogenous} variables, which are not explicitly modeled but which can affect relationships within the model.
    \item $V = \{V_1,\ldots,V_n\}$ is a set of \emph{endogenous} variables, that are deterministically determined by variables in $U\cup V$; in the setting of this paper, one typically chooses $V = \{X,A,Y\}$ or $V = \{X,A,Y,S\}$ to respectively model covariates, treatment, outcome and selection.
    \item $F$ is a set of functions $\{f_1, .., f_n\}$ such that each $f_i$ uniquely determines the value of $V_i\in V$ by the so-called \emph{structural equation} $v_i = f(pa_i, u_i)$, where $PA_i \subset V\backslash\{V_i\}$ are called the \emph{parents} of $V_i$ and $U_i \subset U$.
    \item $P$ is a probability distribution for $U$.
\end{enumerate}
The \emph{causal diagram} corresponding to an SCM is a graph with $V$ as vertices, directed edges from each parent to its children, and undirected dotted edges between vertices $V_i$ and $V_j$ such that $U_i\cap U_j\neq \emptyset$. Alternatively, the $U$ can be explicitly represented, with directed dotted edges from $U_i$ to $V_i$, as in Figure~\ref{fig:scm} which represents the SCM with $V=(X,A,Y)$, $U=(U_x, U_a, U_y)$, and structural equations:
\begin{eqnarray*}
x &\leftarrow& f_x(u_x) \\
a &\leftarrow& f_a(x, u_a),    \\
y &\leftarrow&  f_y(a, x, u_y).    
\end{eqnarray*}
Often, no parametric assumptions is made on $F$ or $P$. The distribution $P(U)$ induces a distribution $P_M(V)$ through $V=F(U)$, and in the case where the causal diagram is a directed acyclic graph and variables in $U$ are independent, then the distribution $P_M(V)$ is a Bayesian network. In particular, the causal diagram encodes the conditional independence relationships among variables in $V$.

\paragraph{Interventions.}
At the core of the SCM framework is the $do$-$operator$ which enables the use of structural equations to represent causal effects and counterfactuals. 
 The $do(A = a_0)$ operation marks the replacements of the mechanism $f_a$ with a constant $a_0$,  while keeping the rest of the model unchanged, resulting in the following post-treatment model for our toy example:
\begin{eqnarray*}
x &\leftarrow& f_x(u_x) \\
a &\leftarrow& a_0  \\
y &\leftarrow& f_y(a, x, u_y)    
\end{eqnarray*}
In the causal graph, this corresponds to deleting all incoming arrows in $A$  (Figure \ref{fig:post_int}).
We denote $Q = P(Y\mid do(A = a_0))$ the post-intervention distribution, i.e., the distribution of a random variable $Y$ after a manipulation on $A$. From this distribution, the ATE can be written as: 
\begin{eqnarray*}
\tau &=& \EE[Y \mid  do(A=a_1) ] -  \EE[Y \mid  do(A=a_0) ]  \\ &=& \sum_{y}  y \left(P(Y=y \mid  do(A=a_1)) -  P(Y=y \mid  do(A=a_0)) \right). 
\end{eqnarray*}
Note that the post-intervention distribution can also be denoted in counterfactual
notation as $P(Y=y\mid do(A=a))  =  P(Y(a) = y)$.
The distinction between 
$P(Y \mid A = a)$ and $P(Y \mid do(a))$ corresponds  in the PO framework to the difference between
$P(Y\mid A = a)$ and $P(Y (a))$.

\paragraph{D-separation.}
Conditional independences between variables can be read from the DAG induced by an SCM using a graphical criterion known as $d$-$separation$. This criterion will be useful in identifying the causal effect. 

\begin{definition}[d-separation]
A set $X$ of nodes is said to block a
path $p$ if either
\begin{itemize}
\item $p$ contains at least one arrow-emitting node that is in $X$, or
\item $p$ contains at least one collision node that is outside $X$ and has no descendant
in $X$.
\end{itemize}
\end{definition}
If $X$ blocks all paths from set $A$ to set $Y$, it is said to ``d-separate $A$ and $Y$'' and then it can be shown that $A \independent Y \mid X $. As an illustration,
let us consider a path with $ A \rightarrow D \leftarrow B \rightarrow C$. Since $B$ emits arrows on that path, it blocks the path between $A$ and $C$, and $A \independent C \mid  B$. $D$ is a collider (two arrows incoming) and consequently it blocks the path without conditioning $A \independent C$; but  conditioning on $D$ would open the path and thus would imply that $A  \not\independent C \mid  D$. Furthermore, in the SCM framework it is generally assumed that \textit{faithfulness} holds, i.e., that all conditional independences are encoded in the graph, allowing to infer dependencies from the graph structure \citep{peters2017elements}. In other words, if the Global Markov property (i.e., $d$-separation implies conditional independence), and faithfulness hold, then the resulting equivalence between conditional independences and $d$-separation allows to move back and forth between the graphical and the probabilistic model.

\paragraph{Identifiability.}
We are interested in answering the \emph{identifiability} question: \textit{can the post-intervention distribution $Q$ be estimated using observed data (such as pre-intervention distribution)?}
\begin{definition}[identifiability]
A causal query $Q$ is identifiable from distribution $P(y)$ compatible with a causal graph $G$, if for any two (fully specified) models $M_1$ and $M_2$ that satisfy the assumptions in $G$, we have
\begin{eqnarray*}
P_1(V)= P_2(V) &\Longrightarrow& Q(M_1) = Q(M_2).
\end{eqnarray*}
\end{definition}
Specifically, if a causal query $Q$ in the form of a $do$-expression can be $reduced$ to an expression no longer containing the $do$-$operator$ (i.e, containing only estimable expressions using nonexperimental, observed data) by iteratively applying the inference rules of $do$-$calculus$, then $Q$ is identifiable. The language of $do$-$calculus$ is proved to be $complete$ for queries in the form $Q = P(Y=y\mid do(A=a), X=x)$ meaning that if no reduction can be obtained using these rules, $Q$ is not identifiable.\\

\begin{figure}[tb!]
\center
\includegraphics[width=0.5\textwidth]{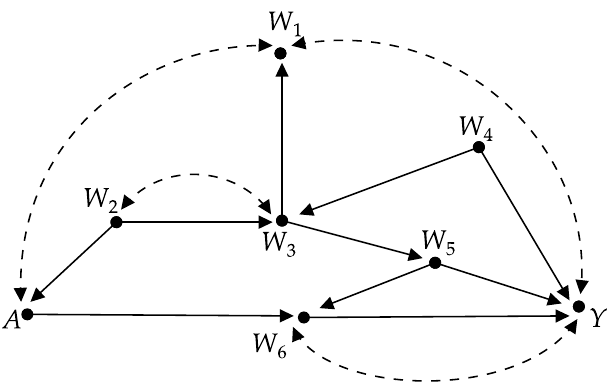}
\caption{\textbf{Application of the backdoor criterion in large graphs.} Based on the admissible set definition \ref{def:admissible}, \eqref{eq:possibleadjustements} present all the following sets that are admissible and can be used for adjustment. For example, the set $\{W_2, W_3\}$ blocks all backdoor paths between $A$ and $Y$. $W_2$ block the path $A \leftarrow W_2 \rightarrow W_3 \rightarrow W_5 \rightarrow Y$.}
\label{fig:back}
\end{figure}

The application of  previous rules and the backdoor criterion in the graph of Figure~\ref{fig:back} allows to list all possible admissible adjustment sets for identifying
$P(y \mid  do(a))$: 
\begin{multline}
X = \{W_2\}, \{W_2, W_3\}, \{W_2, W_4\}, \{W_3, W_4\}, 
\{W_2, W_3, W_4\}, \{W_2, W_5\}, \{W_2, W_3, W_5\}, \\ \{W_4, W_5\},
\{W_2, W_4, W_5\}, \{W_3, W_4, W_5\}, \{W_2, W_3, W_4, W_5\}
\label{eq:possibleadjustements}
\end{multline}

The analyst can select from this list which is preferable. Note that conditioning on $W_1$ would induce bias as it is a collider. \\

\subsubsection{Confounding bias}
In order to estimate the causal effect $P(Y \mid do(A=a))$ using only available observational data, following the observational distribution $P(A,X,Y)$, the idea is to identify---on the basis of the causal graph---a set of \textit{admissible variables} such that measuring and adjusting for these variables removes any bias due to confounding. 
The \textit{backdoor criterion} defined below provides a graphical method for selecting admissible sets for adjustment. 

 \begin{definition}[Admissible sets -  the backdoor criterion]
  Given an ordered pair of treatment and outcome variables $(A, Y )$ in a causal DAG $G$, a set
 $X$ is backdoor admissible if it blocks every path between $A$ and $Y$ in the graph  $G_{\underline{A}}$,  with $G_{\underline{A}}$  the graph that is obtained when all edges emitted
 by node $A$ are deleted in $G$.
 \label{def:admissible}
 \end{definition}
The backdoor criterion can be seen as the counterpart of unconfoundedness in Assumption \ref{asump:unconfounded}: If a set
$X$ of variables satisfies the backdoor condition relative to $(A, Y )$, then  $Y(a) \independent A \mid  X$. Identifying backdoor admisible variables is important because it allows to estimate causal effects from observational data as follows:
 \begin{theorem}[Backdoor adjustment criterion]\label{theo:back}
 If a set of variables satisfies the
backdoor criterion relative to $(A, Y )$, the causal effect of $A$ on $Y$ can be identified
from observational data by the adjustment formula:
$$
P(Y=y \mid  do(A=a)) = \sum_x P(Y=y \mid  A=a, X=x)P(X=x)\,.
$$
\end{theorem}
The adjustment formula can be seen as part of the identifiability formula in Equation \ref{eq:idregobs}.

\begin{figure}[h!]
\centering
\includegraphics[width=.7\textwidth]{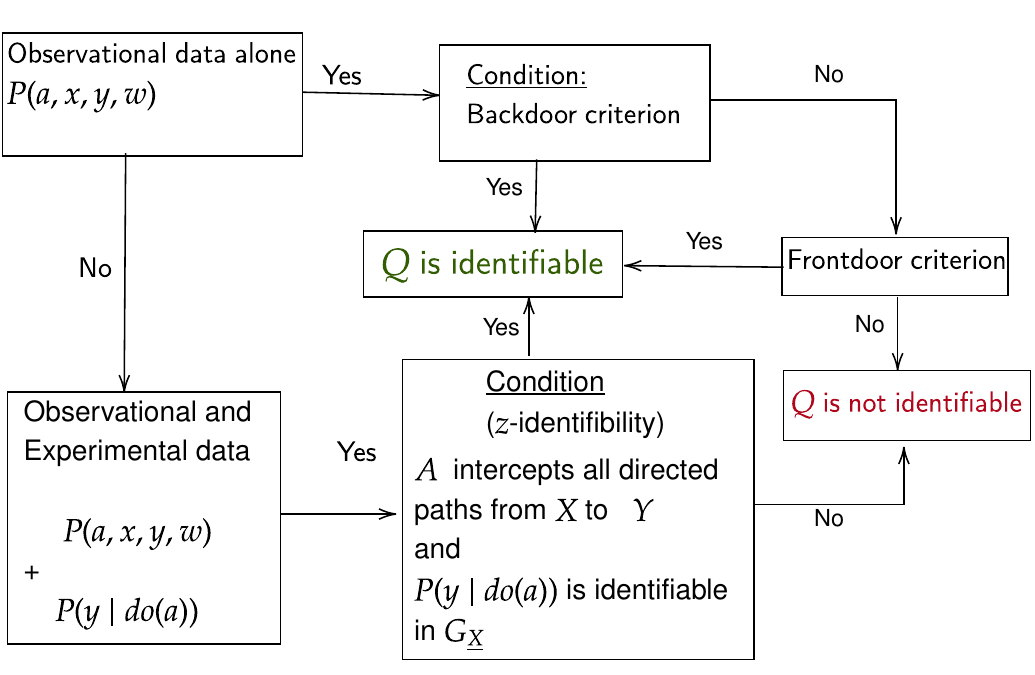}
\caption{\textbf{Summary of identifiability results to control for confounding bias}: If there exists a set of observed variables that satisfies the backdoor criterion, then the causal effect of $A$ on $Y$ can be identified using nonexperimental data alone. In the case where no set of observed variables satisfies the backdoor condition but the effect of $A$ can be mediated by an observed variable $M$ (mediator),  if there exists a set of observed variables that satisfies the frontdoor criterion, then the causal effect if also identifiable from observational data alone. If none of these conditions holds, the query is not identifiable. If, in addition to observational data, RCTs through surrogate experiments are available, the $z$-identifiability condition is sufficient to determine if the query is identifiable or not.}
\label{fig:confounding}
\end{figure}

The backdoor criterion is one of the graphical methods for identifying admissible sets. In cases where it is not applicable, an extended definition called the \textit{frontdoor criterion} can be applied using mediators in the graph.
Figure~\ref{fig:confounding} provides a summary of the identifiability conditions when the available data is either observational data or data from surrogate experiments. 

\subsubsection{Sample selection bias}\label{eq:SCMselectionbias}
To tackle sample selection bias, i.e., preferential selection of units, the authors consider an indicator variable $S$ such that $S=1$ identifies units in the sample. The data at hand can be seen as  $P(A,Y,X \mid  S=1)$ and the target is $P(Y \mid  do(A=a))$. 
\begin{figure}[tbh!]
\begin{minipage}{.3\linewidth}
\caption{\textbf{Cases with sample selection bias}: $A$ is the treatment and $Y$ the outcome, $S$ is the selection process and the aim is to estimate $P(y\mid  do(a))$ when data available come from $P(a,y \mid  S=1)$ in (a) and (b).}
\label{fig:selectionbias}
\end{minipage}%
\hfill%
\begin{minipage}{.72\linewidth}
    \includegraphics[width=\textwidth]{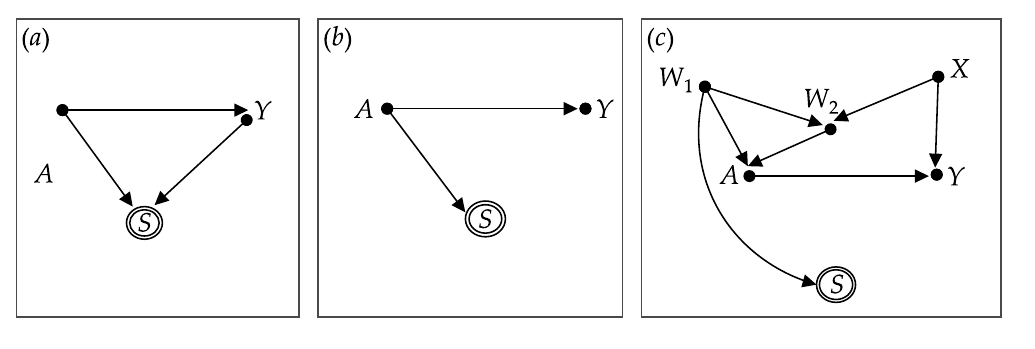}
\end{minipage}%
\end{figure}
Figure~\ref{fig:selectionbias} (b) presents a case where the selection process is $d$-separated (definition in Appendix~\ref{sec:appendixscm}) from $Y$ by $A$, then $P(y\mid  a) = P(y \mid a,   S=1)$; since $A$ and $Y$ are unconfounded, $P(y\mid  do(a)) = P(y\mid  a)$ so that the experimental distribution is recoverable from observed data. This is not the case for Figure \ref{fig:selectionbias} (a) without further assumptions. 
When both confounding bias and selection bias are present in the data (Figure \ref{fig:selectionbias} (c)), the graphical framework can help selecting among the list of adjustment sets, $\{W_1, W_2\}$, $\{W_1, W_2,X\}$, $\{W_1,X\}$, $\{W_2,X\}$, and $X$,  (these sets control for confounding),
the one that can be used as available from biased data; here it will be $X$ as it is the only one separated from $S$, leading to $P(y\mid  do(a)) =\sum_x P(y\mid  a,x, S=1)P(x\mid  S=1)$. This ability 
to select relevant covariates for identifiability is presented as an important advantage of the SCM framework.

\paragraph{Combined biased and unbiased data.}
Note that the previous examples in Figure \ref{fig:selectionbias} concern only one set of data but 
the approach is extended to   
combine data, biased (with a selection) data, and unbiased data (for example covariates from the target population) as follows.  
To do so, \cite{bareinboim2016causal} define the \textbf{$S$-backdoor admissible criterion} which is a sufficient condition but not necessary. It states that if
$X$ is backdoor admissible,
$A$ and $X$ block all paths between $S$ and $Y$, i.e.
$Y \independent S \mid  A, X$, and
 that $X$ is measured in both population-level
data and biased data, then,  the causal effect can be identified as
$$P(Y \mid  do(A=a)) = \sum_x P(Y \mid  do(A=a), X=x, S=1)P(X=x),$$
where $P(X=x)$ denotes the probability in the target population.
If the set X contains post-treatment covariates, then this formula is generally wrong. Indeed S-ignorability is rarely satisfied in that case, as illustrated with several examples by \citet{pearl2015findings}.
This formula is called the post-stratification formula, to define this action of re-calibrate or re-weight \citep{pearl2015findings}.
This expression shows that one can generalize what is observed on the selected sample  by reweighting or recalibrating by $P(X=x)$ that is available from the target population (unbiased data). 
More complex setting can be handled, such as dealing with post-treatment variables. In such a case, they show that generalizibility can be obtained by another weighting strategy (not by $P(X=x)$), which can also be seen as a benefit of this framework.

\subsection{Proof of the transport formula (\ref{eq:transport})\label{sec:transportproof}}
We compute: 
  \begin{eqnarray*}
     P(Y \mid  do(A=a))
    &=&\sum_x  
      P(Y \mid  do(A=a), X=x) P(X=x \mid  do(A=a)) \\
       &=&\sum_x  
      P(Y \mid  do(A=a), X=x, S=1) P(X=x \mid  do(A=a)) \\
      &=& \sum_x  P(Y \mid  do(A=a), X=x, S=1) P(X=x)  \,,
    \end{eqnarray*} 
where the first equation follows by conditioning, the second one by $S$-admissibility assumption of $X$, and the third one from the fact $X$ are pre-treatment variables.

\section{Additional simulation results}
\label{sec:appendixsimulations}

This section follows Section \ref{sec:simulations} and provides additional results for the simulations.

\subsection{Distributional shift between RCT and observational samples}

\begin{figure}[tbh!]
    \centering
    \includegraphics[width= 0.5 \textwidth]{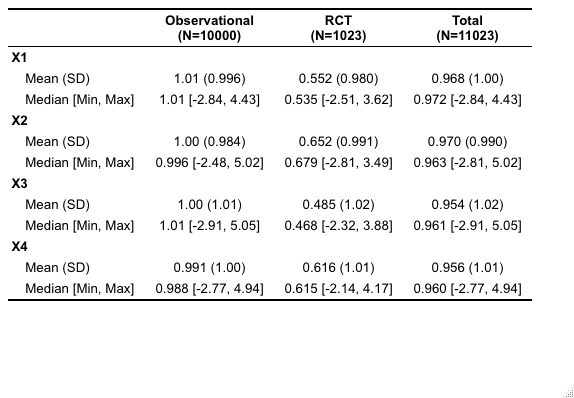} \hspace{0.5cm} 
    \includegraphics[width= 0.45 \textwidth]{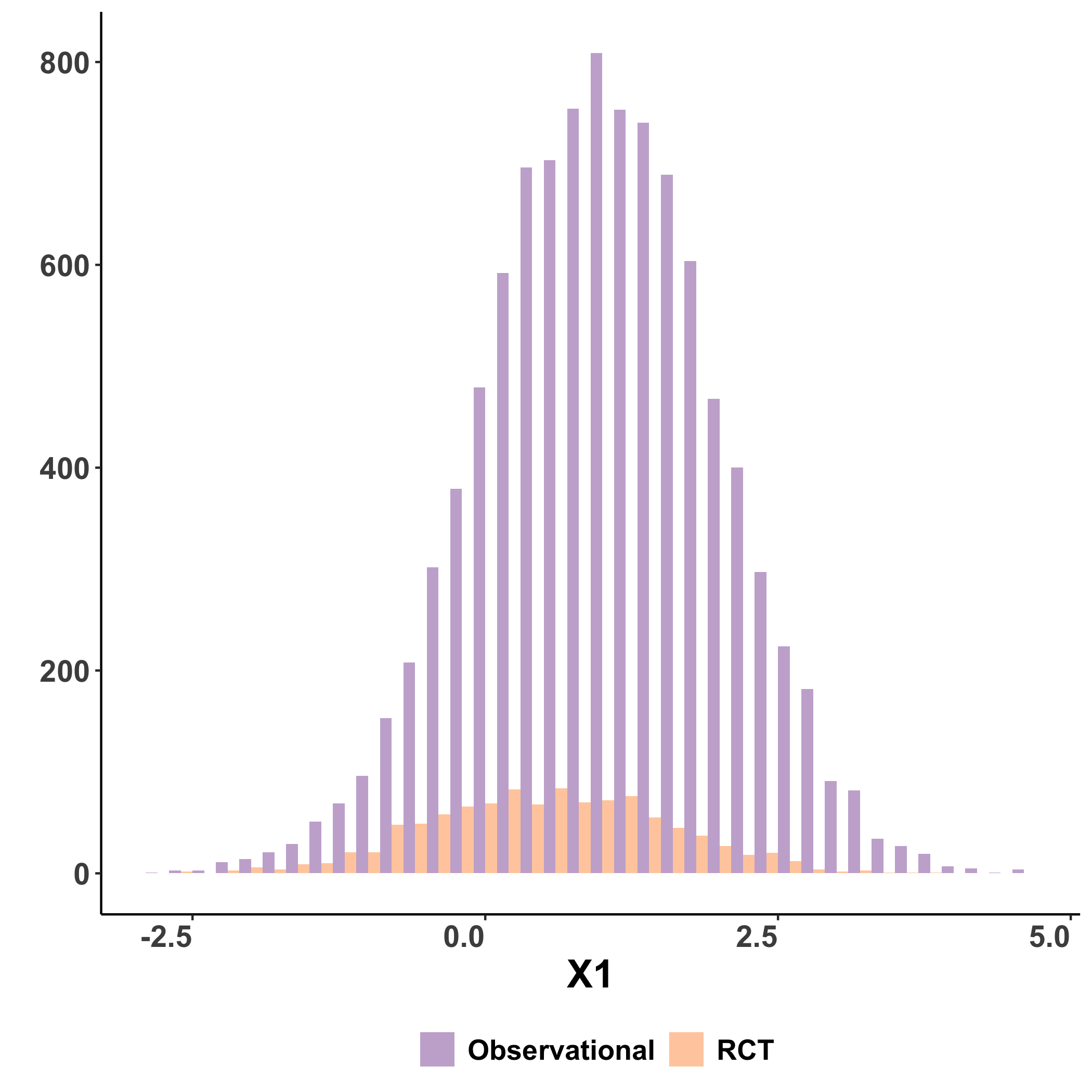}
    \caption{\textbf{Covariates distributions differences} between experimental sample and observational sample when simulating according to \eqref{eq:rctmodel} as detailed in Section \ref{sec:simulations} (left), with a focus on the $X_1$ distributional shift with histograms overlap for the two samples (right).}
    \label{fig:simbaseline}
\end{figure}

The simulation design proposed simulates a situation where the RCT data reveals a distributional shift with the observational sample. In the RCT all the covariates tend to have lower values than in the observational sample. Still, the overlap assumption (Assumption \ref{a:pos}) is valid as each observation in the target sample has a non-zero probability to be included in the experimental sample. Summary statistics  obtained for a simulation with $\sim 1000$ observations in the RCT and $10\,000$ observations in the observational sample is given on Figure \ref{fig:simbaseline}, in addition with an histogram illustrating overlaps and the distributional shift for the covariate $X_1$.

\begin{figure}[h]
 \begin{minipage}[t]{.65\linewidth}\vspace{0pt}
  \includegraphics[width=0.9\linewidth]{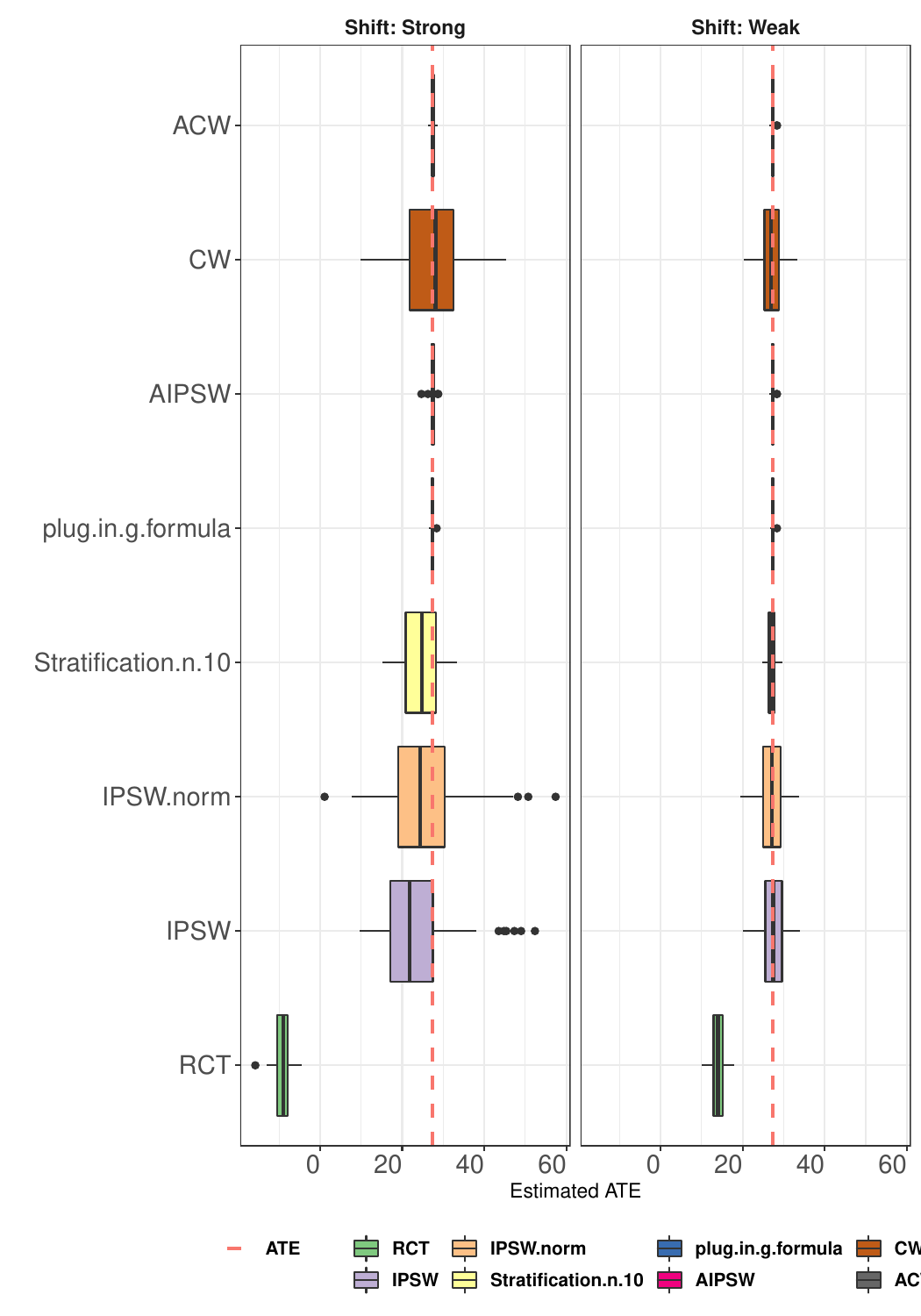}
  \label{fig:image1}
 \end{minipage}
 \hfill
 \begin{minipage}[t]{.35\linewidth}\vspace{0pt}\raggedright
  \begin{minipage}[t]{\linewidth}\vspace{0pt}\raggedright
   \includegraphics[width=0.7\linewidth]{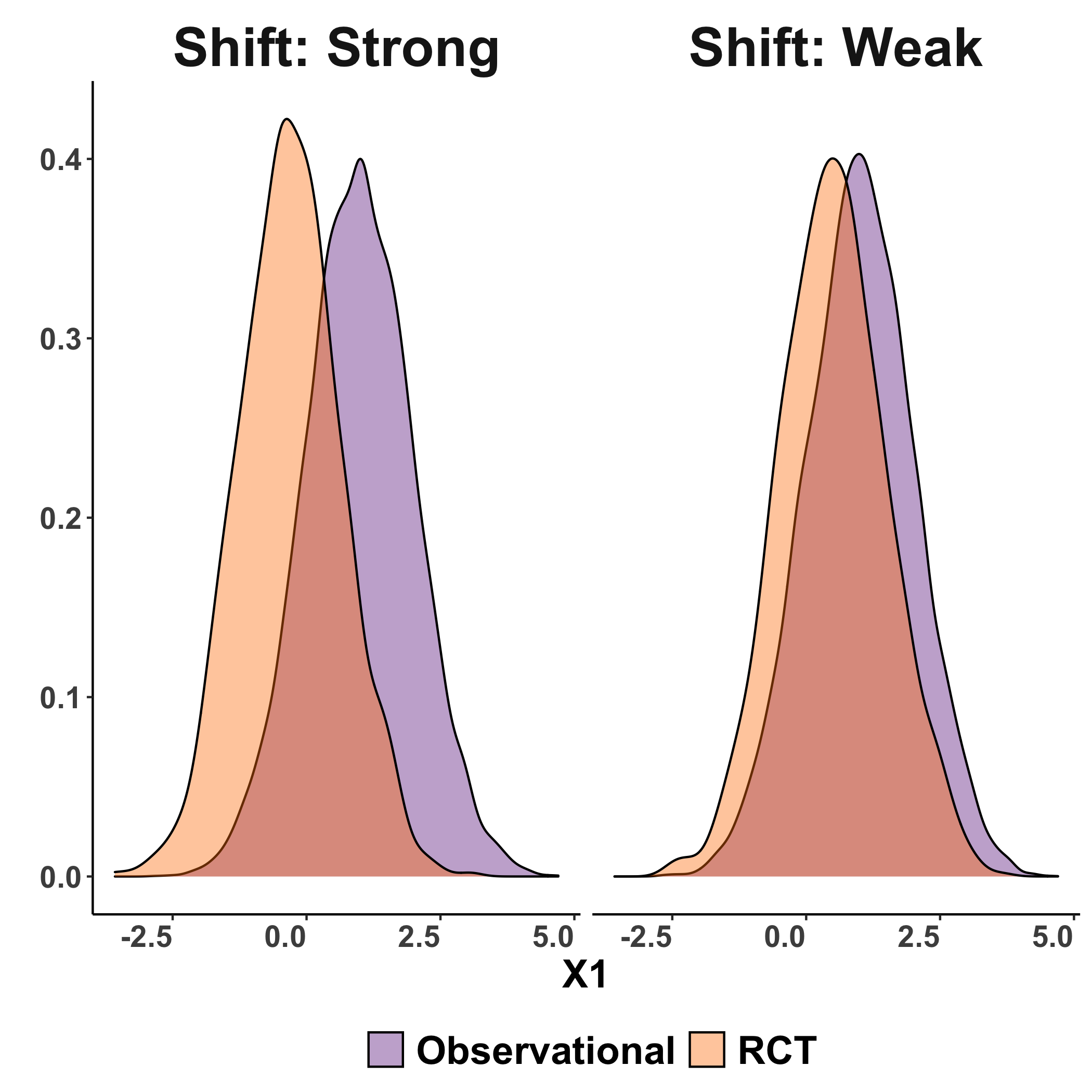}
   \label{fig:image2}
  \end{minipage}
  \begin{minipage}[t]{\linewidth}\vspace{20pt}\raggedright
   \caption{\textbf{Weak versus strong distributional shift between experimental and observational data} with estimated ATE when RCT is weakly or strongly shifted from the target population distribution. Estimators used being IPSW (IPSW and IPSW.norm; Def.~\ref{def:ipsw}), stratification (with 10 strata; Def.~\ref{def:strat}), g-formula (Def.~\ref{def:g-formula}), calibration weighting (CW; Def.~\ref{def:cbss0}), augmented IPSW (AIPSW; Def.~\ref{def:aipsw}), \modif{and ACW (Def.~\ref{def:taudc-2})} over 100 simulations.}%
    \label{fig:sim-strong-shift}
  \end{minipage}
 \end{minipage}
\end{figure}

The sampling propensity score model used to generate the simulated data \eqref{eq:rctmodel} implies a weak covariate shift between the RCT sample and the observational sample. A stronger shift can be obtained, at least on covariate $X_1$, swapping the coefficient $-0.5X_{1}$ with $-1.5X_{1}$. 
Figure \ref{fig:sim-strong-shift} shows that the variance of the weighted and CW estimators have increased in the setting with a stronger covariate shift.  


\subsection{Stratification}

Within the weighted estimators, the stratification estimator (Section \ref{sec:ipsw}) supposes to choose an additional parameter being the number of strata used. Simulations are launched with the number of strata varying from 3 to 15, and the results are presented on Figure \ref{fig:simstrata}. We observed that the number of strata has an impact on the results, the higher the number of strata used, the better the prediction.

\begin{figure}[tbh!]
 \begin{minipage}{.30\linewidth}
  \caption{\textbf{Effect of strata number} Estimated ATE obtained while varying the number of strata $L \in \{3, 5, 7, 9, 11, 13, 15\}$ with 100 repetitions each time. All others simulation parameters being the same as the standard case described in \ref{sec:simulations} and in Figure \ref{fig:sim-simple-100}.}
     \label{fig:simstrata}
    \end{minipage}%
    \hfill%
    \begin{minipage}{.75\linewidth}
    \begin{center}
           \includegraphics[width= 0.7 \textwidth]{fig/strata-100.png}
     \end{center}
    \end{minipage}
\end{figure}

\subsection{Impact of a hidden treatment effect modifier}\label{sec:sim-hidden}

\begin{center}
\begin{figure}[tb!]
    \begin{minipage}{.3\linewidth}
    \caption{\textbf{Impact of the treatment-effect modifiers} Estimated ATE when IPSW estimator includes all covariates, only $X_1$, or all covariates except $X_1$ (IPSW; Section \ref{sec:ipsw}), with g-formula (Section \ref{sec:gformula}) presented as a control, over 100 simulations. Simulations are still performed with \eqref{eq:rctmodel} for RCT eligibility and \eqref{eq:Ymodel} for outcome modeling.}
    \label{fig:sim-treat-effect-modifier}
    \end{minipage}%
    \hfill%
    \begin{minipage}{.7\linewidth}
    \begin{center}
       \includegraphics[width=0.8\textwidth]{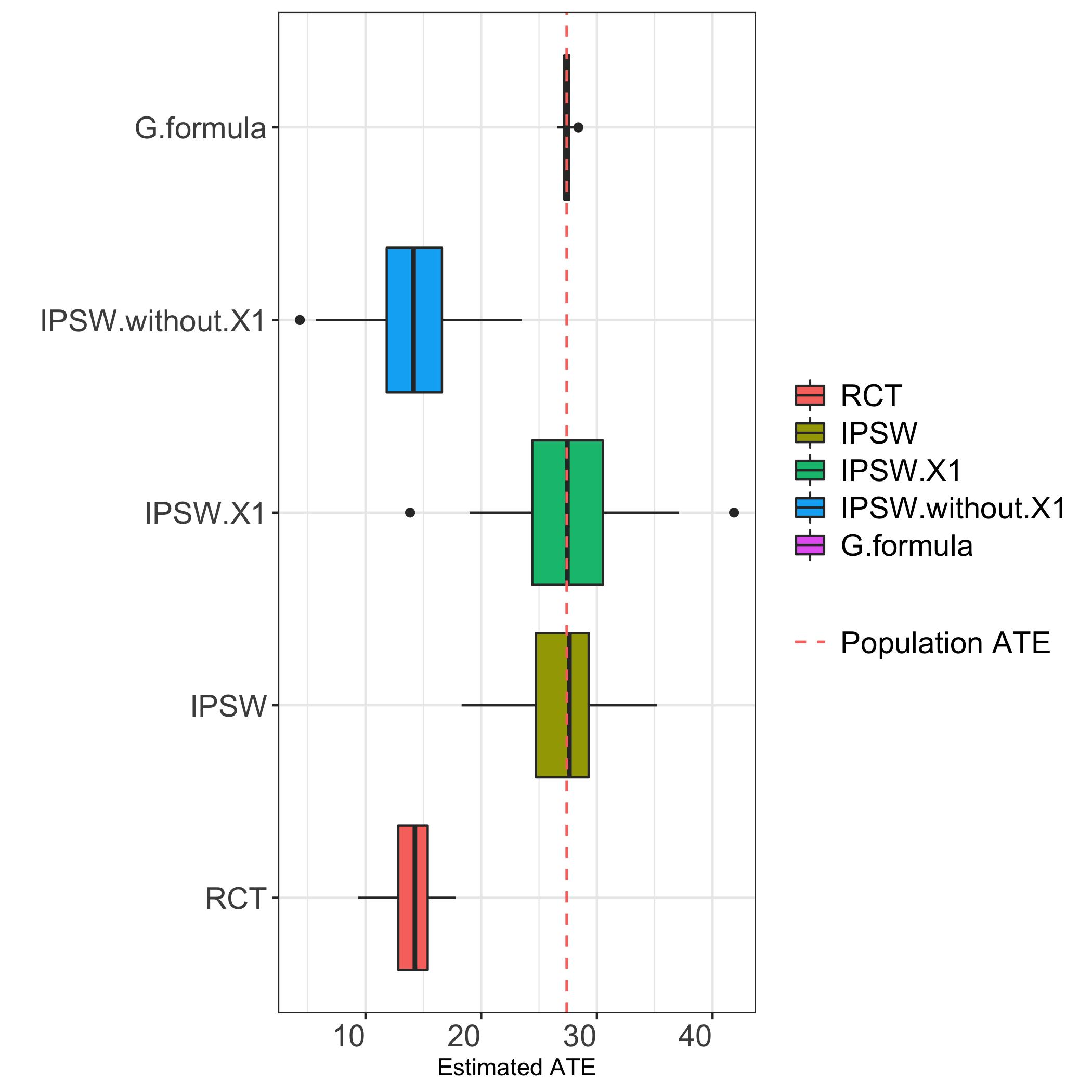}
    \end{center}
    \end{minipage}
\end{figure}
\end{center}

In this part, we consider a heterogeneous treatment effect setting where $X_1$  impacts the RCT sampling while also being a treatment effect modifier. We consider the IPSW estimator and its variations without using $X_1$ (labeled as IPSW.without.X1) and using only $X_1$ (labeled as IPSW.X1). As shown in Figure \ref{fig:sim-treat-effect-modifier}, IPSW.X1 is still unbiased when using only $X_1$ in the sampling propensity score estimation, as it is the only covariate being the shifted treatment effect modifier. However, if $X_1$ is missing, the resulting estimator IPSW.without.X1 is strongly biased. Therefore, by including all variables that affect both sampling and outcome one can ensure identifiability. 
\sout{We also conjecture that including outcome predictors that do not affect sampling may increase the efficiency of the estimator.}\modiff{A recent work suggests to add non-shifted treatment effect modifier for precision \citep{Colnet2022Reweighting}.} 
 
Note also that if the treatment effect were homogeneous (does not depend on $X_1$), then the estimated ATE on the RCT would be unbiased (as shown Figure \ref{fig:sim-homogeneous-treat-effect} in the section below, Section~\ref{sec:homo}) so in this setting there is no need to use the observational data and associated methods to transport the ATE from the trial to the target population \modiff{as the causal effect investigated is on the absolute different scale.} 

\subsection{Homogeneous treatment effect}\label{sec:homo}

\begin{figure}[H]
 \begin{minipage}{.30\linewidth}
   \caption{\textbf{Homogeneous treatment effect} Estimated ATE with a homogeneous treatment effect $ Y(a) \mid X= -100 + X_{1}+13.7 X_{2}+13.7 X_{3}+13.7 X_{4}+ 27.4 a + \epsilon $. All others simulation parameters being the same as the standard case described in \eqref{sec:simulations} and in Figure \ref{fig:sim-simple-100}.}
    \label{fig:sim-homogeneous-treat-effect}
    \end{minipage}%
    \hfill%
    \begin{minipage}{.80\linewidth}
        \begin{center}
           \includegraphics[width= 0.7 \textwidth]{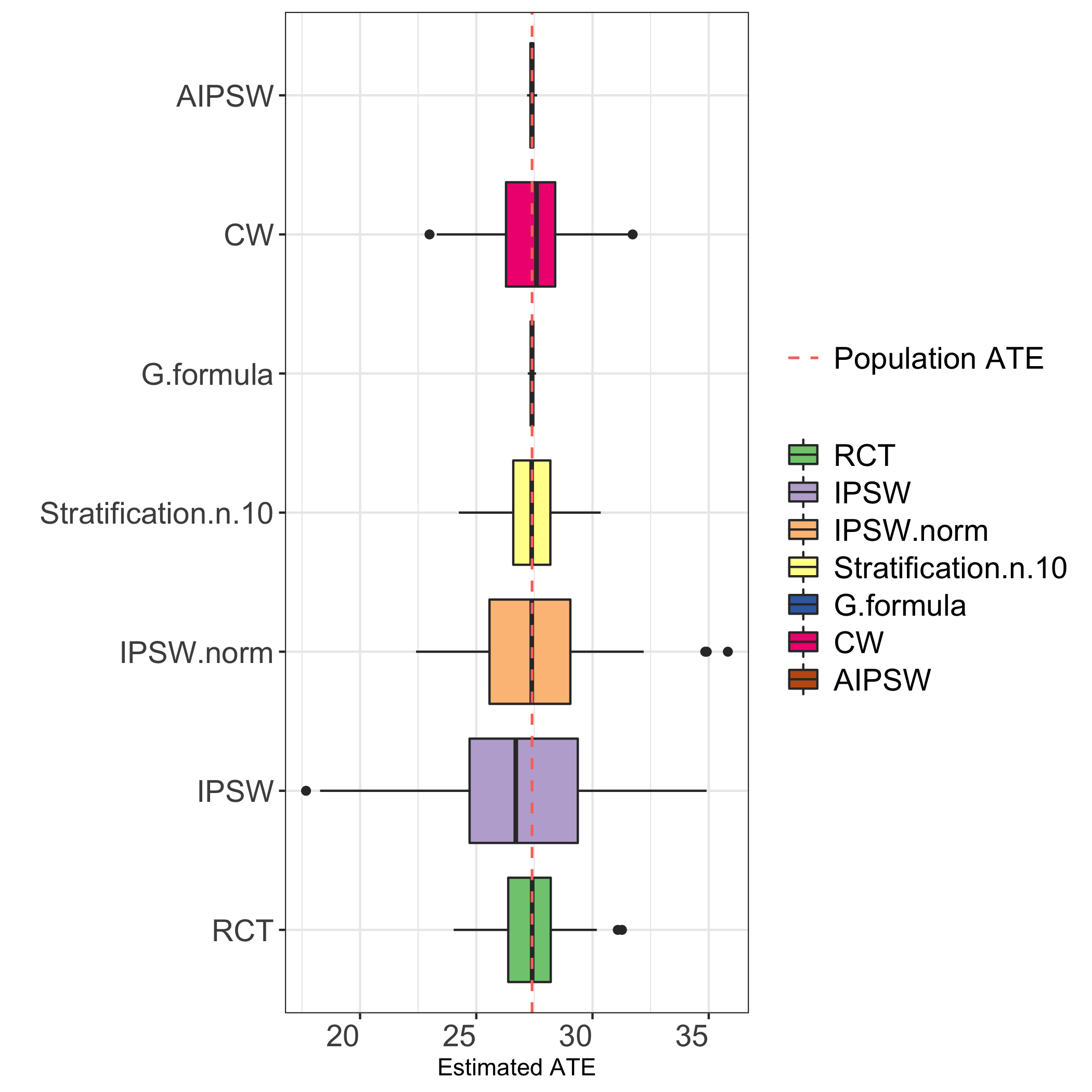}
           \end{center}
    \end{minipage}
\end{figure}

It is always interesting to note that in the case of an homogeneous treatment effect the RCT sample contains all the information to estimate the population ATE, in other words $\tau_1$ is a consistent estimator of the ATE. We performed simulation with an homogeneous treatment effect (results are presented on Figure \eqref{fig:sim-homogeneous-treat-effect}) such as: $$ Y(a) \mid X= -100 + X_{1}+13.7 X_{2}+13.7 X_{3}+13.7 X_{4}+ 27.4 a + \epsilon $$

\section{Supplementary information on Traumabase and CRASH-3}
\label{sec:appendixdataanalysis}

\subsection{Additional information on the Traumabase}\label{appendix:sup-TB}

\subsubsection{Missing values}

\begin{figure}[H]
\centering
\includegraphics[width=0.8\textwidth]{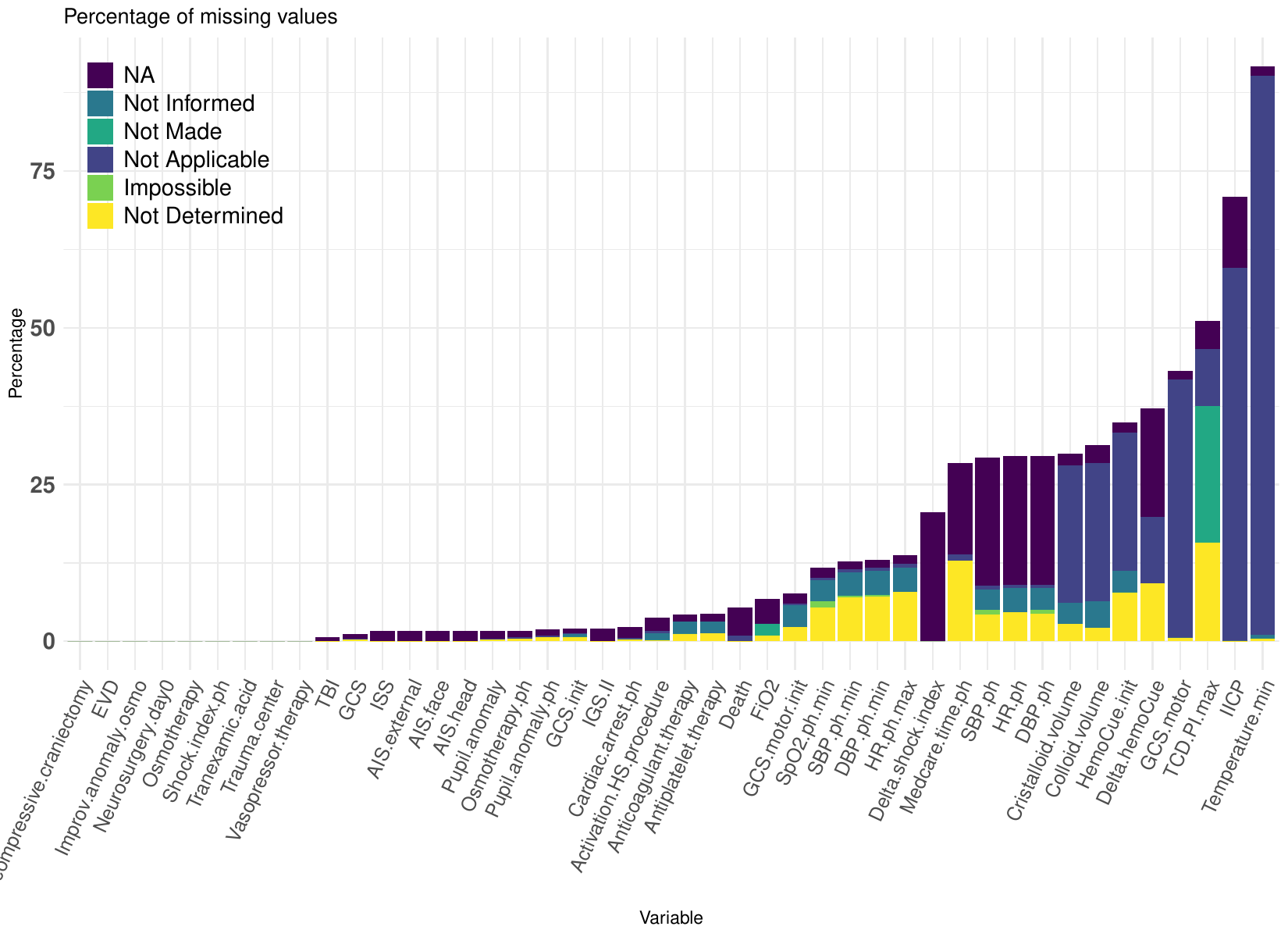}
\caption{\textbf{Missing values} Percentage of missing values for a subset of Traumabase variables relevant for traumatic brain injury. Different encodings of missing values are available such as: \textit{NA} (not available), but also \textit{not informed}, \textit{not made}, \textit{not applicable}, \textit{impossible}.}
\label{fig:tbi-txa-NAs}
\end{figure}

The problem of missing values is ubiquitous in data analysis practice and particularly present in observational data, as they are not necessarily collected for research purposes. The Traumabase is a high-quality data set but, nevertheless, missing values occur. 
Figure \ref{fig:tbi-txa-NAs} represents the percentage of missing values for the covariates selected by the medical doctors from the Traumabase. It varies from 0 to nearly 60\% for some features. In addition, there are different codes for missing values giving hints on the reason of their occurrence, e.g., not available (NA), impossible (imp), not made (NM), etc. Some of these values can be seen as missing completely at random (MCAR), e.g., the information has not been recorded simply because the form was not filled out, but they can be informative and missing not at random (MNAR), e.g., when the state of the patient is such that it was impossible to take a measurement.

There is an abundant literature available on how to deal with missing values in a general context and \cite{mayer2019r} identify more than 150 R \citep{CRAN} packages available on the topic. Missing values add a layer of complexity to conducting causal analyses as they require coupling conventional hypotheses of causal effect identifiability in the complete case with hypotheses about the mechanism that generated the missing data \citep{rubin1976inference}, or defining new hypotheses, to establish conditions of causal effect identifiability with missing data. \citet{mayer2019doubly} survey available works, classify the methods in three families that differ with respect to the different assumptions and provide associated estimators to estimate the ATE from an observational data set with missing values in the covariates. 
More precisely, they advocate the use of  multiple imputation \citep{vanbuuren_2018}  by IPW or doubly robust estimators when missing values can be considered to be missing (completely) at random (M(C)AR) and the classical unconfoundedness assumption (Assump. \ref{asump:unconfounded}) holds \citep{seaman2014inverse}. As an alternative, they recommend using a doubly robust estimator adapted to missing values. More specifically an estimator that makes use of random forests with a missing incorporate in attributes splitting criterion \citep{twala_etal_2008,josse2019consistency} to estimate the generalized propensity scores \citep{rosenbaum1984reducing} and the regression function with missing values\footnote{This doubly robust method is implemented in the R package grf \citep{athey_etal_AS2019}.}; this approach does not require a particular missing values mechanism but an adapted unconfoundedness hypothesis with missing data. Finally, when covariates can be seen as noisy incomplete proxies of true confounders, latent variable models can be a solution to estimate causal effect with missing values \citep{kallus2018causal, louizos2017causal}. Note that for the generalization task, IPSW weights are also computed after imputation in \cite{susukida2016representativeness}.

\subsubsection{Covariate adjustment}

\begin{figure}[ht!]
\centering
\includegraphics[width=.9\textwidth]{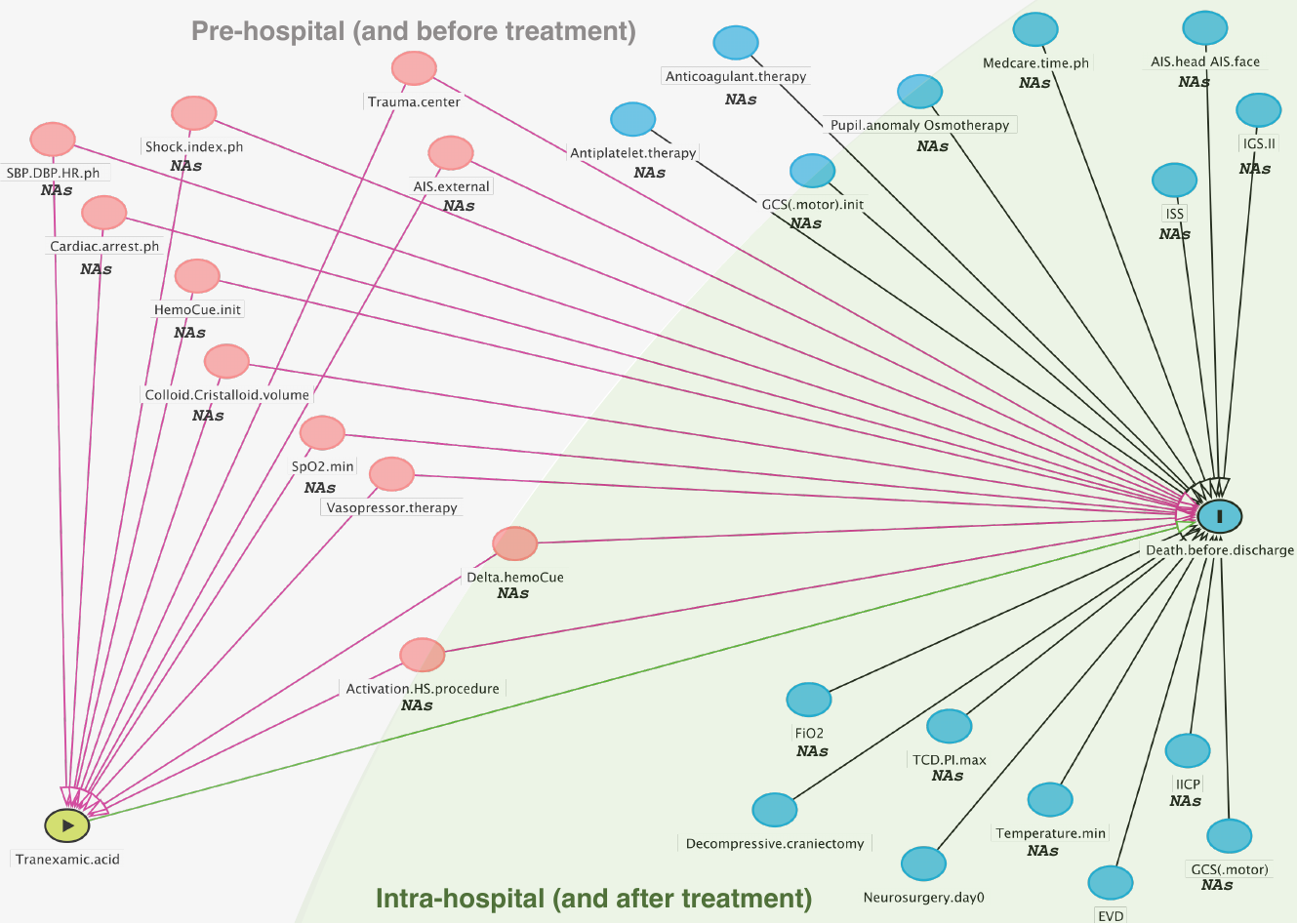}
\caption{\textbf{Causal graph} representing treatment, outcome, confounders and other predictors of outcome (Figure generated using DAGitty \citep{textor_etal_2011}; \texttt{NAs} indicates variables that  have missing values).}
\label{fig:tbi-txa-dag}
\end{figure}

Since the Traumabase is an observational registry, straightforward treatment effect estimation on these data is not possible due to confounding. The causal graph in Figure~\ref{fig:tbi-txa-dag} is the result of a two-stage Delphi method \citep{delphimethod} in which six anesthetists and resuscitators specialized in critical care---and therefore familiar with the allocation process for TXA---first select covariates related to either treatment or outcome or both, and second classify these covariates into confounders and predictors of only  outcome. Even though it is not possible to test for unobserved confounding, this Delphi procedure is an attempt to gather as much expert knowledge about the studied question as possible to manually identify possible confounders and qualitatively assess the plausibility of the unconfoundedness assumption. Note that this approach is an explicit example where we leverage the advantages of the SCM and PO frameworks: the causal graph helps to select relevant variables during the conception phase of the study and to assess identifiability of the target estimand, and the treatment effect analysis uses different estimation methods from the PO framework.

\subsection{Common covariates description between CRASH-3 and Traumabase}\label{sec:combine-crash3-traumabase-covariates}

In the following, we discuss definitions of common variables, outcome, treatment, and designs in order to leverage both sources of information. We recall the causal question of interest: ``What is the effect of the TXA on head-injury related death in patients suffering from TBI?'' This part is important for the alignment of the study protocol.

\paragraph{Treatment exposure.}
The treatment protocol of CRASH-3 precisely frames the timing and mean of administration (a first dose given by intravenous injection shortly after randomization, i.e., within 3 hours of the accident, and a maintenance dose given afterwards \citep{crash3protocol}). For consistency with the original CRASH-3 study described above, we also only keep observations from the RCT with administration within 3 hours.
The Traumabase study being a retrospective analysis, this level of granularity concerning TXA is not available. Neither the exact timing, nor the type of administration are specified for patients who received the drug. However, the expert committee agreed that the assumption of treatment within 3 hours of the accident is plausible since this drug is administered in pre-hospital phase or within the first 30 minutes at the hospital.

\paragraph{Outcome of interest.}
The CRASH-3 trial defines its primary outcome as head injury related death in hospital within 28 days of injury. For the Traumabase data we also look at death in hospital within 28 days but with a wider range of possible causes of death, namely TBI, brain death, multiple organ failure, brain death, or withdrawal of life-sustaining therapy. 

\paragraph{Multi-centered design.}
Both studies are multi-centered, but while the Traumabase is a French registry with over 20 participating Trauma Centers, the CRASH-3 trial enrolled patients in various countries on different continents. This large spectrum of participating centers is likely to contribute to external validity of the CRASH-3 trial, it should nevertheless be noted that more than 65\% of the patients included are from developing countries; regions of the world that differ from developed countries by a prolonged pre-hospital care period, limited access to brain imaging tests and neurosurgery within short periods of time, and the absence of expert centers for major trauma and neuro-intensive care. Thus, on top of the restrictive inclusion criteria of the RCT, this aspect of large heterogeneity in the participating Trauma centers motivates the combination of both studies to estimate the effect for a population with access to a specific high level of care, here represented by the French Trauma centers.

\paragraph{Covariates accounting for trial eligibility.}
In total, four criteria depending on five variables determined inclusion in the CRASH-3 trial: age (only adults were eligible), presence of TBI (defined as presence of intracranial bleeding on the CT scan, or a GCS of less than 13 in the case of no available CT scan), absence of major extracranial bleeding (defined explicitly in CRASH-3 and defined via the number of packed red blood cells transfused in the first 6 hours of admission or by colloid injection in the Traumabase), and delay of less than 8 hours (later reduced to 3 hours) between the injury and the randomization. The necessary variables are also available in the Traumabase, either exactly or in form of close proxies, which allows the estimation of the trial inclusion model on the combined data.

\paragraph{Additional covariates.}
Note that other covariates are available in both data sets, while not directly related to trial inclusion according to CRASH-3 investigators. But as they could be covariates moderating the treatment effect, we include them. According to the two studies, we can add three of them: sex (binary), systolic blood pressure (continuous), and pupils reactivity (categorical, ranging from $0$ to $2$, being the number of active pupils). Note that these three covariates are all mentioned as baselines for the CRASH3 study \citep{crash32019}, where the authors argue that they are likely to impact the outcome.

\subsection{Additional analysis}

This part proposes additional analysis to the data analysis part (Section~\ref{sec:dataanalysis}). We first propose additional visualization of the distributional shift between CRASH-3 and the Traumabase, then we present a principal component analysis of the combined database. Propensity scores obtained either with the logistic regression or the forest are analyzed with histograms and scatter plots. Finally, a focus on the different patients strata, based on the severity of the injury, is presented.

\subsubsection{Distributional shift between CRASH-3 and Traumabase}\label{sec:shift-crash-tb}
\begin{figure}[H]
    \centering
    \includegraphics[width=10cm]{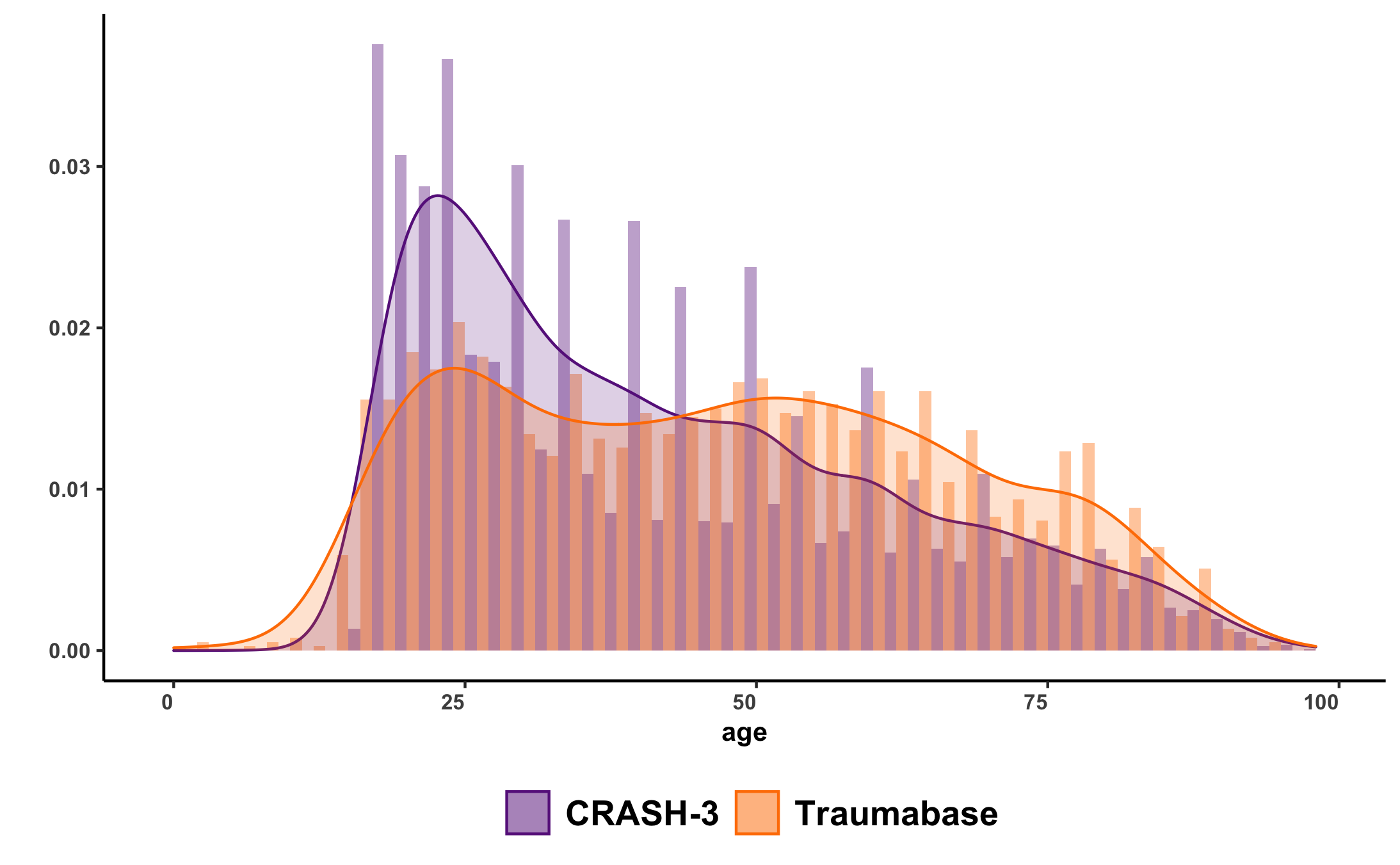}
    \caption{\textbf{Distributional shift of Age} between the Traumabase and the CRASH-3 studies.}
    \label{fig:histogramsshiftage}
\end{figure}

\begin{figure}[H]
    \centering
    \includegraphics[width=10cm]{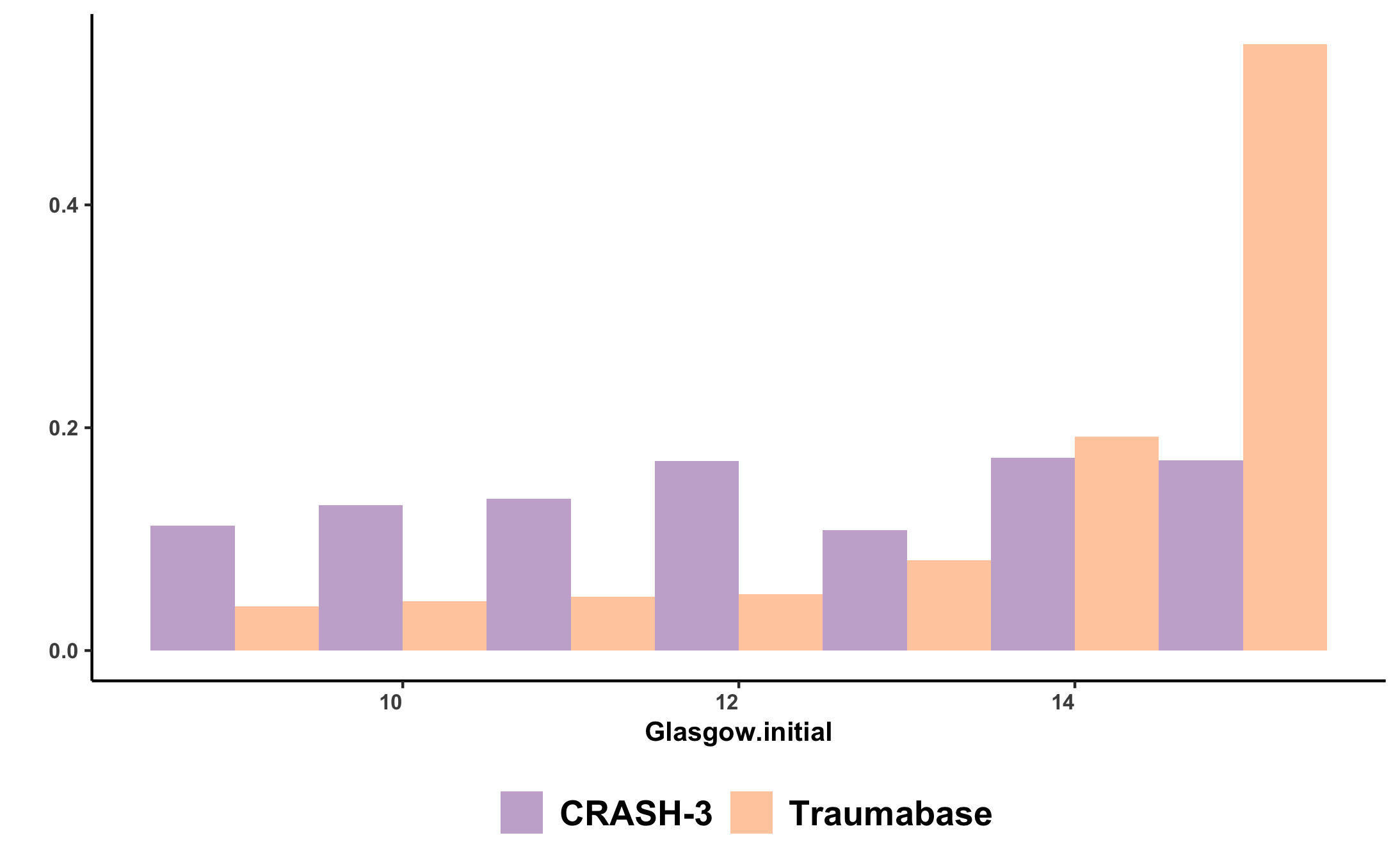}
    \caption{\textbf{Distributional shift of the Glasgow score} between the Traumabase and the CRASH-3 studies.}
    \label{fig:histogramsshiftGCS}
\end{figure}

\begin{figure}[H]
    \centering
    \includegraphics[width=10cm]{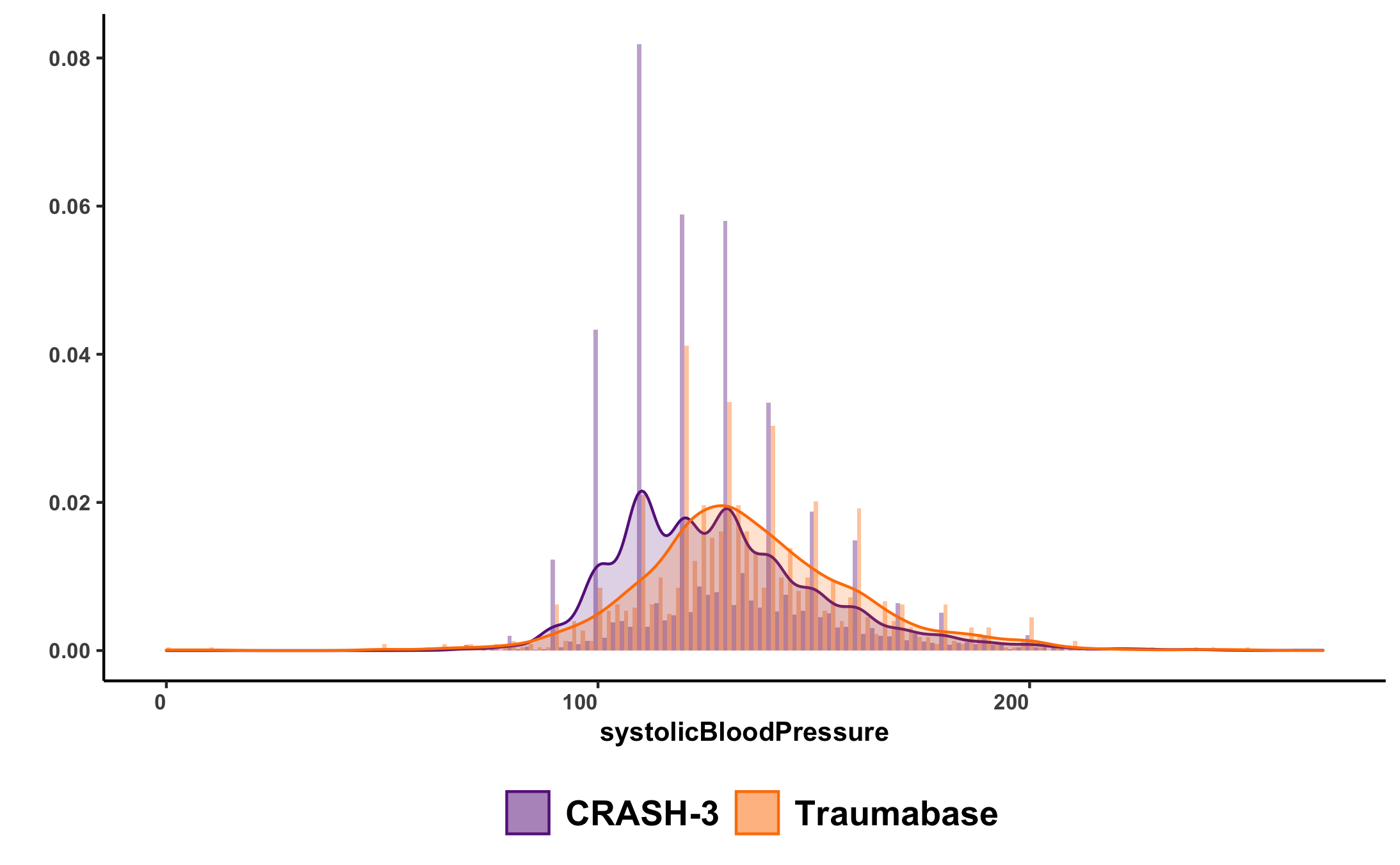}
    \caption{\textbf{Distributional shift of the systolic blood pressure} between the Traumabase and the CRASH-3 studies.}
    \label{fig:histogramsshiftsystolic}
\end{figure}

\begin{figure}[H]
    \centering
    \includegraphics[width=10cm]{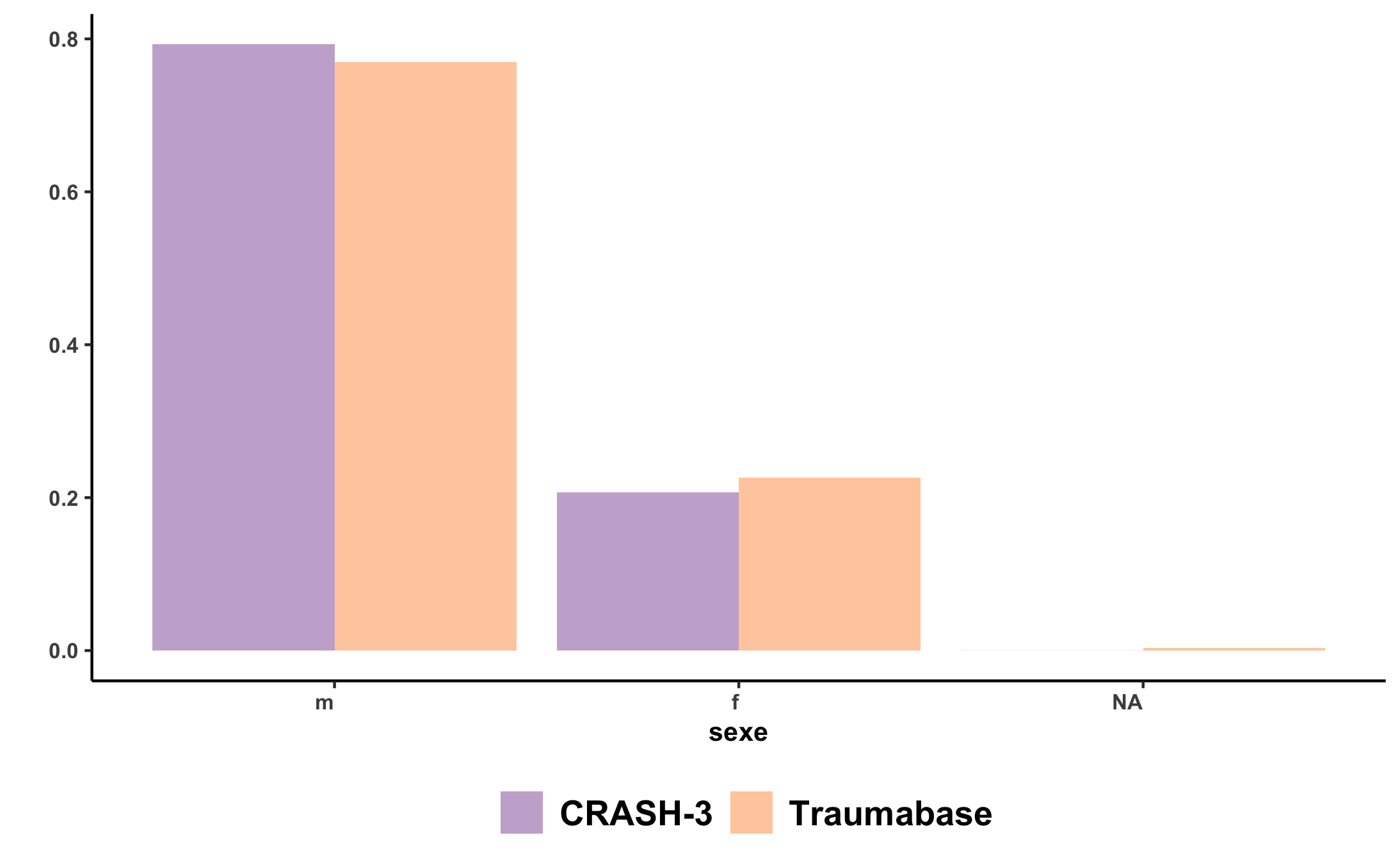}
    \caption{\textbf{Distributional shift of the sex} between the Traumabase and the CRASH-3 studies.}
    \label{fig:histogramsshiftsexe}
\end{figure}

\begin{figure}[H]
    \centering
    \includegraphics[width=10cm]{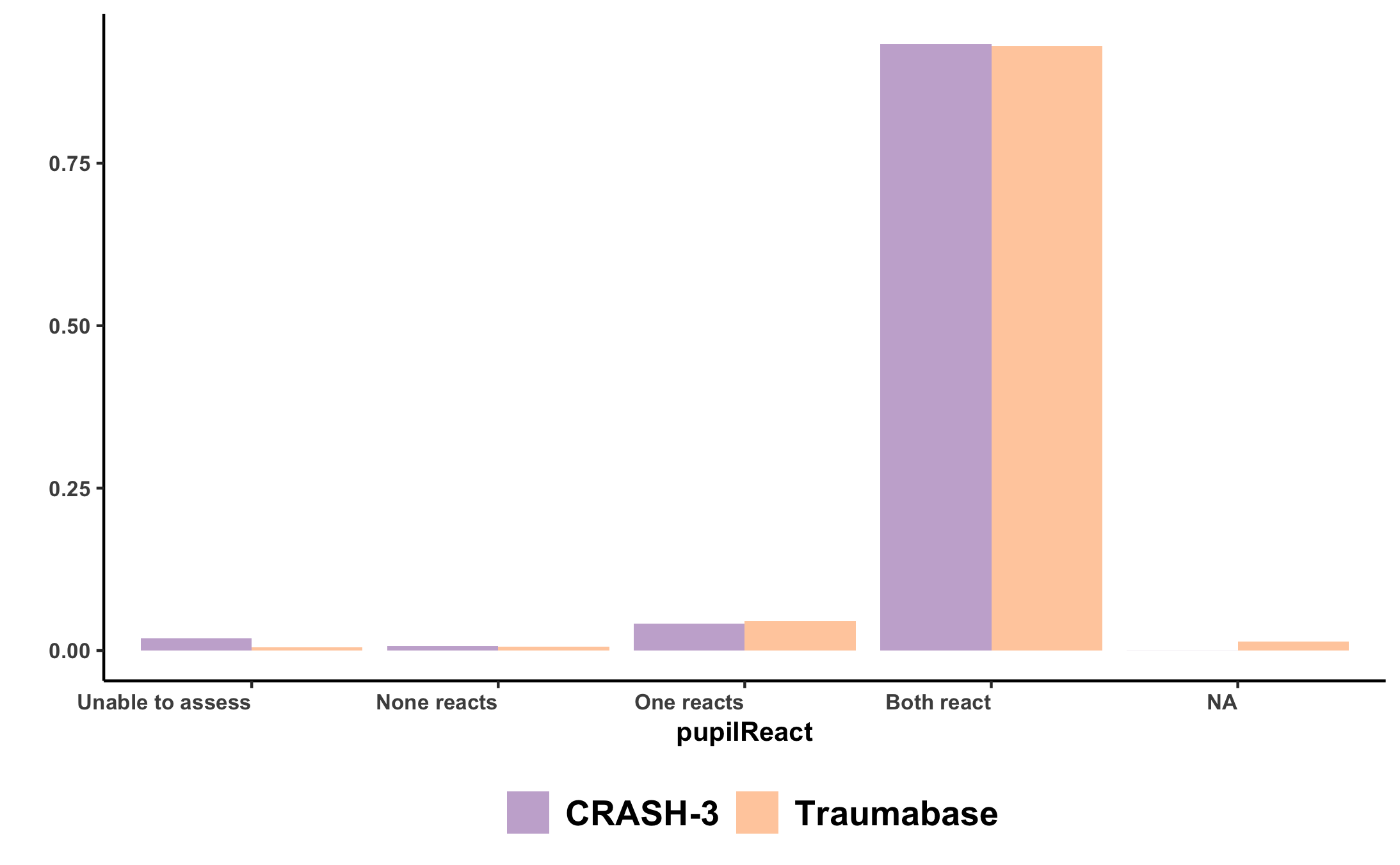}
    \caption{\textbf{Distributional shift of the pupils reactivity} between the Traumabase and the CRASH-3 studies.}
    \label{fig:histogramsshiftpupil}
\end{figure}

Distributional shift between CRASH-3 and the Traumabase data can be illustrated with histograms. Figures \ref{fig:histogramsshiftage} -- \ref{fig:histogramsshiftpupil} presents the empirical distribution shift between the Traumabase and CRASH-3 for age, Glasgow score, systolic blood pressure, sex and pupils reactivity (respectively). Differences can be observed, and for example the fact that the CRASH-3 study contains more young patients, while the Traumabase contains more moderate case (corresponding to a high Glasgow score). It is interesting to notice that the overlaps assumption seems to hold in our situation.


\subsubsection{Principal component analysis}
\begin{figure}[H]
    \centering
    \includegraphics[width = 10cm]{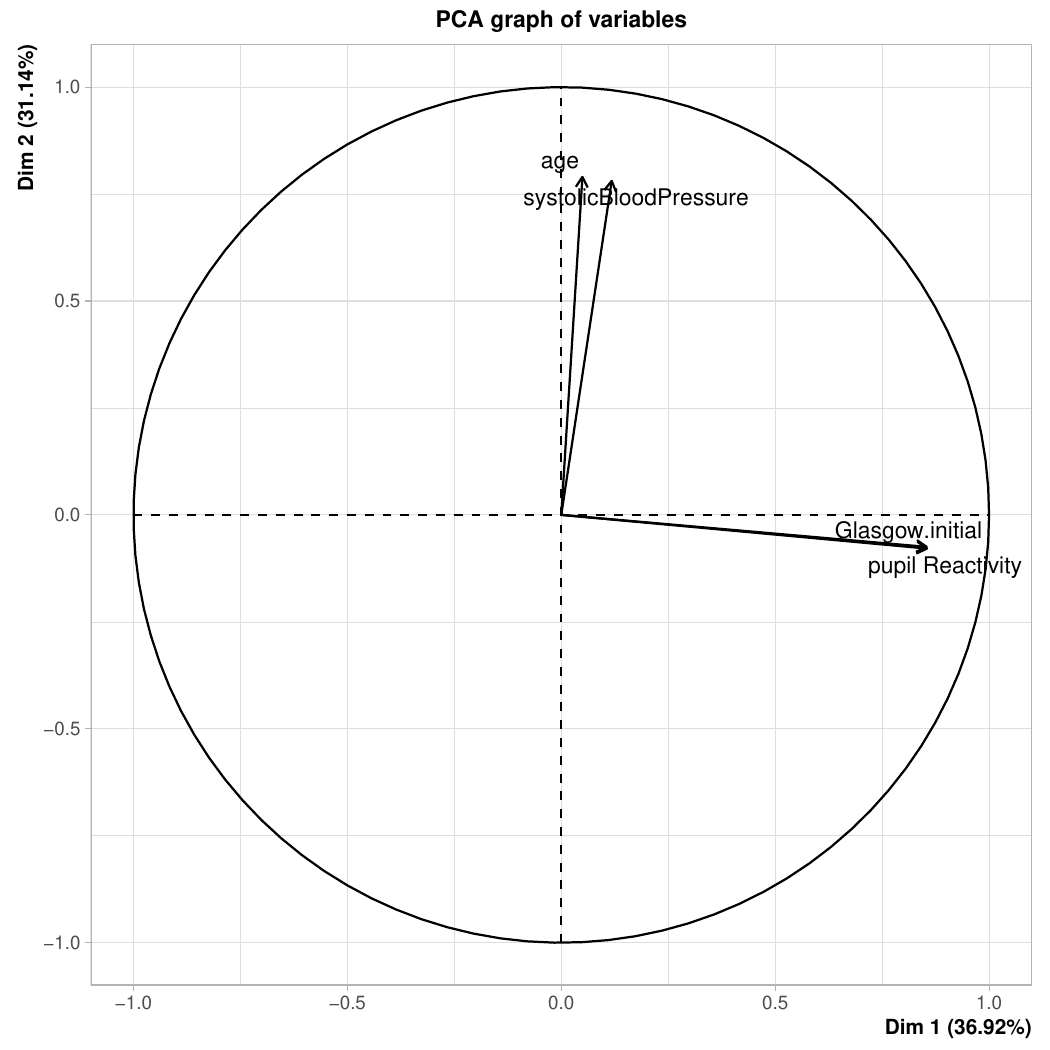}
    \caption{\textbf{Principal Components Analysis (PCA)} of the data set combining CRASH-3 and Traumabase data.}
    \label{fig:pca-dataanalysis}
\end{figure}

A principal component analysis is performed on the combined data set for the Traumabase and the CRASH-3 data using the \texttt{FactoMineR} package \citep{FactoMineR}, results are presented on Figure \ref{fig:pca-dataanalysis}. As expected the Glasgow coma scale score and the pupils reactivity are related (paralysis of the cranial nerves leading to pupillary anomalies being closely related to the presence of an intracranial lesion, itself linked to the state of consciousness encoded in the Glasgow.). Additionally, the link between age and systolic blood pressure can be explained by the fact that atherosclerosis of the arteries is the source of an increase in blood pressure and is related to age.

\subsubsection{\sout{Sampling propensity scores}\modiff{Conditional odds}}

\begin{figure}[H]
    \centering
    \includegraphics[width = 13cm]{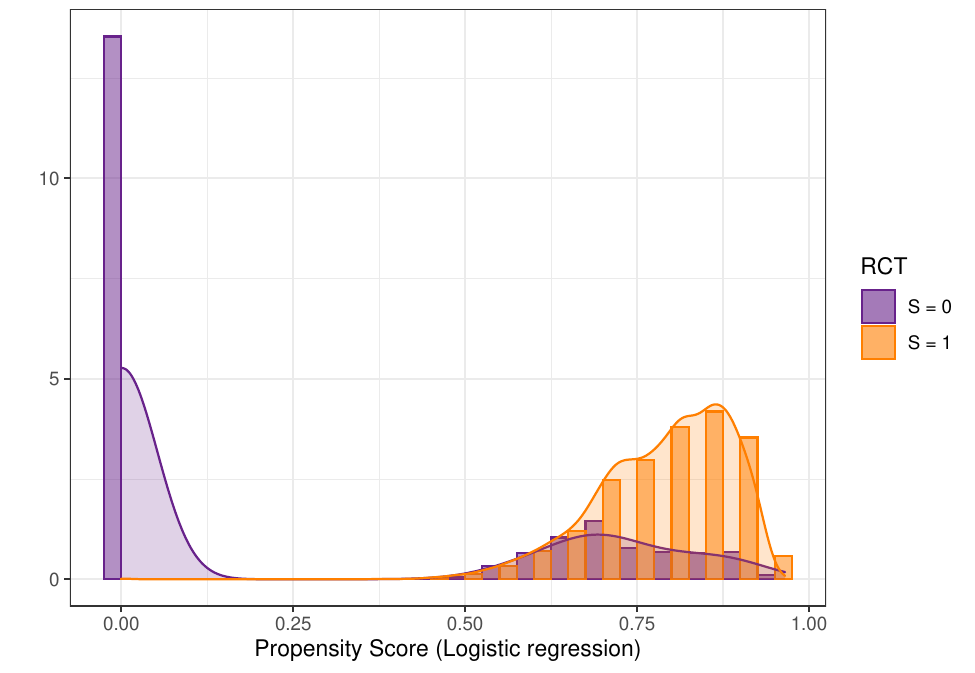}
    \caption{\textbf{Conditional odds histogram (\texttt{glm})} obtained with the \texttt{misaem} R package.}
    \label{fig:propensityscores_glm}
\end{figure}

\begin{figure}[H]
    \centering
    \includegraphics[width = 13cm]{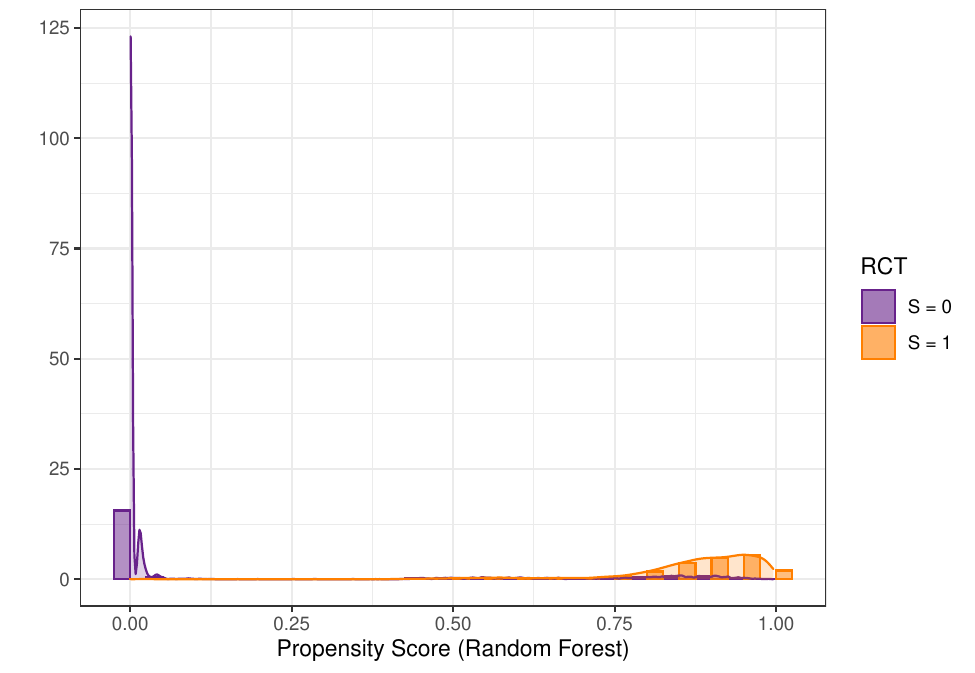}
    \caption{\textbf{Conditional odds histogram (\texttt{grf})} obtained with random forests.}
    \label{fig:propensityscores_grf}
\end{figure}

\begin{figure}[H]
    \centering
    \includegraphics[width = 13cm]{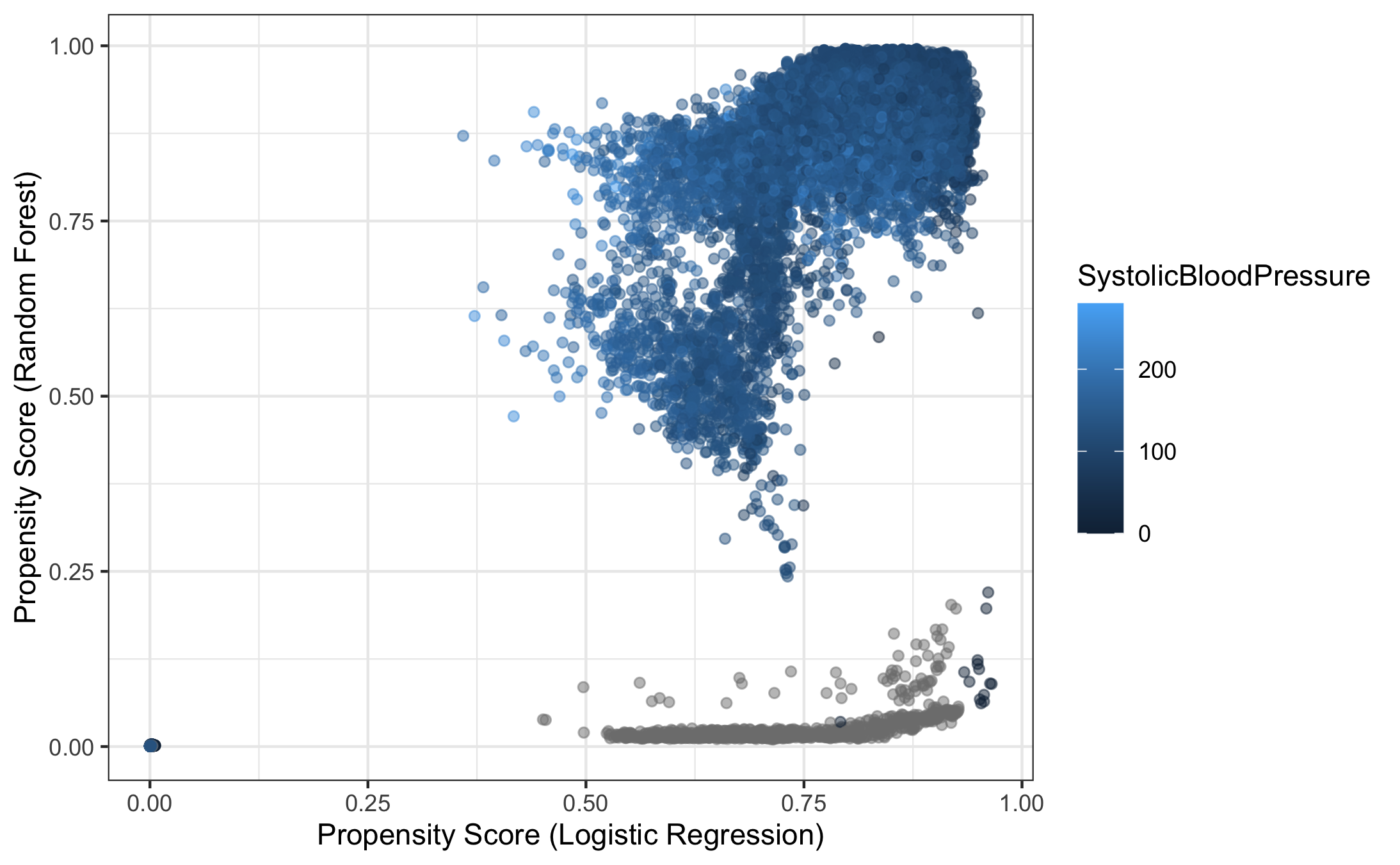}
    \caption{\textbf{Scatter plot of the two conditional odds obtained} with \texttt{glm} in x-axis and \texttt{grf} in the y-axis. Color is set according to the systolic blood pressure covariate values (while missing values are in grey).}
    \label{fig:scatter_plot_propensity_score}
\end{figure}

The \sout{sampling propensity scores}\modiff{conditional odds} obtained while performing the generalization from the CRASH-3 patients to the observational data are presented on Figures~\ref{fig:propensityscores_glm} (logistic regression) and \ref{fig:propensityscores_grf} (forest). We observe that extreme coefficient values are obtained, and that the forest \texttt{grf} strengthens this trend. We can further investigate the differences in between the two methods to infer the propensity scores noticing that the forest method uses the \texttt{NAs} from the Traumabase to learn the propensity scores model. Figure~\ref{fig:scatter_plot_propensity_score} shows that the \texttt{NAs} present in the systolic blood pressure covariate are used by the random forest to predict $S$, leading to more extreme values at the end. This importance of different missing values patterns when combining two data sets are of importance and highlight the need for a better understanding of the impact of missing values in the present framework. 

\subsection{Evidence on other patient strata} \label{app:trauma_strata}

The data analysis part only focuses on all the patients from the two studies CRASH-3 and Traumabase. This part proposes a focus on different patients type, based on the severity of the brain trauma (measured either with the Glasgow score or the pupils reactivity). 

\subsubsection{Traumabase: evidence on different strata}

When stratifying along different criteria of severity as in the CRASH-3 study, namely pupil reactivity and the Glasgow Coma Scale as illustrated in Table~\ref{tab:tbi_txa_results_full} with Mild/moderate and Severe strata, the two studies provide different evidence: no average treatment effect in any of the strata for the Traumabase, while the CRASH-3 study finds a beneficial effect for mild forms of TBI.

\begin{table}[h!] 
\caption{\textbf{ATE estimations from the Traumabase} for TBI-related 28-day mortality. Red cells conclude on deteriorating effect, white cells conclude on no effect.}
\label{tab:tbi_txa_results_full}
\centering
\footnotesize
\begin{tabular}{|l|cccc|cc||c|}
\cline{2-8}
\multicolumn{1}{l|}{\multirow{3}{*}{}}     & \multicolumn{4}{c|}{\small Multiple imputation (MICE)} & \multicolumn{2}{c||}{\small MIA} & \multirow{3}{*}{\makecell{ Unad-\\justed\\ATE\\ $\times 10^{2}$}} \\ \cline{2-7}
\multicolumn{1}{l|}{}                      & \multicolumn{2}{c|}{\makecell{\small IPW\\(95\% CI)\\ $\times10^{2}$}} & \multicolumn{2}{c|}{\makecell{\small AIPW\\(95\% CI)\\ $\times10^{2}$}} & \multicolumn{1}{c|}{\makecell{\small IPW\\(95\% CI)\\$\times10^{2}$}} & \makecell{\small AIPW\\(95\% CI)\\$\times10^{2}$}                    &                   \\ \cline{2-5}
\multicolumn{1}{l|}{}                      & \multicolumn{1}{c|}{GLM} & \multicolumn{1}{c|}{GRF} & \multicolumn{1}{c|}{GLM} & \multicolumn{1}{c|}{GRF} &         \multicolumn{1}{c|}{}        &                &   \\ \hline
\makecell[l]{Total\\ ($n=8248$)} & \multicolumn{1}{l|}{\cellcolor{red!05}\makecell{\small15\\(6.8, 23)}} & \multicolumn{1}{l|}{\cellcolor{red!05}\makecell{\small11\\(6.0, 16)}}  & \multicolumn{1}{l|}{\makecell{\small3.4\\(-9.0, 16)}} & \makecell{\small-0.1\\(-4.7, 4.4)} & \multicolumn{1}{l|}{\cellcolor{red!05}\makecell{\small9.3\\(4.0, 15)}} & \makecell{\small-0.4\\(-5.2, 4.4)}                      &  \cellcolor{red!05}  16               \\ \hline
\makecell[l]{Mild/moderate\\($GCS>8$,\\$n=5228$)} & \multicolumn{1}{l|}{\makecell{\small17\\(-7.9, 42)}} & \multicolumn{1}{l|}{\cellcolor{red!05}\makecell{\small11\\(3.3,18)}} & \multicolumn{1}{l|}{\makecell{\small15\\(-47, 77)}} & \makecell{\small2.1\\(-8.5, 13)}  &  \multicolumn{1}{l|}{\cellcolor{red!05}\makecell{\small6.8\\(2.6, 11)}} & \makecell{\small-0.1\\(-4.9, 4.7)}                      &  \cellcolor{red!05} 8.7                \\ \hline
\makecell[l]{Severe\\($GCS\leq 8$,\\$n=2855$)} & \multicolumn{1}{l|}{\makecell{\small10\\(-7.0, 27)}} & \multicolumn{1}{l|}{\makecell{\small7.7\\(-6.6, 22)}} & \multicolumn{1}{l|}{\makecell{\small2.2\\(-14, 18)}}&  \makecell{\small-1.3\\(-14, 11)} & \multicolumn{1}{l|}{\makecell{\small7.1\\(-1.0, 15)}} & \makecell{\small-0.3\\(-4.6, 4.0)}        &      \cellcolor{red!05}  9.5           \\ \hline 
\end{tabular}
\end{table}

\subsubsection{CRASH-3: evidence on different strata}

The CRASH-3 trial presents a significant treatment effect only on some strata (in particular on less severe injured patients). As the brain-injury gravity has an effect on the outcome---most patients
with TBI with a GCS score of 3 (corresponding to a coma or
unconsciousness state) and those with bilateral non-reactive pupils have
a very poor prognosis regardless of treatment---, the treatment effect is
likely to be biased towards the null. Therefore the CRASH-3 authors
observe the maximal treatment effect and statistical strength on mild to
moderate injured patients, which is what we retrieve from the data. This evidence is computed from the data, with a link between the risk ratio (RR) and the average treatment effect (ATE) on Table~\ref{tab:crash3-results}.

\begin{table}[h!]
\begin{center}
\caption{\textbf{Results reproduction for CRASH-3}, with four possible stratifications based on the gravity level of the injury. Results are both presented as risk ratio (in accordance with \cite{crash32019}) and as ATE (in accordance with our framework, Section \ref{sec:basic setup}).}
\label{tab:crash3-results}
\begin{tabular}{l|c|c||c|c|}
\cline{2-5}
\multirow{2}{*}{}      & \multicolumn{2}{c||}{Relative risk} & \multicolumn{2}{c|}{Average Treatment Effect}       \\
    &   RR    &  95\% CI & ATE & 95\% CI  \\ \hline
\multicolumn{1}{|l|}{Total (within 3 hours)} &  0.94  &   (0.855, 1.02)    & -0.12 & ($-$0.28, 0.004) \\ \hline
\multicolumn{1}{|l|}{$GCS>3$ or at least 1 pupil reacts} &   0.90    &   (0.78, 1.01) & -0.02 & ($-$0.03, 0.0005) \\ \hline
\multicolumn{1}{|l|}{Mild/moderate ($GCS>8$)} &  0.78   &   (0.59, 0.98)  & $-$0.2 & ($-$0.03, $-$0.003)   \\ \hline
\multicolumn{1}{|l|}{Severe ($GCS\leq8$)} &    0.99   &    (0.91, 1.07)   & $-$0.004 & ($-$0.04, 0.03)   \\ \hline
\multicolumn{1}{|l|}{Both pupils react} &   0.87    &   (0.74, 1.00) & -0.015 & ($-$0.03, $-$0.001) \\ \hline
\end{tabular}
\end{center}
\end{table}

\begin{table}[h!]
\begin{center}
\caption{\textbf{Sample sizes for both studies} and different strata along the Glasgow Coma Scale. \#maj.Ex corresponds to the number of patients with a major extracranial bleeding.}
\footnotesize
\label{tab:sample_sizes_full}
\begin{tabular}{l|c|c|c|c||c|c|c|c|}
\cline{2-9}
\multirow{2}{*}{}      & \multicolumn{4}{c||}{Traumabase} & \multicolumn{4}{c|}{CRASH-3}       \\
    &   m    &   \#treated    &  \#death  & \#maj.Ex    & n & \#treated &   \#death & \#maj.Ex \\ \hline
    
\multicolumn{1}{|l|}{Total (within 3 hours)} &  8248  &   683    &   1411  & 5583 & 9168 & 4632 & 1745 & 5 \\ \hline

\multicolumn{1}{|l|}{Mild/moderate ($GCS>8$)} &   5456   &   535   &    527 & 3392 & 5844 & 3075 & 600 & 0 \\ \hline

\multicolumn{1}{|l|}{Severe ($GCS\leq8$)} &   3083    &   596    &      1322 & 2224 & 3717 & 1985 & 1601 & 5 \\ \hline
\end{tabular}
\end{center}
\end{table}

\subsubsection{Generalizing treatment effect on patient strata}\label{sec:limitations-crash-tb}
As found by the CRASH-3 study, the group with potential benefit from TXA seems to be mild to moderate TBI patients (Table \ref{sec:basic setup}), defined as patients with a Glasgow Coma Scale between 9 and 15, while the evidence obtained from the Traumabase has not found a significant treatment effect for this group. However, in this stratum, for the CRASH-3 study, none of the patients has major extracranial bleeding, leading to a constant variable for this group. Conversely, in the Traumabase, in this stratum, only four patients without major extracranial bleeding are treated (while 1867 are not treated with TXA). Since the practitioners are interested in the treatment effect transported on patients with mild to moderate TBI and with major extracranial bleeding, we cannot restrict the target population to those patients without major extracranial bleeding. The current methodology does not allow to satisfy the necessary assumptions for transporting the effect using the presented estimation strategies and defining a clinically relevant target population. Further methodological investigations are required to transport the effect on the stratified subpopulations (see Table~\ref{tab:sample_sizes_full} for the corresponding sample sizes).  

\begin{figure}[H]
\centering
    \includegraphics[width=.8\textwidth]{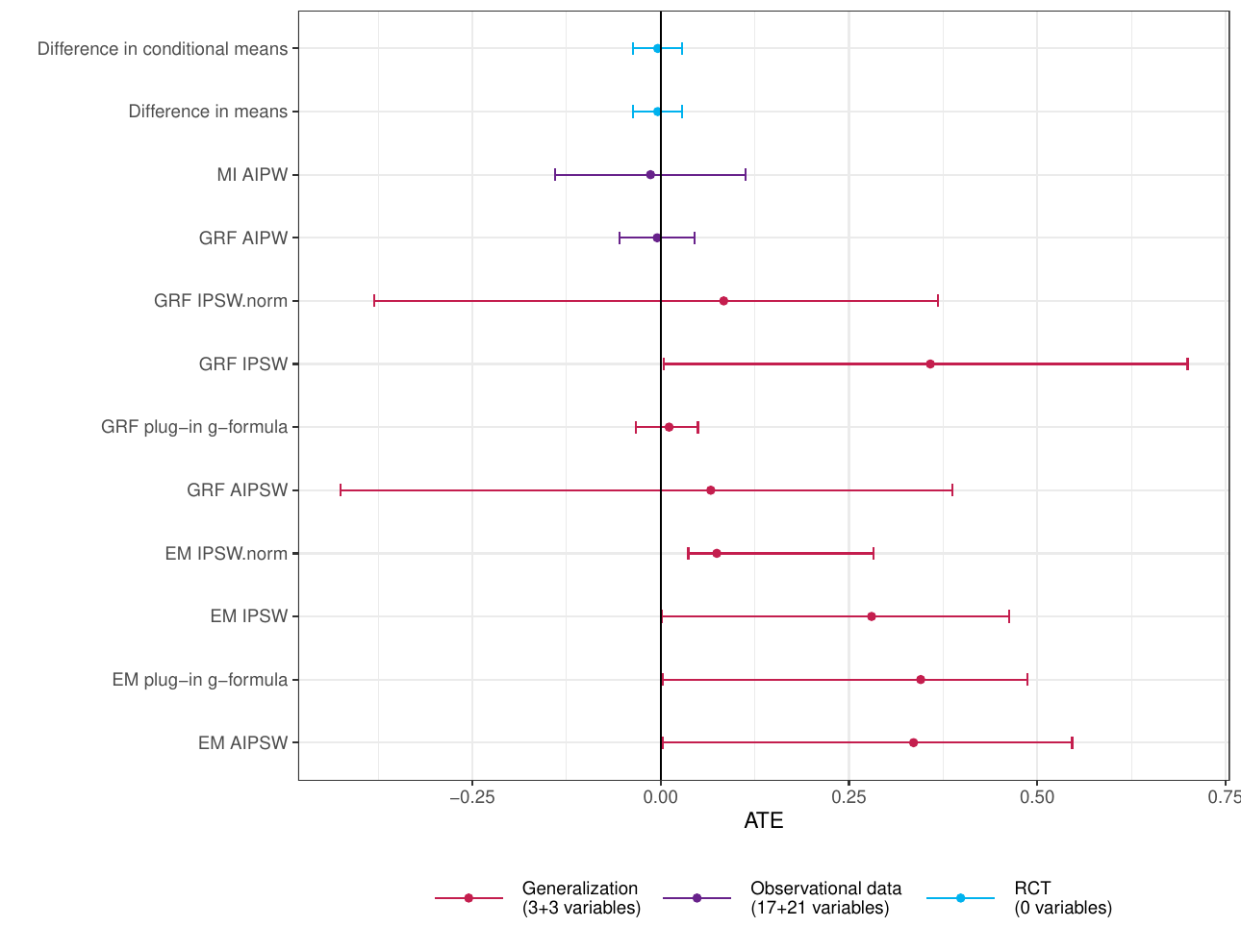}
    \caption{\textbf{Juxtaposition of different estimation results for target population corresponding to the severe Traumabase patients} with ATE estimators computed on the Traumabase (observational data set), on the CRASH-3 trial (RCT), and transported from CRASH-3 to the Traumabase target population (severe TBI patients). Number of variables used in each context is given in the legend.}
    \label{fig:results_crash3_traumabase_severe}
\end{figure}

This issue does not apply to the complementary stratum of severe TBI patients (corresponding to a low Glasgow score, $GCS\leq8$). We can thus provide the results for this stratum in Figure~\ref{fig:results_crash3_traumabase_severe}. We observe that on this strata discrepancies between the solely Traumabase estimators and the generalized estimators are presents. The generalization supports either no-effect or a deleterious effect, while the RCT and the observational estimators support the no-effect hypothesis.

\include{sections/notation}

\end{document}